\documentclass[preprint,11pt]{elsarticle}

\usepackage[letterpaper, margin=1.in]{geometry}
\usepackage{amsmath,amsthm,amsbsy,amsfonts,amssymb}
\usepackage{graphicx}
\usepackage{xcolor}
\usepackage{mathtools}
\usepackage{subfiles}
\usepackage{blindtext}
\usepackage{tikz-cd}

\usepackage{hyperref}
\hypersetup{bookmarks=true,colorlinks=true, linkcolor=blue,urlcolor=red,linktoc=all}
\usepackage{listings}

\usepackage{tcolorbox}
\usepackage{blkarray}
\usepackage{multirow}
\usepackage{numprint}
\usepackage[]{algorithm2e}
\usepackage{stmaryrd}
\usepackage{comment}
\usepackage{siunitx}

\usepackage[utf8x]{inputenc}
\usepackage{amsmath}
\usepackage{bbm}
\usepackage{comment}
\usepackage{color}
\usepackage{graphicx}
\graphicspath{{figs/}}
\usepackage{caption}
\usepackage{subcaption}
\usepackage{tikz}
\usepackage{bm}
\usepackage{cancel}

\usepackage{pgfplots}
\usepackage{pgfplotstable}
\usepackage{filecontents}
\pgfplotsset{compat=1.12}

\usetikzlibrary{shapes,arrows}

\newcommand{\ZJ}[1]{{\textcolor{blue}{\footnotesize\sf[ZJ: #1]}}}
\newcommand{\KL}[1]{{\textcolor{green}{\footnotesize\sf~[KL: #1]}}}
\newcommand{\XT}[1]{{\textcolor{purple}{\footnotesize\sf[XT: #1]}}}

\def\V{\mathbf V}
\def\E{\mathbf E}
\def\j{\mathbf j}
\def\B{\mathbf B}

\DeclareMathAlphabet\mathbfcal{OMS}{cmsy}{b}{n}

\newtheorem{prop}{Proposition}
\newtheorem{remark}{Remark}

\usepackage{enumitem}
\usepackage{pifont}
\usepackage{amsthm}
\newtheorem{theorem}{Theorem}[section]
\newtheorem{lemma}[theorem]{Lemma}
\tikzset{
    cross/.pic = {
    \draw[rotate = 45] (-#1,0) -- (#1,0);
    \draw[rotate = 45] (0,-#1) -- (0, #1);
    }
}
\pgfplotsset{compat=1.16}

\begin{document}

\begin{frontmatter}

\title{A mimetic finite difference based quasi-static magnetohydrodynamic solver for force-free plasmas in tokamak disruptions}

\author[inst1]{Zakariae Jorti\fnref{LANLThanks}}

\affiliation[inst1]{organization={Theoretical Division, Los Alamos National Laboratory},
            city={Los Alamos},
            postcode={87545}, 
            state={New Mexico},
            country={U.S.A.}}

\author[inst1]{Qi Tang\fnref{LANLThanks}}
\author[inst1]{Konstantin Lipnikov\fnref{LANLThanks}}
\author[inst1]{Xian-Zhu Tang\fnref{LANLThanks}}

\fntext[LANLThanks]{This work was jointly supported by the
  U.S. Department of Energy through the Fusion Theory Program of the
  Office of Fusion Energy Sciences and the SciDAC partnership on
  Tokamak Disruption Simulation between the Office of Fusion Energy
  Sciences and the Office of Advanced Scientific Computing.  
  It was also partially supported by Mathematical Multifaceted Integrated 
  Capability Center (MMICC) of Advanced Scientific Computing Research. Los
  Alamos National Laboratory is operated by Triad National Security,
  LLC, for the National Nuclear Security Administration of
  U.S. Department of Energy (Contract No. 89233218CNA000001).  
  }

\begin{abstract}
  Force-free plasmas are a good approximation where the plasma
  pressure is tiny compared with the magnetic pressure, which is the
  case during the cold vertical displacement event (VDE) of a major
  disruption in a tokamak.  On time scales long compared with the
  transit time of Alfv\'{e}n waves, the evolution of a force-free
  plasma is most efficiently described by the quasi-static
  magnetohydrodynamic (MHD) model, which ignores the plasma inertia.
  Here we consider a regularized quasi-static MHD model for force-free
  plasmas in tokamak disruptions and propose a mimetic finite
  difference (MFD) algorithm. The full geometry of an ITER-like
  tokamak reactor is treated, with a blanket module region, a vacuum
  vessel region, and the plasma region.  Specifically, we develop a parallel, fully
  implicit, and scalable MFD solver based on PETSc and its DMStag data
  structure for the discretization of the five-field quasi-static
  perpendicular plasma dynamics model on a 3D structured mesh. The MFD
  spatial discretization is coupled with a fully implicit DIRK
  scheme. The algorithm exactly preserves the divergence-free
  condition of the magnetic field under the resistive Ohm’s law. The
  preconditioner employed is a four-level fieldsplit preconditioner,
  which is created by combining separate preconditioners for
  individual fields, that calls multigrid or direct solvers for
  sub-blocks or exact factorization on the separate fields. The
  numerical results confirm the divergence-free constraint is strongly
  satisfied and demonstrate the performance of the fieldsplit
  preconditioner and overall algorithm. The simulation of ITER VDE
  cases over the actual plasma current diffusion time is also
  presented.
\end{abstract}



\end{frontmatter}

\clearpage
\tableofcontents


\section{Introduction}
\label{sec:intro}

The macroscopic plasma motion can be described by the plasma momentum
equation in the magnetohydrodynamic (MHD) model~\cite{Goedbloed-Poedts-MHD-2004}
\begin{align}
\rho \left(\frac{\partial {\V}}{\partial t} +
     {\V}\cdot\nabla{\V}\right) = {\j}\times{\B} - \nabla p - \nabla\cdot\boldsymbol{\pi}, \label{eq:momentum}
\end{align}
where $\rho$ is the plasma mass density,
${\V}$ the flow field,
$p$ the plasma pressure,
$\boldsymbol{\pi}$ the viscosity tensor,
${\j}\times{\B}$ the Lorentz force with ${\B}$ the magnetic field,
and $\mu_0 {\j} = \nabla\times{\B}$ the plasma current
with $\mu_0$ the vacuum magnetic permeability.
A large class of practical problems considers
low-beta plasmas in which the plasma pressure is low compared with the
magnetic pressure. 
The limiting case is $\nabla p \approx 0,$ so after
a possible transient period over which the plasma flow is damped by
viscosity, the plasma will settle into a steady state that is
approximately force-free,
\begin{align}
{\j} \times {\B} = 0, \label{eq:force-free-constraint}
\end{align}
and ${\V} \approx 0$ (hence $\boldsymbol{\pi}\approx 0$). This is a~\emph{force-free plasma} 
where the plasma current supports a force-free
magnetic field satisfying Eq.~(\ref{eq:force-free-constraint}).  The
general solution to Eq.~(\ref{eq:force-free-constraint}) is~\cite{ck-apj-1957}
\begin{align}
{\j} = \lambda {\B}. \label{eq:force-free-field}
\end{align}
A quasi-neutral plasma has $\nabla\cdot{\j}=0.$ Combining this
constraint with Eq.~(\ref{eq:force-free-field}), and making use of
$\nabla\cdot{\B}=0,$ one finds
\begin{align}
{\B}\cdot\nabla\lambda = 0, \label{eq:BdotLambda=0}
\end{align}
which says that $\lambda$ is a constant along a magnetic field line.
If the magnetic field is integrable with an irrational winding
number~\cite{boozer-rmp-2004}, $\lambda$ will be a function of the
flux surface label $\psi,$ i.e., $\lambda(\psi).$ This can be the case in
which the magnetic field obeys geometrical symmetry, say the toroidal
symmetry in a tokamak. For a three-dimensional magnetic field with no geometrical symmetry,
the magnetic field line is generally non-integrable, so the stochastic
field line fills an ergodic sub-volume in space, for which
Eq.~(\ref{eq:BdotLambda=0}) implies a constant $\lambda.$ If the
magnetic field produces globally stochastic field lines, one would
have $\lambda$ a global constant.

It is of interest to note that ${\j}=\lambda {\B}$ with $\lambda$ a
global constant is the celebrated Taylor state, which is the minimum
energy state for a zero-beta plasma under the constraint of conserved
magnetic
helicity~\cite{woltjer-pnas-1958,taylor-prl-1974,taylor-rmp-1986}. The
connection between magnetic field line topology and $\lambda$ indeed
underlies the dynamical processes by which the so-called magnetic or
Taylor relaxation is realized (see Ref.~\cite{tb-pop-2004b} for instance). Namely, when the
plasma becomes unstable to macroscopic MHD instabilities, flux
surfaces would be broken so that the magnetic field lines can become globally
stochastic, which then relax $\lambda$ to be a global
constant. For weak stochasticity, even small perpendicular current
associated with the plasma inertia, can produce significant modulation
in $\lambda$ along the magnetic field line due to the
Pfirsch-Schl\"{u}ter effect~\cite{tb-pop-2003}. As the MHD
instabilities die down and the flux surfaces reheal, like those in
laboratory confinement experiments with reversed field
pinch~\cite{Bonfiglio-etal-prl-2013}, spheromaks~\cite{tb-pop-2008},
spherical tokamak~\cite{tb-pop-2005a,tb-pop-2006a,tb-pop-2007}, and
tokamak disruptions~\cite{Izzo-pop-2021}, or in the solar
corona~\cite{Wiegelmann-Sakurai-LRSP-2012} and radio
lobes~\cite{tang-apj-2008}, one can end up with a force-free plasma
with $\lambda(\psi)$ a function of the flux surface label $\psi,$ also know
as a nonlinear force-free magnetic
field~\cite{Pevtsov-etal-ApjL-1994,Regnier-etal-AA-2002,DeRosa-etal-AA-2009,Valori-SolarPhysics-2012},
which despite the deviation from the constant $\lambda$ Taylor state,
retains the key feature of magnetic self-organization via a resonant
coupling between the helicity injection source and global magnetic
configuration~\cite{tb-prl-2005a,tb-prl-2005b,tang-prl-2007}. Further
evolution of such a nonlinear force-free plasma, for example, a
post-thermal-quench tokamak plasma undergoing a cold vertical
displacement event (VDE), is governed by the slow transport process,
namely the resistive decay of the plasma current. By slow, we refer
to a time scale much longer than the transit time of the Alfv\'{e}n
waves, which sets the time scale for the zero-beta plasma to
re-establish force-balance. This condition is easily satisfied in a
plasma with Lundquist number much greater than unity, which applies to
most cases of practical interest.
It is of interest to note that during the cold VDE after the plasma thermal quench,
the Ohmic heating power by a decaying plasma current, is mostly balanced by
radiative cooling, so the zero-beta or force-free plasma remains a good approximation
throughout the VDE~\cite{McDevitt_2023}.

The drastic time scale separation between Ohmic decay of the plasma
current and re-establishment of force-balance in a zero-beta plasma by
Alfv\'{e}n waves, suggests the utility of a quasi-static force-free
description of the plasma dynamics on transport time scale as opposed
to the full Alfv\'{e}n wave dynamics. 
 The most obvious and naive form is
\begin{align}
  \left(\nabla\times{\B}\right)\times{\B} & = 0 \label{eq:ff-default} \\
  \frac{\partial {\B}}{\partial t} & = - \nabla\times{\E} \label{eq:faraday-law} \\
  {\E} & = -{\V}\times{\B} + \frac{\eta}{\mu_0} \nabla\times{\B} \label{eq:Ohms-law}
\end{align}
with $\eta$ the plasma resistivity.
As shown later in~\ref{sec:nonuniqueness} as well as by many others 
(see~\cite{priest2012solar} for detailed discussions on force-free fields in astrophysics), 
 this model however is
not well posed without a proper regularization. The solution to such a
quasi-static model has long benefited from a MHD relaxation method
that Chodura and Schl\"{u}ter first introduced to find a 3D MHD
equilibrium~\cite{Chodura-Schluter-JCP-1981}.  The idea is to
introduce a fictitious drag coefficient $\epsilon$ so the
force-balance equation is regularized as
\begin{align}
\epsilon {\V} = \frac{1}{\mu_0}\left(\nabla\times{\B}\right)\times{\B} - \nabla p. \label{eq:ff-relaxation}
\end{align}
This is to be solved in combination with the induction equation and the ideal Ohm's law
\begin{align}
  \frac{\partial {\B}}{\partial t} & = - \nabla\times{\E} \\
  {\E} & = -{\V}\times{\B}
\end{align}
and the constraint $\nabla\cdot{\B} = 0.$ Setting $\nabla p =0$ would
recover the force-free magnetic field as ${\V}$ is damped to zero by
the fictictious drag toward a steady state.  This ends up to be a
popular numerical model for studying force-free coronal magnetic
fields in solar
physics.~\cite{mikic-mcclymont-1994,McClymont-etal-SolarPhysics-1997,Roumeliotis-ApJ-1996,valori-etal-AA-2005}

Recent interests in the quasi-static MHD model for magnetic
confinement fusion have focused on the tokamak disruption modeling,
particularly the force-free evolution of a cold VDE after the initial
thermal quench. The basic physics is that once the thermal quench
drives the plasma beta to approximately zero, the vertical
force-balance of the plasma column is lost on the time scale that the
vertical position control coil current can be adjusted to provide the counter-acting vertical magnetic fields. The vertical
displacement of the entire current-carrying plasma column is driven by the
Ohmic current decay rate, the physics of which is described by a
finite $\eta$ in Eq.~(\ref{eq:Ohms-law}). This is a problem quite
amendable for a quasi-static treatment, and the MHD relaxation method
of Chodura and Schl\"{u}ter has been employed by replacing
Eq.~(\ref{eq:ff-default}) with Eq.~(\ref{eq:ff-relaxation}) in
Ref.~\cite{zakharov-li-pop-2015} and~\cite{kiramov-breizman-pop-2018}.

Here we revisit the formulation of the quasi-static force-free MHD
model and its numerical solution by implicit time
stepping. Specifically we will analyze the role of the fictitious
drag force in the regularization of the model and its physical
implication on the solution.  This will be followed by proposing an alternative
regularization of the model, which we find to have a more
straightforward physics interpretation.  In terms of spatial
discretization, a focus is on ensuring the divergence-free constraint
of the magnetic field, as failure to do so is known to spoil the
numerical solution~\cite{NicolaidesWang,JIANG1996104}.  Various 
approaches have been proposed in the literature, such as starting from
existing convergent discretizations and enhancing their numerical
accuracy by introducing appropriate divergence-free reconstructions
\cite{BALSARA20095040}, or employing divergence-free methods, like
divergence-free Discontinuous Galerkin methods \cite{COCKBURN2004588},
and stable finite element method \cite{HuMaXu}.
A particularly relevant and common approach in the compressible MHD literature is 
the constrained transport approach originally proposed in~\cite{evans1988simulation}, 
where a staggered formulation of the electric and magnetic fields is proposed 
to create specific mimetic finite difference operators that
result in a magnetic field which is divergence-free regarding a specific discrete divergence operator. 
The recent work attempts to generalize it to a non-staggered formulation such as 
finite volume~\cite{rossmanith2006unstaggered}, high-order finite difference~\cite{christlieb2014finite, christlieb2016high},
and the extension to a mapped curvilinear grid~\cite{christlieb2018high}.
There also exists abundant literature on spatial discretizations for the MHD
equations, see \cite{Schtzau2004MixedFE, CHACON2004143,
  SOVINEC2004355, SHADID20161, BADIA2013399, tang2022adaptive} and the references therein.

For the linear solver part, there has been a lot of research work
focusing on the design of efficient and scalable preconditioners for
solving linear systems stemming from MHD models.  Ref.~\cite{Cyr}
 proposes scalable block preconditioners for Newton-Krylov
solver that rely on the approximate block factorization (ABF) approach
and the recursive approximation of the Schur complement.
Ref.~\cite{Chacon2008, tang2022adaptive} follow the same line of thinking by employing the
physical-based ABF technique to devise a preconditioner for
fully-implicit Newton-Krylov solvers.  By expressing the matrix in
blocks corresponding to different unknowns, and approximating the
resulting Schur complements, they carry out a parabolization which
transforms ill-conditioned hyperbolic systems that are difficult to
solve into well-conditioned diagonally dominant parabolic operators
for which multigrid (MG) methods perform very well.  More recently, a
new family of recursive block LU preconditioners is proposed
in~\cite{BADIA2014562} for solving the thermally coupled inductionless
MHD equations, whereas in~\cite{MaHuHuXu} robust block
preconditioners, which satisfy the divergence-free condition exactly
when used in Krylov iterative methods, are developed for the
structure-preserving discretization of the incompressible MHD system.

In this paper, we use a Mimetic Finite Difference (MFD) method~\cite{LIPNIKOV20141163} for the
discretization of a quasi-static perpendicular plasma dynamics model
on a 3D structured grid.  There is a rich literature on the usage of this
method for solving diffusion
\cite{BrezziLipnikovShashkovSimoncini2007}, advection-diffusion
\cite{CangianiManziniRusso2009, BeiraodaVeigaDroniouManzini2011},
elasticity \cite{BeiraodaVeiga2010}, Stokes
\cite{BeiraodaVeigaGyryaLipnikovManzini2009} and porous media flow
problems \cite{LipnikovMoultonSvyatskiy2008}. As for the
time-integration, we use a fully implicit L-stable second order
diagonally implicit Runge--Kutta (DIRK) scheme \cite{PareschiRusso}. The preconditioner employed in
this article is a four-level block preconditioner, which is
created by combining separate preconditioners for individual fields
(as many splits as fields), that calls multigrid methods or exact
factorization on the separate fields.


The rest of this article is structured as follows.  Section~\ref{sec:model}
introduces the quasi-static perpendicular plasma dynamics model.
Section~\ref{sec:mfd} presents some basic elements of the mimetic
discretization methodology and presents the discrete equations of the
quasi-static model.  The numerical results of the mimetic
finite difference method applied to the quasi-static perpendicular
plasma dynamics model are shown in Section~\ref{sec:numres}.
It is followed by the conclusion section of Section~\ref{sec:conc}.
Some mathematical aspects of the models, such as well-posedness 
and the energy dissipation law, are included in the appendix.

\section{Quasi-static force-free MHD model~\label{sec:model}}

\subsection{Regularized quasi-static force-free model\label{sec:regularization}}

The concept of a quasi-static force-free MHD model is based upon the
idea that any force imbalance introduced by the resistive decay of
plasma current in the Ohm's law of Eq.~(\ref{eq:faraday-law}), is
quickly removed by the Alfv\'{e}n wave dynamics, so at any given instance,
the plasma is approximately in a force-free state.  By dropping the
plasma inertia in Eq.~(\ref{eq:ff-default}), the Alfv\'{e}n wave
dynamics is deliberately removed, so maintaining a force-free
magnetic field comes from the solution of the perpendicular flow
\begin{align}
  {\V}_\perp \equiv {\V} - {\V}\cdot{\B}/|{\B}|
\end{align}
from
Eq.~(\ref{eq:Ohms-law}), subjected to the force-free constraint of
Eq.~(\ref{eq:ff-default}). The Faraday's law,
Eq.~(\ref{eq:faraday-law}), connects the constraint of
Eq.~(\ref{eq:ff-default}) to the solution of ${\V}_\perp$ from
Eq.~(\ref{eq:Ohms-law}). In other words, the Ohm's law is the equation
from which ${\V}_\perp$ is solved.  One obvious implication of this is
that since this is a time-dependent partial differential equation with
a constraint, one would need to solve the time-dependent equations with
implicit time stepping.

A more subtle implication, although well-known in the constrained
optimization problem such as saddle point problems~\cite{benzi2005numerical, bochev2005finite},
is the need for regularization. In the specific
case of the quasi-static force-free MHD model, the coupled system of
Eqs.~(\ref{eq:ff-default}, \ref{eq:faraday-law}, \ref{eq:Ohms-law})
has a null space in the solution of ${\V}_\perp.$ One can see this by
noting that if ${\V}_{\perp}$ is a solution, then ${\V}_{\perp} +
\delta{\V}_{\perp}$ is also a solution with
\begin{align}
\delta {\V}_{\perp} = - \frac{\nabla\varphi\times{\B}}{B^2} \label{eq:Delta-V}
\end{align}
for any $\varphi$ that satisfies
\begin{align}
{\B}\cdot\nabla\varphi = 0.
\end{align}
The underlying physics is that the electrostatic field, which can be
written as $-\nabla\varphi,$ does not contribute to $\nabla\times{\E}$
and hence has no effect on magnetic field evolution.  This results in
a degeneracy of the mathematical formulation that cannot be inverted
for $\V_{\perp}$ that is required in quasi-static evolution.
The formal derivation of such a null space is given in~\ref{sec:nonuniqueness}.

To gain insights into how the degeneracy can be removed,
we introduce an explicit treatment of the electrostatic potential $\Phi$
via a Helmholtz decomposition of the electric field,
\begin{align}
{\E} = - \nabla\Phi + \nabla\times \mathbf{h} = -\nabla\Phi + \boldsymbol{\tau}. 
\end{align}
Substituting this form of ${\E}$
into Eq.~\eqref{eq:Ohms-law}, we find
\begin{align}
  \boldsymbol{\tau} = \nabla\Phi - {\V}_{\perp}\times {\B}
  + \frac{\eta}{\mu_0} \nabla\times{\B}. \label{eq:tau-def}
\end{align}
The Faraday's law, Eq.~\eqref{eq:faraday-law}, is now rewritten as
\begin{align}
\frac{\partial{\B}}{\partial t} =  - \nabla\times\boldsymbol{\tau}.
\end{align}
The final step is to come up with another equation to solve for $\Phi,$
which can be done in the usual way by taking the divergence of the electric field,
\begin{align}
  \nabla^2\Phi = - \nabla\cdot\left[
    - {\V}_{\perp}\times {\B} + \frac{\eta}{\mu_0} \nabla\times{\B}
    \right] \label{eq:qsp-Phi}
\end{align}
The boundary condition for $\Phi,$ in the case of ITER configuration, is simply
\begin{align}
\Phi |_{\partial\Omega} = 0, \label{eq:qsp-Phi-bdy}
\end{align}
where the boundary $\partial\Omega$ is the outer vacuum vessel wall,
which is assumed to be perfectly conducting on the time scale of a major
disruption.  So in all, we add one more unknown ($\Phi$) and one
additional equation, Eq.~\eqref{eq:qsp-Phi}, with its boundary
condition, Eq.~\eqref{eq:qsp-Phi-bdy}, to the quasi-static force-free
model.

A special null space in an axisymmetric tokamak plasma is particularly relevant. 
For an axisymmetric tokamak plasma with flux surface label $\psi,$ a
pure radial electric field
\begin{align}
{\E} = - \nabla \varphi(\psi) = -
\frac{\partial\varphi}{\partial\psi}\nabla\psi \label{eq:radial-E-field}
\end{align}
is unconstrained by Eq.~(\ref{eq:qsp-Phi}).  To see this, one can
substitute ${\V}_\perp+\delta {\V}_\perp$ with $\delta{\V}_\perp$ given in
Eq.~(\ref{eq:Delta-V}) having $\varphi=\varphi(\psi),$ for ${\V}_\perp$ in
Eq.~(\ref{eq:tau-def}) and Eq.~(\ref{eq:qsp-Phi}). The result is
\begin{align}
\boldsymbol{\tau} & = \nabla\left(\Phi - \varphi(\psi)\right) - {\V}_{\perp}\times {\B}
+ \frac{\eta}{\mu_0} \nabla\times{\B} \\
  \nabla^2\left(\Phi - \varphi(\psi)\right) & = - \nabla\cdot\left[
    - {\V}_{\perp}\times {\B} + \frac{\eta}{\mu_0} \nabla\times{\B}
    \right], 
\end{align}
which says that $\Phi-\varphi(\psi)$ is also a solution if $\Phi$ is a
solution.  This null space for the solution of ${\V}_\perp$ or
degeneracy of the force-free model with respect to a pure radial
electric field of the form in Eq.~(\ref{eq:radial-E-field}) needs to be
removed for numerical computation. The fictitious viscous drag in
Eq.~(\ref{eq:ff-relaxation}), first introduced by Chodura and
Schl\"{u}ter~\cite{Chodura-Schluter-JCP-1981}, precisely provides such
a regularization. Specifically the value of the fictitious viscous
drag coefficient $\epsilon$ picks a particular $\varphi(\psi).$ In other
words, the regularization of the quasi-static force-free model sets a
radial electrical field that is not constrained by the MHD model.
Although this peculiarity does not affect the force-free magnetic
field during the quasi-static evolution, one needs to be aware of the
regularization-induced radial electric field $\varphi(\psi).$ For
example, should one be interested in advancing the particle motion
using the electromagnetic field from this regularized quasi-static
model, the component of the pure radial electric field should be
removed as it is not {\em physically} constrained in the MHD model.

This artificialness in radial electric field motivates a more careful
look at the widely used $\epsilon {\V}$ regularization approach.  The
physical origin of the collisional damping of the flow field in single
fluid MHD, which is to be mimicked by $\epsilon {\V},$ is not collisional
friction, but the viscosity in Eq.~(\ref{eq:momentum}). This suggests that a more
physically sound regularization is to simply retain the viscosity while
ignoring the inertia, so
\begin{align}
\left(\nabla\times{\B}\right)\times{\B} = - \mu_0 \nabla\cdot\boldsymbol{\pi}.
\end{align}
For simplicity, we adopt the approximate form
\begin{align}
  \nabla\cdot\boldsymbol{\pi} = \nu \nabla^2{\V},
\end{align}
so the alternatively regularized force-free constraint is
\begin{align}
\left(\nabla\times{\B}\right)\times{\B} = - \mu_0 \nu \nabla^2{\V}.\label{eq:viscous-regularize}
\end{align}
The formal derivation of the regularization term removing the null space is given in~\ref{sec:uniqueness}.
In addition, note that an energy dissipation law can be derived with this regularization term, 
see the derivation in~\ref{sec:magNRJ}.  
If ones desires, part of the plasma inertia, which is quadratic
in ${\V},$ can be retained as well,
\begin{align}
\left(\nabla\times{\B}\right)\times{\B} = \mu_0 \nu \nabla^2{\V} + \rho {\V}\cdot\nabla{\V}.\label{eq:viscous-inertial-reg}
\end{align}
For ${\V}$ with small amplitude, which scales with $\eta,$ the quadratic inertia term
has very little effect.
Finally, we should note that the force-free MHD model does not constrain ${\V}\cdot{\B}$ at all,
the regularized quasi-static force-free MHD model is supplemented with the constraint,
\begin{align}
{\B}\cdot{\V} = 0.
\end{align}

In summary, we consider two forms of regularized quasi-static force-free MHD model in the current work.
One uses the fictitious drag of Chodura and Schl\"{u}ter for regularization,
\begin{align}
\left(\nabla\times{\B}\right)\times{\B} = \epsilon {\V},\label{eq:drag-regularize}
\end{align}
while the other invokes the viscous damping regularization,
Eq.~(\ref{eq:viscous-regularize}) or
Eq.~(\ref{eq:viscous-inertial-reg}).  One of these regularized
force-balance
equations~(\ref{eq:viscous-regularize}, \ref{eq:viscous-inertial-reg}, \ref{eq:drag-regularize})
will be solved in tandem with
\begin{align}
  \frac{\partial{\B}}{\partial t} & = - \nabla \times \boldsymbol{\tau}, \\
  \boldsymbol{\tau} & = \nabla\Phi - {\V}\times{\B} + \frac{\eta}{\mu_0}\nabla\times{\B}, \\
  \nabla^2\Phi & = - \nabla\cdot\left[
    - {\V}\times {\B} + \frac{\eta}{\mu_0} \nabla\times{\B}
    \right], \\
  {\V}\cdot{\B} & = 0.
\end{align}

\subsection{Coupling to ITER blankets and vacuum vessel}
\label{sec:wall_model}

\begin{figure}[ht]
\begin{center}
\includegraphics[trim={10cm 0 1cm 0},clip, width=0.5\textwidth]{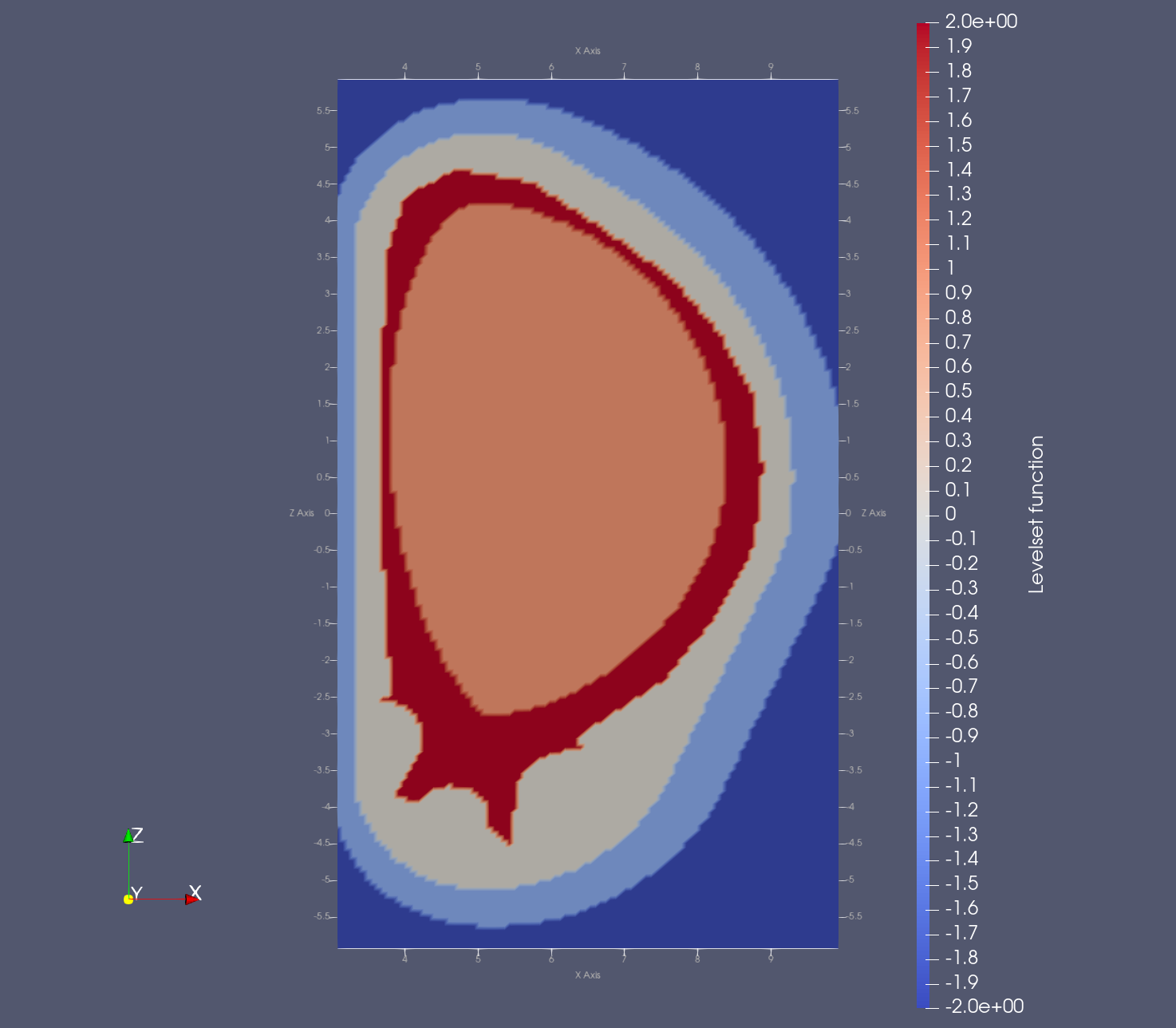}
\caption{ITER tokamak sub-domains in the poloidal plane. The level-set function shown in the color-map indicates different physical domains: 1 in the inner area of the plasma chamber, 2 between the separatrix and the wall, 0 at the rigid wall, -1 at the vacuum vessel and -2 outside the vacuum vessel. The above configuration is used in the simulations of the numerical section.} \label{fig:5layer}
\end{center}
\end{figure}

The plasma in the ITER tokamak reactor is enclosed by a chamber wall,
behind which are blanket modules secured on a stainless steel vacuum
vessel.  See Figure~\ref{fig:5layer} for the ITER's poloidal cross section. 
The vacuum vessel (in light blue) is continuous toroidally and poloidally, so
it is a good flux conserver with a wall time of about \SI{500}{\milli\second}. The
blanket modules (in ivory white) are attached to the vacuum vessel, and they are
constructed and arranged in such a way that a net toroidal current is
impeded. To a reasonable approximation, we will approximate the
entire vacuum vessel as toroidally symmetric conductor with a constant
resistivity, so that the wall time is \SI{500}{\milli\second}. This simplification ignores
the neutron shielding materials embedded in the vacuum vessel.
The electromagnetic field is evolved inside the vacuum vessel with the
standard Ohm's law of constant resistivity,
\begin{align}
  \frac{\partial {\B}}{\partial t} & = - \nabla\times{\E},\\
  {\E} & = \eta_{\rm vv} {\j} = \frac{\eta_{\rm vv}}{\mu_0} \nabla\times{\B}. 
\end{align}
Since the wall time depends on inductance as well, so we numerically compute the
current decay time in the vacuum vessel and match the \SI{500}{\milli\second} wall time to an effective
resistivity $\eta_{\rm vv}$ for ITER's vacuum vessel.


For the first wall and blanket module section, we will deploy an anisotropic
resistivity that has the toroidal resistivity $\eta_t$ much greater than the
poloidal resistivity $\eta_p,$ so
\begin{align}
  \frac{\partial {\B}}{\partial t} & = - \nabla\times{\E}, \label{eqn:eq34} \\
       {\E} & =
  \frac{\eta_t}{\mu_0} \left(\nabla\times{\B}\right)_\phi +
  \frac{\eta_p}{\mu_0} \left[\nabla\times{\B} -
    \left(\nabla\times{\B}\right)_\phi\right]   \label{eqn:eq35}
\end{align}
with $()_\phi$ denoting the toroidal component.  The ratio of
$\eta_t$ and $\eta_p$ is chosen so that the toroidal current in the
blanket is suppressed and the halo current can flow poloidally in the
blankets to enter the vacuum vessel, where the electrical current can
have a strong toroidal component.

\subsection{Quasi-static perpendicular plasma dynamics model and its interface conditions}
For tokamak simulations, we consider the cylindrical coordinate of $(R,\phi,Z)$ for a direct mapping from the Cartesian coordinate $(x, y, z)$.  The structured staggered mesh under the cylindrical coordinate is  used in the current work.
The tokamak computational domain is $\Omega =  [R_{\min}, R_{\max}]\times [0, 2\pi]\times [Z_{\min}, Z_{\max}]$, which can be decomposed into two sub-domains: 
\begin{align*}
    \Omega := \Omega^{P} \cup \Omega^{W},
\end{align*}
where $\Omega^{P}$ corresponds to the tokamak’s plasma chamber whereas $\Omega^{W}$ includes the rigid wall region, the vacuum vessel and the area outside it (see Figure~\ref{fig:5layer} for details: $\Omega^{P}$ comprises the areas where the level-set function is positive, $\Omega^{W}$ corresponds to non-positive values of the level-set function). We use $\Gamma^{PW}$ to denote the interface between the two subdomains $\Omega^{P}$ and $\Omega^{W}$.
Note that in $\Omega^{W}$, there is no plasma and thus the plasma density ($n$) and velocity equations become
\begin{align*}
\frac{\partial n}{\partial t} = & ~ 0 \qquad \text{in} \quad \Omega^{W}, \\
  {\V}_\perp = & ~ \mathbf{0} \qquad \text{in} \quad \Omega^{W},
\end{align*}
while the fields satisfy the diffusion equation as discussed in Section~\ref{sec:wall_model}.

In the sequel, we consider the model introduced in
Section~\ref{sec:regularization} for the plasma region $\Omega^P$ with
some slight adjustments:
\begin{itemize}
    \item When compared to the perpendicular component, the parallel velocity component can be considered as negligible. Therefore, it is neglected in the density equation and not solved in the velocity equation.  
    \item For stabilization purposes, a viscosity term is added to the velocity equations in $R$ and $Z$ directions (see~\ref{sec:nonuniqueness}, \ref{sec:uniqueness} and~\ref{sec:magNRJ} for some discussions on its impact).
    \item The resistivity is assumed to be a constant in each sub-domain. 
\end{itemize}

An interface condition is needed for such a multi-domain interface problem.
For the fields, the following jump conditions should be naturally satisfied,   
\begin{alignat}{3}
  &[{\B} \cdot \mathbf{n}]=0 \qquad  &\text{on} \quad \Gamma^{PW} , \label{eqn:edge-face-continuity1} \\
  &[{\E} \times \mathbf{n}]=\mathbf{0}  \qquad & \text{on} \quad \Gamma^{PW}, \label{eqn:edge-face-continuity2} 
\\
  &\V_{i\perp} = \mathbf{0} \qquad & \text{on} \quad \Gamma^{PW}, \label{eqn:vertex-continuity2}
\end{alignat}
where $[\cdot]$ stands for the jump operator along the interface. 
These conditions are consistent with the absence of surface charge/current and the fact that there is no jump in the electrical field.

 Finally, the multi-domain quasi-static plasma model along with its interface condition that is considered in the current work is summarized as follows: 
\begin{itemize}
\item In the plasma region:
\begin{align}
  \frac{\partial n}{\partial t} + \nabla\cdot\left(n{\V}_\perp\right) = & ~0 \qquad \text{in} \quad  \Omega^{P}, \label{eq:4.565} \\
  -\left( \nu n_0 m_i \nabla^2 \V_{i\perp} + \dfrac{1}{\mu_0} (\nabla\times{\B})\times{\B} \right) \cdot \mathbf{e}_R= & ~0 \qquad \text{in} \quad \Omega^{P} , \label{eq:4.566} \\
  {\V}_\perp \cdot {\B} = & ~ 0 \qquad \text{in} \quad \Omega^{P}, \label{eq:4.567} \\
  -\left( \nu n_0 m_i \nabla^2 \V_{i\perp} + \dfrac{1}{\mu_0} (\nabla\times{\B})\times{\B} \right) \cdot \mathbf{e}_Z = & ~ 0 \qquad \text{in} \quad \Omega^{P}, \label{eq:4.568} \\
  - \nabla^2 \Phi - \nabla\cdot\left[ - {\V}_\perp\times {\B} + \frac{\eta}{\mu_0}\left(\nabla\times{\B}\right) \right] = & ~ 0 \qquad \text{in} \quad \Omega^{P}, \label{eq:4.569} \\
  \boldsymbol{\tau} - \nabla\Phi + {\V}_\perp\times {\B} - \frac{\eta}{\mu_0}\left(\nabla\times{\B}\right) = & ~ \mathbf{0} \qquad \text{in} \quad \Omega^{P}, \label{eq:4.570} \\
  \frac{\partial{\B}}{\partial t} + \nabla\times\boldsymbol{\tau} = & ~ \mathbf{0} \qquad \text{in} \quad \Omega^{P}. \label{eq:4.571}
\end{align}
\item In the wall region:
\begin{align}
\frac{\partial n}{\partial t} = & ~ 0 \qquad \text{in} \quad \Omega^{W}, \label{eq:4.572} \\
  {\V}_\perp = & ~ \mathbf{0} \qquad \text{in} \quad \Omega^{W} , \label{eq:4.573} \\
  - \nabla^2 \Phi - \nabla\cdot\left[ \frac{\eta}{\mu_0}\left(\nabla\times{\B}\right) \right] = & ~ 0 \qquad \text{in} \quad \Omega^{W} \label{eq:4.574} \\
  \boldsymbol{\tau} - \nabla\Phi - \frac{\eta}{\mu_0}\left(\nabla\times{\B}\right) = & ~ \mathbf{0} \qquad \text{in} \quad \Omega^{W}, \label{eq:4.575} \\
  \frac{\partial{\B}}{\partial t} + \nabla\times\boldsymbol{\tau} = & ~ \mathbf{0} \qquad \text{in} \quad \Omega^{W} . \label{eq:4.576}
\end{align}
\item At the wall/plasma interface:
\begin{alignat}{3}
  &{\V}_\perp = ~ 0 \qquad &&  \text{on} \quad  \Gamma^{PW} ,\label{eq:4.577} \\
  &\boldsymbol{\tau} \times \mathbf{n} \qquad  &&  \text{continous across} \quad \Gamma^{PW},  \label{eq:4.579} \\
  &{\B} \cdot \mathbf{n} \qquad && \text{continous across} \quad \Gamma^{PW} . \label{eq:4.578} 
\end{alignat}
\item At the outer rectangular boundary $\partial \Omega$:
\begin{alignat}{3}
   \Phi & = ~ {0},  \label{eq:4.580} \\
   \boldsymbol{\tau} \times \mathbf{n} & = ~ \mathbf{0} .  \label{eq:4.580tau}  
\end{alignat}
\end{itemize}
Note that Equations~\eqref{eq:4.565} and ~\eqref{eq:4.572} express the plasma density continuity equation and that the field of $\boldsymbol{\tau}$ and $\Phi$ are considered instead of the electric field due to the reason described in Section~\ref{sec:regularization}.

\def\hh{\hspace*{-3mm}}
\def\DIV{{\mathrm{Div}}_h}
\def\DIVt{\widetilde{\mathrm{Div}}_h}
\def\GRAD{{\mathrm{Grad}}_h}
\def\GRADt{\widetilde{\mathrm{Grad}}_h}
\def\CURL{{\mathrm{Curl}}_h}
\def\CURLt{\widetilde{\mathrm{Curl}}_h}

\section{Mimetic finite difference discretization}
\label{sec:mfd}
This section focuses on the MFD formulation and the full multi-domain quasi-static plasma model.
We consider a structured orthogonal mesh in 3D with hexahedral cells $c$ that form a subdivision of the computation domain. Let $|c|$, $|f|$ and $|e|$ be the volume of cell $c$, the area of face $f$ and the length of edges $e$, respectively.  
We use $\mathcal{N}_h$, $\mathcal{E}_h$, $\mathcal{F}_h$ and
$\mathcal{C}_h$ to denote the discrete node, edge, face and cell spaces,
respectively.  The MFD employed here is a staggered mesh method; the
velocity and electrostatic potential unknowns are defined at mesh
vertices ($\mathbf{V}_h, \Phi_h \in \mathcal{N}_h$), the electric
field unknowns are defined on mesh edges ($\boldsymbol{\tau}_h \in
\mathcal{E}_h$), the magnetic field unknowns are defined on mesh faces
($\mathbf{B}_h \in \mathcal{F}_h$) while the ion number density
unknowns are defined on mesh elements (${n_i}_h \in \mathcal{C}_h$).
We first introduce the primary and dual mimetic operators acting between the discrete spaces:
$$
\begin{array}{cccccccc}
\mathcal{N}_h & \longrightarrow & \mathcal{E}_h & \longrightarrow & \mathcal{F}_h & \longrightarrow & \mathcal{C}_h\\[-0.5ex]
& \hh\GRAD\hh && \hh\CURL\hh && \hh\DIV\hh & \\[1.8ex]
\mathcal{N}_h & \longleftarrow & \mathcal{E}_h & \longleftarrow & \mathcal{F}_h & \longleftarrow & \mathcal{C}_h \\[-0.5ex]
& \hh\DIVt\hh && \hh\CURLt\hh && \hh\GRADt\hh &
\end{array}
$$
and then use them to discretize
the quasi-static perpendicular plasma dynamics model.

\subsection{Primary and derived mimetic operators}
The MFD framework approximates  first-order operators using
coordinate-invariant formulas. We write the Stokes theorem for a finite-size
mesh object, either cell $c$ or face $f$ or edge $e$. For instance,
$$
  \frac{1}{|c|} \int_c \nabla\cdot \mathbf{B} \, {\rm d} V
  = \frac{1}{|c|} \oint_{\partial c} \mathbf{B} \cdot \mathbf{n}\, {\rm d} S,
$$
where $\mathbf{n}$ is the unit normal vector to the surface $\partial c$. 
This gives us the following definition of the primary mimetic divergence operator
$$
{\mathrm{Div}_h} \mathbf{B}_h := \frac{1}{|c|} \sum\limits_{f \in \partial c} \alpha_{c,f} |f|\, B_f,
  \qquad
  \text{where } B_f := \frac{1}{|f|} \int_f \mathbf{B}\cdot \mathbf{n}\, {\rm d} S.
$$
Here $|c|$ is the cell volume that has different formula in different coordinate systems.
Similarly, $|f|$ is the face area. $\alpha_{c,f}$ is the orientation factor, $\pm 1$.


The discrete gradient and curl operators are defined in a similar fashion:
$$
  {\mathrm{Grad}_h} v_h := \frac{1}{|e|} \int_e \nabla v\, {\rm d} L
   = \frac{1}{|e|} (v_{2} - v_{1}),
$$
where vertices $v_1$ and $v_2$ are the endpoints of $e$
and
$$
  {\mathrm{Curl}_h} \mathbf{E}_h := \frac{1}{|f|} \int_f \nabla \times \mathbf{E}\, {\rm d} S
   = \frac{1}{|f|} \sum\limits_{e \in \partial f} \alpha_{f,e}\, |e|\, E_e,
  \qquad
  \text{with } E_e := \frac{1}{|e|} \int_e \mathbf{E}\cdot \boldsymbol{\tau}\, {\rm d} L,
$$
where $\boldsymbol{\tau}$ is the unit vector tangent to an edge $e$ and $\alpha_{f,e}$ is the orientation factor, $\pm 1$.
The three primary mimetic operators satisfy the following discrete identities:
\begin{equation}
  {\mathrm{Div}_h}\, {\mathrm{Curl}_h} = 0 
  \quad\mbox{and}\quad
  {\mathrm{Curl}_h}\, {\mathrm{Grad}_h} = 0.
  \label{eqn:prim_disc_ident}
\end{equation}

A set of dual operators can be defined by discrete analogs of integration by parts.
The dual operators preserve the duality property by design.
To simplify the presentation, we consider the Green's formulas in functional spaces 
where the boundary integrals are zeros.
An dual (injective) operator $\widetilde{\mathrm{Div}_h} : \mathcal{E}_h \rightarrow \mathcal{N}_h$ is (uniquely) defined via the 
discrete duality relationship:
$$
  [{\mathrm{Grad}_h}\, p_h,\, \mathbf{v}_h]_{\mathcal{E}_h} = -[p_h,\, \widetilde{\mathrm{Div}_h}\,\mathbf{v}_h]_{\mathcal{N}_h} ,
$$
where the brackets denote inner products 
in the aforementioned discrete spaces. 
The other dual operators $\widetilde{\mathrm{Grad}_h} : \mathcal{C}_h \rightarrow \mathcal{F}_h$ and $\widetilde{\mathrm{Curl}_h} : \mathcal{F}_h \rightarrow \mathcal{E}_h$ are defined similarly:
\begin{align*}
  [{\mathrm{Div}_h}\, \mathbf{u}_h ,\, p_h]_{\mathcal{C}_h} & = -[\mathbf{u}_h,\, \widetilde{\mathrm{Grad}_h}\,p_h]_{\mathcal{F}_h}, \\
  [{\mathrm{Curl}_h}\, \mathbf{u}_h,\, \mathbf{v}_h]_{\mathcal{F}_h} &= [\mathbf{u}_h,\, \widetilde{\mathrm{Curl}_h}\,\mathbf{v}_h]_{\mathcal{E}_h}.
\end{align*}
The dual operators also satisfy discrete identities:
$$
  \widetilde{\mathrm{Div}_h}\, \widetilde{\mathrm{Curl}_h} = 0 
  \quad\mbox{and}\quad
  \widetilde{\mathrm{Curl}_h}\, \widetilde{\mathrm{Grad}_h} = 0.
$$

\subsection{MFD discretization of the quasi-static perpendicular plasma dynamics model}
\label{sec:mfd_discretization}
To discretize the model above, we first need to define some projections and reconstructions between the spaces $\mathcal{N}_h$, $\mathcal{E}_h$, $\mathcal{F}_h$ and $\mathcal{C}_h$. 
For vectors that are discrete representations of continuous functions, integer indices, e.g., $(i,j,k)$, in the three directions correspond to values associated with vertices; integer indices in two directions, e.g. $(i,j+\frac{1}{2},k)$ correspond to values associated with edges; integer indices in one direction, e.g. $(i+\frac{1}{2},j,k+\frac{1}{2})$ correspond to values associated with faces; whereas non-integer indices in all directions, e.g. $(i+\frac{1}{2},j+\frac{1}{2},k+\frac{1}{2})$ correspond to values associated with cells. 

 For any vertex-based vector $\mathbf{W}$, face-based vector $\mathbf{X}$, cell-based vector $\mathbf{Y}$ and edge-based vector $\mathbf{Z}$, we define the cell-to-face projection ${\mathcal{P}}_{\mathrm c\rightarrow f}$ by:
\begin{align}
  {\mathcal{P}}_{\mathrm{c\rightarrow f}}(\mathbf{Y})_{i+\frac{1}{2},j+\frac{1}{2},k} = & ~ \frac{1}{2}\left((\mathbf{Y}_{i+\frac{1}{2},j+\frac{1}{2},k+\frac{1}{2}} + \mathbf{Y}_{i+\frac{1}{2},j+\frac{1}{2},k-\frac{1}{2}}) \cdot \mathbf{e}_Z \right) \mathbf{e}_Z, \label{eq:face-av-cell_1} \\
  {\mathcal{P}}_{\mathrm{c\rightarrow f}}(\mathbf{Y})_{i+\frac{1}{2},j,k+\frac{1}{2}} = & ~ \frac{1}{2}\left((\mathbf{Y}_{i+\frac{1}{2},j+\frac{1}{2},k+\frac{1}{2}} + \mathbf{Y}_{i+\frac{1}{2},j-\frac{1}{2},k+\frac{1}{2}})\cdot \mathbf{e}_{\phi} \right) \mathbf{e}_{\phi}, \label{eq:face-av-cell_2} \\
  {\mathcal{P}}_{\mathrm{c\rightarrow f}}(\mathbf{Y})_{i,j+\frac{1}{2},k+\frac{1}{2}} = & ~ \frac{1}{2}\left((\mathbf{Y}_{i+\frac{1}{2},j+\frac{1}{2},k+\frac{1}{2}} + \mathbf{Y}_{i-\frac{1}{2},j+\frac{1}{2},k+\frac{1}{2}}) \cdot \mathbf{e}_R \right) \mathbf{e}_R, \label{eq:face-av-cell_3}
\end{align}
the edge-to-vertex projection ${\mathcal{P}}_{\mathrm{e\rightarrow v}}$ by:
\begin{align}
  {\mathcal{P}}_{\mathrm{e\rightarrow v}}(\mathbf{Z})_{i,j,k} = ~ & \dfrac{1}{2} \left( ( \mathbf{Z}_{i+\frac{1}{2},j,k} + \mathbf{Z}_{i-\frac{1}{2},j,k} ) \cdot \mathbf{e}_{R} \right)\mathbf{e}_{R} + \dfrac{1}{2} \left( (\mathbf{Z}_{i,j-\frac{1}{2},k} + \mathbf{Z}_{i,j+\frac{1}{2},k} ) \cdot   \mathbf{e}_{\phi} \right) \mathbf{e}_{\phi} \nonumber \\
  + ~ & \dfrac{1}{2} \left( (\mathbf{Z}_{i,j,k-\frac{1}{2}} + \mathbf{Z}_{i,j,k+\frac{1}{2}} ) \cdot \mathbf{e}_{Z} \right) \mathbf{e}_{Z}, \label{eq:vertex-av-edge}
\end{align}
the face-to-vertex projection ${\mathcal{P}}_{\mathrm{f\rightarrow v}}$ by:
\begin{align}
  {\mathcal{P}}_{\mathrm{f\rightarrow v}}(\mathbf{X})_{i,j,k} = ~ & \dfrac{1}{4} \left( ( \mathbf{X}_{i,j+\frac{1}{2},k+\frac{1}{2}} + \mathbf{X}_{i,j-\frac{1}{2},k+\frac{1}{2}} + \mathbf{X}_{i,j+\frac{1}{2},k-\frac{1}{2}} + \mathbf{X}_{i,j-\frac{1}{2},k-\frac{1}{2}} ) \cdot \mathbf{e}_{R} \right)\mathbf{e}_{R} \nonumber \\
  + ~ & \dfrac{1}{4} \left( (\mathbf{X}_{i-\frac{1}{2},j,k-\frac{1}{2}} +   \mathbf{X}_{i+\frac{1}{2},j,k-\frac{1}{2}} + \mathbf{X}_{i-\frac{1}{2},j,k+\frac{1}{2}} + \mathbf{X}_{i+\frac{1}{2},j,k+\frac{1}{2}}) \cdot   \mathbf{e}_{\phi} \right) \mathbf{e}_{\phi} \nonumber \\
  + ~ & \dfrac{1}{4} \left( (\mathbf{X}_{i-\frac{1}{2},j-\frac{1}{2},k} + \mathbf{X}_{i+\frac{1}{2},j-\frac{1}{2},k} + \mathbf{X}_{i-\frac{1}{2},j+\frac{1}{2},k} +   \mathbf{X}_{i+\frac{1}{2},j+\frac{1}{2},k}) \cdot \mathbf{e}_{Z} \right) \mathbf{e}_{Z}, \label{eq:vertex-av-face}
\end{align}
the vertex-to-edge reconstruction ${\mathcal{R}}_{\mathrm{v\rightarrow e}}$ by:
\begin{align}
  {\mathcal{R}}_{\mathrm{v\rightarrow e}}(\mathbf{W})_{i+\frac{1}{2},j,k} = & ~ \frac{1}{2} ~ {\Big (} \mathbf{e}_R \cdot (\mathbf{W}_{i+1,j,k} + \mathbf{W}_{i,j,k}) {\Big )} \mathbf{e}_R \label{eq:edge-av-vertex_1} , \\
    {\mathcal{R}}_{\mathrm{v\rightarrow e}}(\mathbf{W})_{i,j+\frac{1}{2},k} = & ~ \frac{1}{2} ~ {\Big (} \mathbf{e}_{\phi} \cdot (\mathbf{W}_{i,j+1,k} + \mathbf{W}_{i,j,k}) {\Big )} \mathbf{e}_{\phi} \label{eq:edge-av-vertex_2} , \\
      {\mathcal{R}}_{\mathrm{v\rightarrow e}}(\mathbf{W})_{i,j,k+\frac{1}{2}} = & ~ \frac{1}{2} ~ {\Big (} \mathbf{e}_Z \cdot (\mathbf{W}_{i,j,k+1} + \mathbf{W}_{i,j,k}) {\Big )} \mathbf{e}_Z, \label{eq:edge-av-vertex_3}
\end{align}
the vertex-to-face reconstruction ${\mathcal{R}}_{\mathrm{v\rightarrow f}}$ by:
\begin{align}
{\mathcal{R}}_{\mathrm{v\rightarrow f}}(\mathbf{W})_{i+\frac{1}{2},j+\frac{1}{2},k} = ~ \dfrac{1}{4} ~ {\Big (} & \mathbf{e}_Z \cdot ( \mathbf{W}_{i,j,k} + \mathbf{W}_{i,j+1,k} + \mathbf{W}_{i+1,j,k} + \mathbf{W}_{i+1,j+1,k}) {\Big )} \mathbf{e}_Z, \label{eq:face-av-vertex2}
\\
{\mathcal{R}}_{\mathrm{v\rightarrow f}}(\mathbf{W})_{i+\frac{1}{2},j,k+\frac{1}{2}} = ~ \dfrac{1}{4} ~ {\Big (} & \mathbf{e}_{\phi} \cdot (\mathbf{W}_{i,j,k} + \mathbf{W}_{i,j,k+1} + \mathbf{W}_{i+1,j,k} + \mathbf{W}_{i+1,j,k+1}) {\Big )} \mathbf{e}_{\phi} , \label{eq:face-av-vertex3}
\\
{\mathcal{R}}_{\mathrm{v\rightarrow f}}(\mathbf{W})_{i,j+\frac{1}{2},k+\frac{1}{2}} = ~ \dfrac{1}{4} ~ {\Big (} & \mathbf{e}_{R} \cdot ( \mathbf{W}_{i,j,k} + \mathbf{W}_{i,j+1,k} + \mathbf{W}_{i,j,k+1} + \mathbf{W}_{i,j+1,k+1}) {\Big )} \mathbf{e}_{R}. \label{eq:face-av-vertex4}
\end{align}
 Note that all these projection and reconstruction operators are second order accurate for the uniform orthogonal mesh considered.
\begin{remark}
These projection and reconstruction operators could be generalized for non-uniform meshes. 
\end{remark}

We use these projection and reconstruction operators as well as the
mimetic operators to discretize the quasi-static model:
\begin{itemize}
\item In the plasma region:
\begin{align}
  \frac{\partial n_{i,h}}{\partial t} + {\mathrm{Div}_h} \left({\mathcal{P}}_{\mathrm{c\rightarrow f}}({n_{i,h}}) {\mathcal{R}}_{\mathrm{v\rightarrow f}}({{\V}}_{i\perp})\right) = & ~0 \quad \text{in} \quad  \Omega^{P}, \label{eq:4.565_disc} \\
  - \nu n_0 m_i \left( \widetilde{\mathrm{Div}_h} ({\mathrm{Grad}_h} \V_{i\perp,R}) - \dfrac{1}{R^2}\V_{i\perp,R} - \dfrac{2}{R} {\mathcal{P}}_{\mathrm{e\rightarrow v}}( {\mathrm{Grad}_h} \V_{i\perp,\phi} ) \cdot \mathbf{e}_{\phi} \right) \quad & \nonumber \\
  - \left( {\mathcal{P}}_{\mathrm{e\rightarrow v}}( \underline{\widetilde{\mathrm{Curl}}}_h {\mathbf{B}}_h) \times {\mathcal{P}}_{\mathrm{f\rightarrow v}}({\mathbf{B}_h}) \right) \cdot \mathbf{e}_R = & ~0 \quad \text{in} \quad \Omega^{P} , \label{eq:4.566_disc} \\
  {\V}_{i\perp} \cdot {\mathcal{P}}_{\mathrm{f\rightarrow v}}( {{\B}}_h)  = & ~ 0 \quad \text{in} \quad \Omega^{P}, \label{eq:4.567_disc} \\
  - \nu n_0 m_i \left( \widetilde{\mathrm{Div}_h} ({\mathrm{Grad}_h} \V_{i\perp,Z}) \right) - \left( {\mathcal{P}}_{\mathrm{e\rightarrow v}}(\underline{\widetilde{\mathrm{Curl}}}_h {\mathbf{B}}_h) \times {\mathcal{P}}_{\mathrm{f\rightarrow v}}({\mathbf{B}_h}) \right) \cdot \mathbf{e}_Z= & ~ 0 \quad \text{in} \quad \Omega^{P}, \label{eq:4.568_disc} \\
  - \widetilde{\mathrm{Div}_h} ({\mathrm{Grad}_h}({\Phi}_h)) - \widetilde{\mathrm{Div}_h} \left[ - {\mathcal{R}}_{\mathrm{v\rightarrow e}} \left({{\V}}_{i\perp}\times {\mathcal{P}}_{\mathrm{f\rightarrow v}}({{\B}}_h)\right) + \widetilde{\mathrm{Curl}_h}({{\B}}_h) \right] = & ~ 0 \quad \text{in} \quad \Omega^{P}, \label{eq:4.569_disc} \\
  \boldsymbol{\tau}_h - {\mathrm{Grad}_h}({\Phi}_h) + {\mathcal{R}}_{\mathrm{v\rightarrow e}} \left({{\V}}_{i\perp}\times {\mathcal{P}}_{\mathrm{f\rightarrow v}}({{\B}}_h)\right) - \widetilde{\mathrm{Curl}_h}({{\B}}_h) = & ~ \mathbf{0} \quad \text{in} \quad \Omega^{P}, \label{eq:4.570_disc} \\
  \frac{\partial{\B}_h}{\partial t} + {\mathrm{Curl}_h}(\boldsymbol{\tau}_h) = & ~ \mathbf{0} \quad \text{in} \quad \Omega^{P}. \label{eq:4.571_disc}
\end{align}
\item In the wall region:
\begin{align}
\frac{\partial n_{i,h}}{\partial t} = & ~ 0 \quad \text{in} \quad \Omega^{W}, \label{eq:4.572_disc} \\
  {\V}_{i\perp} = & ~ \mathbf{0} \quad \text{in} \quad \Omega^{W} , \label{eq:4.573_disc} \\
  - \widetilde{\mathrm{Div}_h} ({\mathrm{Grad}_h}({\Phi}_h)) - \widetilde{\mathrm{Div}_h} \left[ \widetilde{\mathrm{Curl}_h}({{\B}}_h) \right] = & ~ 0 \quad \text{in} \quad \Omega^{W} \label{eq:4.574_disc} \\
  \boldsymbol{\tau}_h - {\mathrm{Grad}_h}({\Phi}_h) - \widetilde{\mathrm{Curl}_h}({{\B}}_h) = & ~ \mathbf{0} \quad \text{in} \quad \Omega^{W}, \label{eq:4.575_disc} \\
  \frac{\partial{\B}_h}{\partial t} + {\mathrm{Curl}_h}(\boldsymbol{\tau}_h) = & ~ \mathbf{0} \quad \text{in} \quad \Omega^{W} . \label{eq:4.576_disc}
\end{align}
\end{itemize}

 As mentioned earlier, the derived mimetic curl operator ${\widetilde{\mathrm{Curl}_h}}$ used in~\eqref{eq:4.569_disc} and~\eqref{eq:4.570_disc} includes the variable coefficient $\dfrac{\eta}{\mu_0}$ as $K$ in~\eqref{eqn:Kcurl}. On the other hand, the other derived mimetic curl operator $\underline{\widetilde{\mathrm{Curl}}}_h$ used in~\eqref{eq:4.566_disc} and~\eqref{eq:4.568_disc} includes only the constant $\frac{1}{\mu_0}$.
The usage of derived mimetic operators in the discrete model above instead of the combination of projections/reconstructions and primary mimetic operators is motivated by the need to preserve important properties of the continuum problem (see \ref{sec:magNRJ} for magnetic energy dissipation), and also to simplify the Jacobian matrix by cancelling some off-diagonal blocks using discrete identities (see Section~\ref{sec:precsolve}). 

 The MFD discretization of the induction equation \eqref{eq:4.571_disc} and \eqref{eq:4.576_disc} lead to the following property.
 \begin{prop}
 Given a divergence-free initial magnetic field, the magnetic field solved from the above MFD system is divergence-free,
 when the primary divergence operator ${\mathrm{Div}_h}$ is considered.
 \end{prop}
 \noindent To see that, we note that for a discretization such as a backward Euler of the above system, the following identity holds
\begin{align*}
  {\mathrm{Div}_h} (\B_h^{n+1}) & = {\mathrm{Div}_h} (\B_h^n) - \Delta t \, {\mathrm{Div}_h} ( {\mathrm{Curl}_h} (\boldsymbol{\tau}_h^{n+1})) \\
  & =  {\mathrm{Div}_h} (\B_h^n).
\end{align*}
That being said,  the divergence-free property needs to satisfy two constraints: the initial magnetic field is divergence-free and 
the iterative solver has to be accurate enough. 
In practice, however, it is not easy to prepare a set of an initial condition with the magnetic field being divergence-free.
It is also not necessary to solve  the nonlinear system excessively accurately just for the divergence free property. 
An extra corrector stage using the solved $\boldsymbol{\tau}_h^{n+1}$ can be performed in order to avoid the impact of the iterative solver.
Such a strategy is commonly used in the constrained transport approach for ideal MHD~\cite{rossmanith2006unstaggered, christlieb2014finite, christlieb2016high}. 
We have experimented such a strategy but found its impact to the qualify of the solution is very minimal. 

 By the mimetic theory, we have the definitions of the derived divergence and curl operators given by Formulas~\eqref{eqn:DDiv-def} and~\eqref{eqn:DCurl-def} where the mass matrices $\mathbb{M}_n$, $\mathbb{M}_e$ and $\mathbb{M}_f$ are in the spaces of vertex-based, edge-based and face-based unknowns, respectively.
We have to think about the discrete operator ${\mathrm{Curl}_h}$ (resp. ${\mathrm{Grad}_h}$ and ${\mathrm{Div}_h}$) as a rectangular matrix
acting form the space of edge-based (resp. vertex-based and face-based) fields to the space of face-based (resp. edge-based and cell-based) fields.
We will present the detailed formulas for all the pieces in the above equations.

\begin{figure}[ht]
\centering
\includegraphics[trim={1cm 0 0 .5cm},clip, width=10cm]{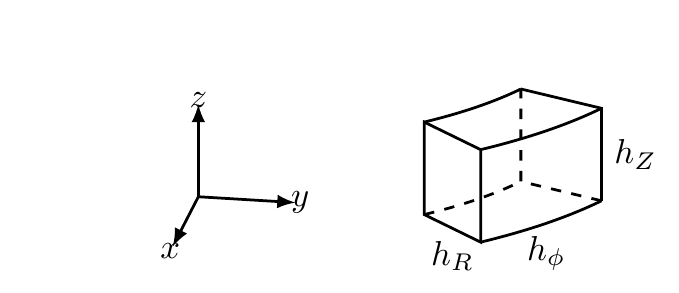}
\caption{A typical computational cell in the configuration space used in the current work.\label{fig:cell3D}} 
\end{figure}

 Consider a cylindrical mesh.
The typical cell $c$ is shown in Figure~\ref{fig:cell3D} and the radius of the cell center is denoted by $R_c$.
 The definition of the primary curl operator for the top face of cell $c$ in Figure~\ref{fig:cell3D} is
$$
  {\mathrm{Curl}_h} \boldsymbol{\tau}_h 
  = \frac{1}{|f|} \sum\limits_{e \in \partial f} \alpha_{f,e}\,|e|\, {\tau}_e
  = \frac{1}{|f|} \left( h_R  \, {\tau}_1 + (R_c+ \frac{h_R}{2}) h_\phi\, {\tau}_2 - h_R  \, {\tau}_3 - (R_c- \frac{h_R}{2}) h_\phi \, {\tau}_4 \right).
$$
The definition of the primary gradient operator for the number $1$ edge in the top face of cell $c$ in Figure~\ref{fig:cell3D} is
$$
  {\mathrm{Grad}_h} {\Phi}_h = \frac{1}{|e|} \left({\Phi}(R_c + \dfrac{h_R}{2}) - {\Phi}(R_c - \dfrac{h_R}{2} ) \right) = \frac{1}{h_R} \left({\Phi}(R_c + \dfrac{h_R}{2}) - {\Phi}(R_c - \dfrac{h_R}{2} ) \right).
$$
The definition of the primary divergence operator for the cell $c$ in Figure~\ref{fig:cell3D} is
\begin{align*}
  &{\mathrm{Div}_h} \mathbf{B}_h = \frac{1}{|c|} \sum\limits_{f \in \partial c} \alpha_{c,f} |f|\, B_f\\ 
  & = \frac{1}{R_ch_R h_z h_\phi} \left( |f_{\rm right}| B_{\rm right} - |f_{\rm left}| B_{\rm left} + |f_{\rm up}| B_{\rm up} - |f_{\rm down}| B_{\rm down} + |f_{\rm front}| B_{\rm front} - |f_{\rm back}| B_{\rm back} \right).
\end{align*}
The mass matrices $\mathbb{M}_n$, $\mathbb{M}_e$ and $\mathbb{M}_f$ are defined similarly. First, we use the
additivity of integration to break the mass matrix into the sum of cell-based
matrices:
$$
  \mathbb{M}_e = \sum\limits_{c} {\cal N}_c \, \mathbb{M}_{e,c} \, {\cal N}_c^T,
$$
where ${\cal N}_c$ is the conventional assembly matrix.
The elemental matrix $\mathbb{M}_{e,c}$ affects the accuracy of the scheme, however, its selection does not break other mimetic properties such as the discrete exact identities.
The MFD framework says that the vector-matrix-vector product should approximate the integral of underlying 3D vector functions.
Recall that material properties (the coefficient $\dfrac{\eta}{\mu_0}$ where the resistivity $\eta$ may vary from a cell to another) are embedded in the derived operator $\widetilde{\mathrm{Curl}_h}$ via mass matrices.
In this case,
$$
  \mathbf{u}_h^T \,\mathbb{M}_{e,c}\, \mathbf{v}_h
  = \frac{\mu_0}{\eta} \int_c \mathbf{u} \cdot \mathbf{v} \, {\rm d} V + O(h) |c|
  = \frac{\mu_0}{\eta} \int_c \mathbf{u} \cdot \mathbf{v} \, r {\rm d} r {\rm d} z {\rm d} \varphi + O(h) |c|.
$$
By taking unitary functions $\mathbf{u}_h$ and $\mathbf{v}_h$ in the formula above, we could select $\mathbb{M}_{e,c}$ the be the diagonal matrix of size $12$, which is the number of edges in a cell, given by: 
\begin{equation}
\mathbb{M}_{e,c} = \mathbb{M}_{e,c}^{(1)} := \frac{\mu_0}{\eta} \frac{|c|}{4} \mathbb{I}_c.
\end{equation} 
Using the same line of thoughts, we conclude that the elemental matrix $\mathbb{M}_{f,c}$ could be also a scalar matrix of size 6, which is the number of faces in a cell, given by: 
\begin{equation}
\mathbb{M}_{f,c} = \frac{|c|}{2} \mathbb{I}_c.
\end{equation} 
Thus, we have for any face-based vector $\mathbf{X}$ and edge-based vector $\mathbf{Y}$: 
\begin{align}
\mathbb{M}_f \mathbf{X} & = \left[ \beta^f_{i} \mathbf{X}_{i} \right]_i ; \quad \beta^f_{i}:= \sum_{c_k \in \mathcal{C}(f_i)} \alpha^f_{c_k} ; \quad \alpha^{f}_{c_k} := \frac{|c_k|}{2} ; \\
\mathbb{M}_e^{(1)} \mathbf{Y} & = {\Big[} \beta^e_{l} \mathbf{Y}_{l} {\Big]}_l ; \quad \beta^e_{l}:= \sum_{c_k \in \mathcal{C}(e_l)} \alpha^e_{c_k} ; \quad \alpha^{e}_{c_k} := \frac{\mu_0}{\eta(c_k)} \frac{|c_k|}{4} ;
\end{align}
where $\mathcal{C}(f_l)$ denotes all cells sharing face $f_l$ and $\mathcal{C}(e_l)$ denotes all cells sharing edge $e_l$. 
Hence, we have all the ingredients to build the discrete derived curl operator $\widetilde{\mathrm{Curl}_h}$. To form the other discrete derived curl operator $\underline{\widetilde{\mathrm{Curl}}}_h$, it suffices to consider the following mass matrix for edge-based vectors:  
\begin{align}
\mathbb{M}_e^{(2)} \mathbf{Y} & = {\Big[} \underline{\beta}^e_{l} \mathbf{Y}_{l} {\Big]}_l ; \quad \underline{\beta}^e_{l} := \sum_{c_k \in \mathcal{C}(e_l)} \underline{\alpha}^e_{c_k} ; \quad \underline{\alpha}^{e}_{c_k} := \mu_0 \frac{|c_k|}{4} ;
\end{align}
As for the discrete derived gradient operator $\widetilde{\mathrm{Div}_h}$, the following mass matrices are needed. For any vertex-based vector $\mathbf{W}$ and edge-based vector $\mathbf{Y}$:  
\begin{align}
\mathbb{M}_n \mathbf{W} & = \left[ \beta^n_{j} \mathbf{W}_{j} \right]_j ; \quad \beta^n_{j} := \sum_{c_k \in \mathcal{C}(n_j)} \alpha^n_{c_k} ; \quad \alpha^{n}_{c_k} := \frac{|c_k|}{8} ; \\
\mathbb{M}_e^{(3)} \mathbf{Y} & = {\Big[} \hat{\beta}^e_{l} \mathbf{Y}_{l} {\Big]}_l ; \quad \hat{\beta}^e_{l} := \sum_{c_k \in \mathcal{C}(e_l)} \hat{\alpha}^e_{c_k}; \quad \hat{\alpha}^{e}_{c_k} := \frac{|c_k|}{4}.
\end{align}



\section{Solver and preconditioning strategy}
\label{sec:precsolve}

This section discusses the details of the solver and block preconditioning strategy. 
In the most outer lever, a nonlinear solver based on Jacobian-free Newton-Krylov (JFNK) and inexact Newton is used. For the inner linear solver, a finite difference coloring Jacobian matrix is formed as a preconditioner 
for the Jacobian matrix inversion. Relying on the finite difference coloring approximated Jacobian turns out to be necessary in this work, since the mimetic formulations, as outlined in the previous section, involve many projections between different bases, and
implementing an analytical Jacobian for such a complicated system is not practical.

The state-of-art approach for inverting such a fully coupled system is to seek an effective preconditioner. 
Here we outline the details of the preconditioning strategy. 
For ease of presentation, continuous operators are used in the following discussion.  
Note that the Jacobian matrix corresponding to the quasi-static perpendicular plasma dynamics model takes the form:

{
\footnotesize
\begin{align}
& \mathbf{J} \, d\mathbf{U} = \nonumber \\
& 
\begin{bmatrix}
  \dfrac{1}{dt}\mathbf{I} + (\nabla \cdot \V_{0}) \mathbf{I} + {\V}_{0} \cdot \nabla & \mathbf{0} & \mathbf{0} & \mathbf{0} & n_{i,0} \nabla \cdot \quad+ (\nabla n_{i,0}) \cdot \quad \smallskip\\
  \mathbf{0} & \mathbf{I} & -\nabla \quad & - \frac{\eta}{\mu_0} (\nabla \times \quad ) + ({\V}_{0} \times \quad ) & - \mathbf{B}_0 \times \smallskip \\
  \mathbf{0} & \mathbf{0} & - \nabla^2 \quad & - \nabla \cdot (\frac{\eta}{\mu_0} \nabla \times \quad) + \nabla \cdot ({\V}_{0} \times \quad) & - \nabla \cdot ( {\B}_0 \times \quad ) \smallskip\\
  \mathbf{0} & \nabla \times \quad & \mathbf{0} & \dfrac{1}{dt}\mathbf{I} & \mathbf{0} \smallskip \\
  \mathbf{0} & \mathbf{0} & \mathbf{0} & \mathbf{C}_2 & \mathbf{C}_1
\end{bmatrix}
\begin{bmatrix}
d n_i \medskip\\
d \boldsymbol{\tau}\medskip\\
d\Phi \bigskip\\
d {\B}\medskip\\
d{\V}_{i\perp}
\end{bmatrix}
\label{eqn:4.581}
\end{align}}
where $\mathbf{C}_1$ and $\mathbf{C}_2$ are different in each domain, given by:
\begin{alignat}{3}
\mathbf{C}_1 := & ~\mathbf{C}_{1,p} \qquad &&\text{in} \quad \Omega^{P}, \label{eq:4.582} \\
\mathbf{C}_1 := & ~\mathbf{C}_{1,w} \qquad &&\text{in} \quad \Omega^{W}, \label{eq:4.583} \\
\mathbf{C}_2 := & ~\mathbf{C}_{2,p} \qquad &&\text{in} \quad \Omega^{P}, \label{eq:4.584} \\
\mathbf{C}_2 := & ~\mathbf{C}_{2,w} \qquad &&\text{in} \quad \Omega^{W}; \label{eq:4.585}
\end{alignat}
and
{\footnotesize
\begin{align}
& \mathbf{C}_{1,p} = ~
\begin{blockarray}{cccc}
d{\V}_r & d{\V}_{\phi} & d{\V}_{z} \medskip\\
\begin{block}{[ccc]c}
  \dfrac{\nu n_0 m_i}{r^2}\mathbf{I} - \nu n_0 m_i \nabla^2 & \dfrac{2 \nu n_0 m_i}{r^2} \dfrac{\partial}{\partial \phi} & \mathbf{0} &  d{\V}_r \medskip\\
  {\mathcal{P}}_{\mathrm{f\rightarrow v}}(\B)_{r,0} \mathbf{I} & {\mathcal{P}}_{\mathrm{f\rightarrow v}}(\B)_{\phi,0} \mathbf{I} & {\mathcal{P}}_{\mathrm{f\rightarrow v}}(\B)_{z,0} \mathbf{I} & d{\V}_{\phi} \bigskip\\
  \mathbf{0} & \mathbf{0} & - \nu n_0 m_i \nabla^2 & d{\V}_{z} \\
\end{block}
\end{blockarray} , \\
& \mathbf{C}_{1,w} =  ~ \mathbf{I} , \\
&\mathbf{C}_{2,p} = ~ \\
&\begin{blockarray}{cccc}
d{\B}_r & d{\B}_{\phi} & d{\B}_{z} \medskip\\
\begin{block}{[ccc]c}
  - \dfrac{\B_{\phi,0}}{r}\dfrac{\partial}{\partial \phi} - \B_{z,0} \dfrac{\partial}{\partial z} & \left( \dfrac{\partial \B_{\phi,0}}{\partial r} - \dfrac{1}{r}\dfrac{\partial \B_{r,0}}{\partial \phi} + \dfrac{2 \B_{\phi,0}}{r} \right) \mathbf{I} + \B_{\phi,0} \dfrac{\partial}{\partial r} & \left( \dfrac{\partial \B_{z,0}}{\partial r} - \dfrac{\partial \B_{r,0}}{\partial z} \right) \mathbf{I} + \B_{z,0} \dfrac{\partial}{\partial r} &  d{\V}_r \medskip\\
  \V_{r,0} \mathbf{I} & \V_{\phi,0} \mathbf{I} & \V_{z,0} \mathbf{I} & d{\V}_{\phi} \bigskip\\
  \left( \dfrac{\partial \B_{r,0}}{\partial z} - \dfrac{\partial \B_{z,0}}{\partial r} \right) \mathbf{I} + \B_{r,0} \dfrac{\partial}{\partial z} & \left( \dfrac{\partial \B_{\phi,0}}{\partial z} - \dfrac{1}{r}\dfrac{\partial \B_{z,0}}{\partial \phi} \right) \mathbf{I} + \B_{\phi,0} \dfrac{\partial}{\partial z} & - \dfrac{\B_{\phi,0}}{r}\dfrac{\partial}{\partial \phi} - \B_{r,0} \dfrac{\partial}{\partial r} & d{\V}_{z} \\
\end{block}
\end{blockarray} , \\
& \mathbf{C}_{2,w} =  ~ \mathbf{0}.
\end{align}}
Note that in the mimetic finite difference, $\mathbf{C}_{2,p}$ as well as its sub-blocks are not square matrices, since $\V$ and $\B$ live on different positions. 
We indicate the corresponding locations of sub-blocks in the full Jacobian using the notations of $d{\B}$ and $d{\V}$.

The linear system resulting from the linearization of the PDE is solved iteratively with a FGMRES solver preconditioned by a four-level block preconditioner that uses PETSc's fieldsplit interface. 
Our strategy is inspired from the recent coupled preconditioning work~\cite{joshaghani2019composable} that considered a four-field system.
In the first level, the following preconditioner is used
\begin{align}
  \mathbf{P}^{-1} &= \begin{bmatrix} \mathcal{KSP}(\mathbf{J}_{n_i}) & \mathbf{0} \\ \mathbf{0} & \mathbf{I}_{\boldsymbol{\tau} \Phi {\B} \V} \end{bmatrix}\begin{bmatrix} \mathbf{I}_{n_i} & - \mathbf{C}_3 \\ \mathbf{0} & \mathbf{I}_{\boldsymbol{\tau} \Phi {\B} \V} \end{bmatrix}\begin{bmatrix} \mathbf{I}_{n_i} & \mathbf{0} \\ \mathbf{0} & \mathcal{KSP}(\mathbf{J}_{\boldsymbol{\tau} \Phi {\B} \V}) \end{bmatrix}.
\end{align}
where $\mathbf{C}_{3}$ is the top right off-diagonal block of the Jacobian matrix
\begin{align}
  \mathbf{C}_{3} = 
    \left[\mathbf{0}\qquad  \mathbf{0} \qquad \mathbf{0} \qquad n_{i,0} \nabla \cdot \quad+ (\nabla n_{i,0}) \cdot \quad \right],
\end{align}
and $\mathcal{KSP}$ is used to denote the linear solver from PETSc.
Since the bottom left block of the Jacobian matrix is zero ($\mathbf{J}_{\boldsymbol{\tau} \Phi {\B} \V, n_i} = \mathbf{0}$), such a factorization is~\emph{exact} as long as we can invert $\mathbf{J}_{n_i}$ and $\mathbf{J}_{\boldsymbol{\tau} \Phi {\B} \V}$ exactly. 
Note that this factorization corresponds to the ``multiplicative'' option of fieldsplit in PETSc. 
In particular, we use the following options for the two remaining linear solvers:
\begin{itemize}
\item
$\mathcal{KSP}(\mathbf{J}_{n_i})$ : GMRES solver preconditioned with a block Jacobi preconditioner;
\item
$\mathcal{KSP}(\mathbf{J}_{\boldsymbol{\tau} \Phi {\B} \V})$ : FGMRES solver preconditioned with $\mathbf{P}^{-1}_{\boldsymbol{\tau} \Phi {\B} \V}$.
\end{itemize}
Now the problem is converted to looking for an efficient preconditioner $\mathbf{P}^{-1}_{\boldsymbol{\tau} \Phi {\B} \V}$ for $\mathbf{J}_{\boldsymbol{\tau} \Phi {\B} \V}$,
which is our second level of the block preconditioner. We use the following preconditioner based on the conventional Schur complement
\begin{align}
  \mathbf{P}^{-1}_{\boldsymbol{\tau} \Phi {\B} \V} &= \begin{bmatrix} \mathbf{I}_{\boldsymbol{\tau}} & - \mathbf{J}_{\boldsymbol{\tau}, \Phi {\B} \V} \\ \mathbf{0} & \mathbf{I}_{\Phi {\B} \V} \end{bmatrix}\begin{bmatrix} \mathcal{KSP}(\mathbf{J}_{\boldsymbol{\tau}}) &  \mathbf{0} \\ \mathbf{0} & \mathcal{KSP}(\mathbf{S}_{\Phi {\B} \V}) \end{bmatrix}\begin{bmatrix} \mathbf{I}_{\boldsymbol{\tau}} & \mathbf{0} \\ - \mathbf{J}_{\Phi {\B} \V, \boldsymbol{\tau}} & \mathbf{I}_{\Phi {\B} \V} \end{bmatrix}.
\end{align}
Note that $\mathbf{J}_{\boldsymbol{\tau}}$ can be trivially inverted since $\mathbf{J}_{\boldsymbol{\tau}}$ is an identity matrix. In addition, $\mathbf{J}_{\boldsymbol{\tau}}$ being an identity ensures that the Schur complement $\mathbf{S}_{\Phi {\B} \V}$  can be~\emph{exactly} formed  through a matrix multiplication. In particular, the Schur complement is 
\begin{equation}
\mathbf{S}_{\{ \Phi {\B} {\V}_{i\perp} \}} \, d\mathbf{U} =
\begin{bmatrix}
  - \nabla^2 \quad & - \nabla \cdot (\frac{\eta}{\mu_0} \nabla \times \quad) + \nabla \cdot ({\V}_{0} \times \quad) & - \nabla \cdot ( {\B}_0 \times \quad ) \smallskip\\
  \mathbf{0} & \dfrac{1}{dt}\mathbf{I} + \nabla \times (\frac{\eta}{\mu_0} \nabla \times \quad) - \nabla \times ({\V}_{0} \times \quad) & \nabla \times ( {\B}_0 \times \quad ) \smallskip \\
  \mathbf{0} & \mathbf{C}_2 & \mathbf{C}_1
\end{bmatrix}
\begin{bmatrix}
d\Phi \medskip\\
d {\B}\medskip\\
d{\V}_{i\perp}
\end{bmatrix} .
\label{eqn:EBV_Schur}
\end{equation}
Note that the zero sub-block in $\mathbf{S}_{\{ \Phi {\B} {\V}_{i\perp} \}}$ is a result of the property of the mimetic operator that guarantees the curl of the gradient operator is zero in the discrete level (up to machine precision). 
This shows another advantage of using the MFD discretization. 
Again, the overall problem is converted to finding an efficient preconditioner for $\mathbf{S}_{\{ \Phi {\B} {\V}_{i\perp} \}}$.
In the third level, the following preconditioner is used
\begin{align}
  \mathbf{P}^{-1}_{\Phi {\B} \V} &= \begin{bmatrix} \mathcal{KSP}(\mathbf{J}_{\Phi}) & \mathbf{0} \\ \mathbf{0} & \mathbf{I}_{{\B} \V} \end{bmatrix}\begin{bmatrix} \mathbf{I}_{\Phi} & -\mathbf{C}_4 \\ \mathbf{0} & \mathbf{I}_{{\B} \V} \end{bmatrix}\begin{bmatrix} \mathbf{I}_{\Phi} & \mathbf{0} \\ \mathbf{0} & \mathcal{KSP}(\mathcal{S}) \end{bmatrix},
\end{align}
where 
\begin{align}
  \mathbf{C}_{4} = 
  \left[- \nabla \cdot (\frac{\eta}{\mu_0} \nabla \times \quad) + \nabla \cdot ({\V}_{0} \times \quad) \qquad - \nabla \cdot ( {\B}_0 \times \quad ) \right],
\end{align}
and 
\begin{align}
 \mathcal{S} = 
 \left[
\begin{matrix}
  \dfrac{1}{dt}\mathbf{I} + \nabla \times (\frac{\eta}{\mu_0} \nabla \times \quad) - \nabla \times ({\V}_{0} \times \quad) & \nabla \times ( {\B}_0 \times \quad )  \\
  \mathbf{C}_2 & \mathbf{C}_1  
\end{matrix}\right].
\end{align}
The preconditioner $\mathbf{P}^{-1}_{\Phi {\B} \V}$ corresponds to the multiplicative option of two sub-block solvers, which is again~\emph{exact} since the bottom left block of ${\mathbf{S}}_{\Phi {\B} \V}$ is zero. 
The following options are used for the two remaining linear solvers
\begin{itemize}
\item $\mathcal{KSP}(\mathbf{J}_{\Phi})$ : GMRES solver preconditioned with BoomerAMG preconditioner from the hypre package. Note here that $\mathbf{J}_{\Phi} = - \nabla^2 $;
\item $\mathcal{KSP}(\mathcal{S})$ : A direct solver (SuperLU\_DIST) is used.
\end{itemize}

The advantage of our preconditioning  is that except for the sub-block solvers for $\mathbf{J}_{n_i}$ and $\mathbf{J}_{\Phi}$, the factorization and solver strategy is~\emph{exact}. 
This is an outcome of the proposed preconditioning strategy and the properties of the mimetic operators. In addition, the solver for $\mathbf{J}_{\Phi}$ is scalable due to algebraic multigrid being used. 
Overall, the nonlinear solver and the preconditioner perform very well in practice, which will be demonstrated in the numerical section. 

As expected,  majority of the computational time comes from the inversion of the operator $\mathcal{S}$. The coupling between $\B$ and $\V$ in $\mathcal{S}$ is non-conventional. In particular, we find that there is a strong off-diagonal coupling in this sub-system. Note that the commonly used physics-based preconditioning strategy through parabolization~\cite{Chacon2008, tang2022adaptive}  does not work since this sub-system does not propagate the wave. We leave it to the future study for an algorithmic scalable solver. Nevertheless, the above preconditioner is sufficient for the problems considered in this work. The typical walltime to take a single time step using the proposed solver is about half a minute on 128 CPUs.  In comparison, a director solver or an iterative solver preconditioned with a generic preconditioner such as block incomplete LU need several hours to invert the same system.

\section{Numerical results\label{sec:numres}}

In this section, numerical simulations are used to assess the
performance of the MFD solver in solving the quasi-static force-free
model (Equations~\eqref{eq:4.565}--\eqref{eq:4.580}) with different
configurations.  The simulations were carried out on NERSC's Cori and
Perlmutter machines.  For our computations on Cori, we used 4 Haswell
nodes for a total of 128 cpus.  For our computations on Perlmutter, we
used 1 CPU-only node for a total of 128 cpus.



For parallelization, we employ the Parallel Extensible Toolkit for
Scientific computing (PETSc) library \cite{petsc-web-page,
  petsc-user-ref, petsc-efficient}.
For the time-integration, we rely on PETSc's TS library that provides a framework for the scalable solvers of ODEs and DAEs arising from the discretization of time-dependent PDEs \cite{AbhyankarEtAl2018}.
The major data structure we build our MFD algorithm upon is DMStag~\cite{Sanan2022}, 
which is PETSc's distributed data structure for a full set of staggered grid representations. 
This allows the degrees of freedom to be associated with all ``strata'' in a logically-rectangular grid, i.e., elements, faces, edges, and vertices.
Note that such representations are more than the commonly used dual mesh setting
such as those in FDTD. 
We use a second-order L-stable DIRK time integrator and JFNK preconditioned by the finite difference coloring Jacobian.
Our major criterion for the solver convergence is the relative tolerance of the Newton iteration, which is always set as \num{1e-4}.
The linear solver tolerance is dynamically controlled by the inexact Newton algorithm,
 which  is used to reduce the total number of linear iterations. 

All the simulations  presented in the current work are run on a cylindrical mesh that has a resolution of
$100 \times 2 \times 200$, i.e., 100 cells in $R$ direction, 2 cells with
periodic boundary conditions in $\phi$ direction and 200 cells in $Z$
direction. A minimal number of grid points in $\phi$ direction are used
since we only consider the axisymmetric case in this work.
However, we maintain a general implementation of full 3D operators, aiming for the whole device 
modeling capability and study of asymmetric cases in the future work.

 The numerical results are organized into five sections. 
 Section~\ref{sec:resistivity} first investigates the impact of resistivity in the vacuum vessel.
 Since the model considered here misses a temperature equation (it will be considered as 
 the immediate follow-up work), the physically relevant resistivity value can be only determined numerically.
 We propose a simple approach to evaluate the model using the diffusion model in the vacuum vessel. 
 Section~\ref{sec:initial} then discusses the steps to prepare the initial condition. 
 As an exactly force-free magnetic field is known to be challenging to find, we propose
 a modeling approach to further reduce the $\j\times\B$ force. 
 The impact of different regularization terms are then studied in Section~\ref{sec:num_regularization}.
 The full VDE simulation is presented in Section~\ref{sec:vde}. 
 Finally, the study of the solver performance is presented in Section~\ref{sec:performance}. 

\subsection{Computing the effective resistivity of ITER vacuum vessel}
\label{sec:resistivity}

The ITER tokamak reactor has an interesting engineering design in
which the vacuum vessel (vv) is to carry the inductive toroidal current as
the response to a dynamically evolving plasma, such as that during a
VDE after a thermal quench.  To enable this, the blanket modules,
which are attached to the vacuum vessel and face the plasma, are
insulated from each other along the toroidal direction so no net
toroidal current is allowed in the blanket modules.  To model the
plasma VDE properly, one must properly account for the image current,
especially the toroidal one, in the vacuum vessel.  The actual ITER
vacuum vessel has complicated structures and a hollow interior for
neutron moderating inserts, the details of which are not necessary in
the usual plasma modeling. The essential property of the vacuum vessel
for plasma modeling is its characteristic time for a wall current to
decay in the absence of a plasma inside the chamber.  This so-called
wall time of the vacuum vessel, $\tau_{\rm vv} \equiv L/R,$ is the ratio
of inductance $L$ and resistance $R$ of the metal structure. The
inductance $L$ is set by the geometry of the conducting structure,
while $R$ has seperate toroidal and poloidal values, corresponding to
the decay of a net toroidal and a poloidal current in the vacuum
vessel. For wall feedback on VDEs, the toroidal one is most important,
and for ITER, the toroidal wall time is known to be around \SI{500}{\milli\second}. The
geometric simplification of the vaccum vessel, shown in light blue
(value of -1 for the levelset function) in Figure~\ref{fig:5layer},
implies that the correct $\tau_{\rm vv}=$~\SI{500}{\milli\second} can be recovered by an
effective electric resistivity $\eta_{\rm vv}$ for the vacuum vessel that
would be somewhat different from $\eta$ of the stainless steel
that is used to construct the vacuum vessel.  This effective
$\eta_{\rm vv}$ is directly computed here to match a given $\tau_{\rm vv}$ as
follows.

The goal here is to consider an isotropic vacuum vessel with uniform
resistivity, and find the resistivity value that would yield the
targeted wall time for ITER setting (around \SI{500}{\milli\second}). For that
purpose, we run initial stand-alone tests for the magnetic diffusion problem where
we take the same high resistivity value (to mimic a vacuum response)
for all the computational domain except the vacuum vessel, and a lower
resistivity value for the vacuum vessel $\Omega^{V}$. The initial
magnetic field has to be set such that the current is nonzero only in
the vacuum vessel. We recall that the magnetic field in the tokamak
can be represented by
\begin{equation}
{\B} = \nabla \phi \times \nabla \psi + g_0 \nabla \phi, 
\label{eqn:Bfrompsi}
\end{equation} 
where $\psi$ is the poloidal flux function and $g_0$ is a scalar constant. 

\begin{figure}[ht]
\begin{center}
\includegraphics[trim={10cm 0 1cm 0},clip, width=0.5\textwidth]{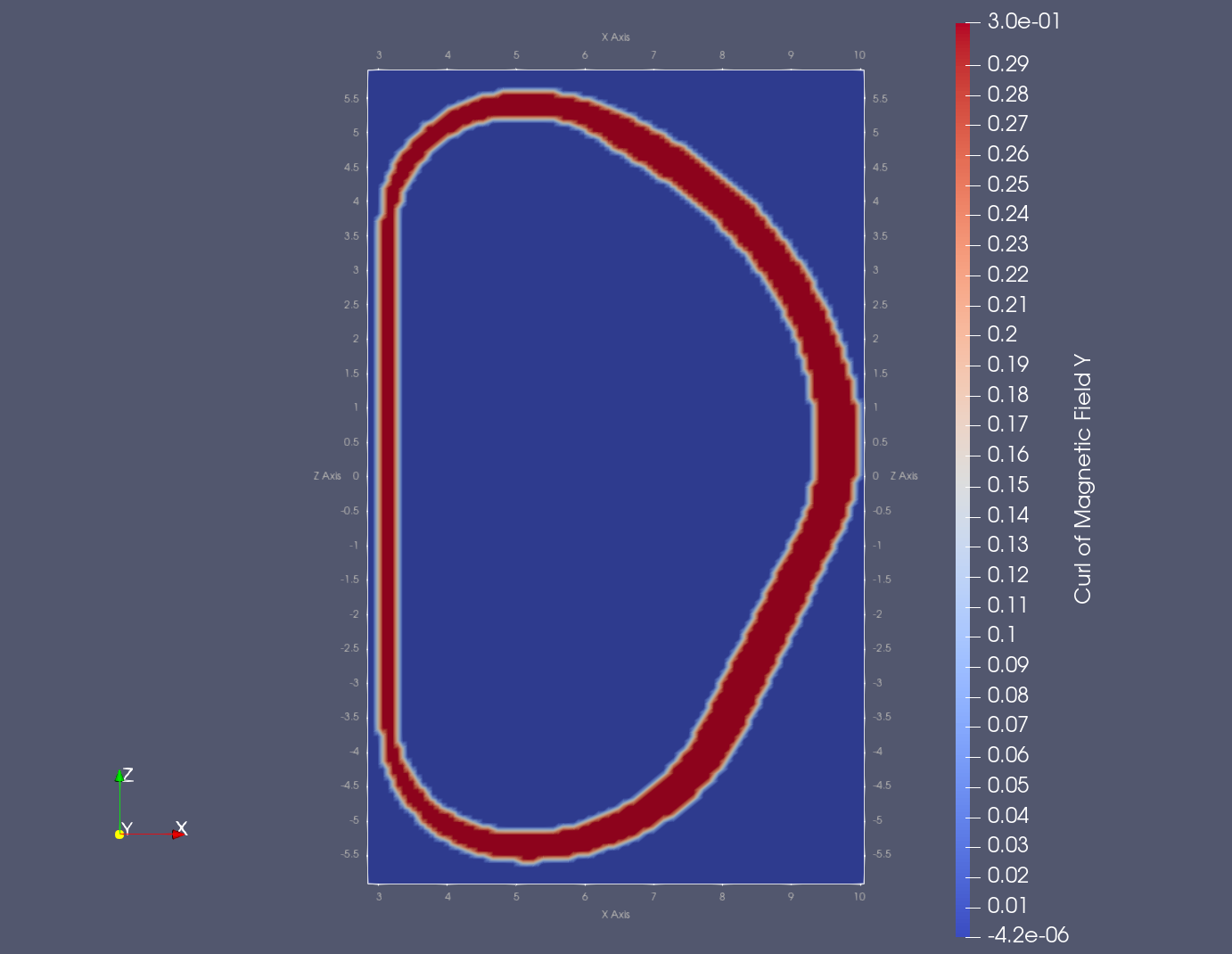}
\caption{The curl of the initial magnetic field in $\mathbf{e}_{\phi}$ direction.} 
\label{fig:5-layer_J0}
\end{center}
\end{figure} 

Enforcing a magnetic field such that the current is nonzero only in
the vacuum vessel amounts to solving the fixed-boundary Grad-Shafranov problem
\begin{equation}\label{eq:fixedbdeq}
  \begin{aligned}
    \frac{1}{\mu_0 R} ~ \Delta^{*} \psi & = J_0 && \text{in} \quad \Omega^{V}, \\
    \frac{1}{\mu_0 R} ~ \Delta^{*} \psi & = 0 && \text{in} \quad \Omega \setminus \Omega^{V}, \\
    \psi & = 0,&& \text{on} \quad \partial \Omega;
  \end{aligned}
\end{equation} 
where the toroidal elliptic operator is defined as   
\begin{equation*}
\Delta^* \psi := \dfrac{\partial^2 \psi}{\partial R^2}-\dfrac{1}{R}\dfrac{\partial \psi}{\partial R}+ \dfrac{\partial^2 \psi}{\partial Z^2},
\end{equation*}
After solving the problem~\eqref{eq:fixedbdeq}, and computing the
magnetic field with~\eqref{eqn:Bfrompsi}, we obtain the current shown
in Figure~\ref{fig:5-layer_J0}.

\begin{figure}[ht]
\centering
\begin{subfigure}[b]{0.5\textwidth}
\centering
\begin{tikzpicture}[scale=.8,>=latex]
\begin{axis}[ymode=log, xlabel=\textsc{{Time (s)}}, ylabel=\textsc{{Current intensity (A)}}, ymajorgrids, xmajorgrids, axis background/.style={fill=gray!4},, legend style={legend columns=2,at={(0.5,-0.2)},anchor=north}]
\addplot+[color=teal,line width=1pt,mark=none] coordinates {(0, 1.55782e+07) (0.5e-9, 3.97973e+06) (1e-9, 1.9874e+06) (1.5e-9, 872492) (2e-9 , 408238) (2.5e-9, 187740) (3e-9, 87194.3) (3.5e-9, 40461.7) (4e-9, 18828.8) (4.5e-9, 8771.88) (5e-9, 4092.83) (5.5e-9, 1911.82)};
\end{axis}
\end{tikzpicture}
\caption{$\eta_{\rm vv} = 10^3 ~\SI{}{\ohm\meter}$}
\label{fig:current_vvdiffusion_S2e-2_2e-5_time}
\end{subfigure}%
~
\begin{subfigure}[b]{0.5\textwidth}
\centering
\begin{tikzpicture}[scale=.8,>=latex]
\begin{axis}[ymode=log, xlabel=\textsc{{Time (s)}}, ylabel=\textsc{{Current intensity (A)}}, ymajorgrids, xmajorgrids, axis background/.style={fill=gray!4},, legend style={legend columns=2,at={(0.5,-0.2)},anchor=north}]
\addplot+[color=teal,line width=1pt,mark=none] coordinates {(0, 1.55782e+07) (5.0e-11, 3.94512e+06) (1e-10, 2.0403e+06) (1.5e-10, 906796) (2e-10 , 433197) (2.5e-10, 202728) (3e-10, 95922.9) (3.5e-10, 45317.9) (4e-10, 21472.3) (4.5e-10, 10183.3) (5e-10, 4836.51)};
\end{axis}
\end{tikzpicture}
\caption{$\eta_{\rm vv} = 10^4 ~\SI{}{\ohm\meter}$}
\label{fig:current_vvdiffusion_S2e-3_2e-5_time}
\end{subfigure}     
\caption{Evolution of the current intensity inside the vacuum vessel over time with two different vacuum vessel resistivities $\eta_{\rm vv}$ and a fixed resistivity elsewhere $\eta= 10^6 ~\SI{}{\ohm\meter}$.}   
\end{figure}
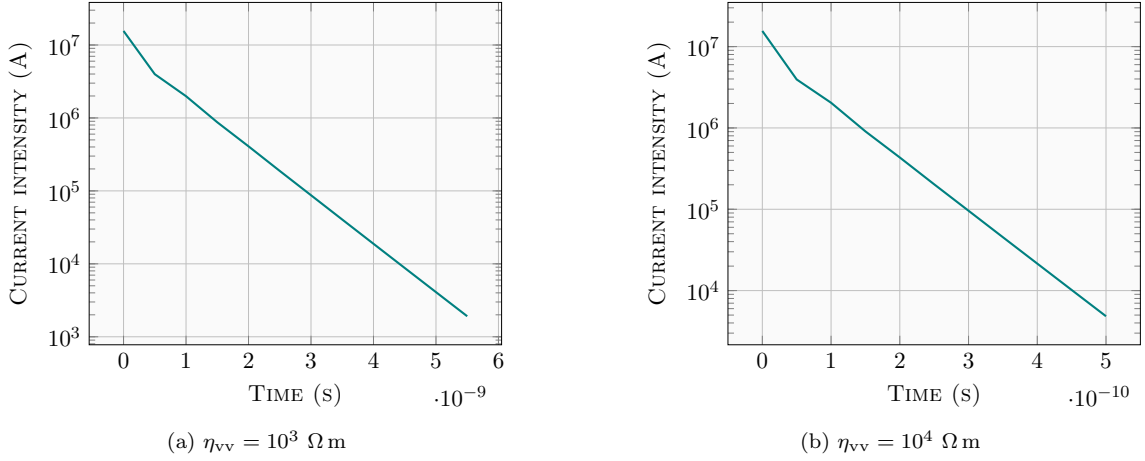

Then, the diffusion model (Equations~\eqref{eq:4.572}--~\eqref{eq:4.576}) is solved.
With a resistivity value of $\eta_{\rm vv} = \SI{1e+3}{\ohm\meter}$
(i.e., a Lundquist number $S_{\rm vv} = \num{2.7646e-2}$ if the resistive diffusion time is normalized by the plasma Alfven time) inside the
vacuum vessel and $\eta = \SI{1e+6}{\ohm\meter}$ ($S =
\num{2.7646e-5}$) elsewhere, we obtain the current evolution depicted
in Figure~\ref{fig:current_vvdiffusion_S2e-2_2e-5_time}.  The slope is estimated as
\begin{equation*}
a_1 := \dfrac{\ln(10^5) - \ln(10^6)}{\num{3e-9} - \num{1.5e-9}} = \SI{-1.53e+9}{\second^{-1}}.
\end{equation*}

With a resistivity value of $\eta_{\rm vv} = \SI{1e+4}{\ohm\meter}$ (i.e., a Lundquist number $S_{\rm vv} = \num{2.7646e-3}$) inside the vacuum vessel and $\eta = \SI{1e+6}{\ohm\meter}$ ($S = \num{2.7646e-5}$) elsewhere, we obtain the current evolution depicted in Figure~\ref{fig:current_vvdiffusion_S2e-3_2e-5_time}. 
The slope 
is now equal to:
\begin{equation*}
a_2 := \dfrac{\ln(10^5) - \ln(10^6)}{\num{3e-10} - \num{1.5e-10}} = -\SI{1.53e+10}{\second^{-1}}.
\end{equation*}

\begin{figure}[ht]
\centering
\begin{tikzpicture}[scale=.8,>=latex]
\begin{axis}[ymode=log, xlabel=\textsc{{Time (s)}}, ylabel=\textsc{{Current intensity (A)}}, ymajorgrids, xmajorgrids, axis background/.style={fill=gray!4},, legend style={legend columns=2,at={(0.5,-0.2)},anchor=north}]
\addplot+[color=teal,line width=1pt,mark=none] coordinates {(0, 1.55781e+07) (0.3858, 3.95356e+06) (0.771606, 1.97043e+06) (1.157409, 860919) (1.543212 , 401278) (1.929015, 183727) (2.314818, 84977.2) (2.700621, 39265.7) (3.086424, 18195.9) (3.472227, 8441.54) (3.858030, 3922.26)};
\end{axis}
\end{tikzpicture}
\caption{Evolution of the current intensity inside the vacuum vessel over time with $\eta_{\rm vv} = \SI{1.30288e-6}{\ohm\meter}$.}
\label{fig:current_vvdiffusion_S2e+7_2e+4_time}
\end{figure}
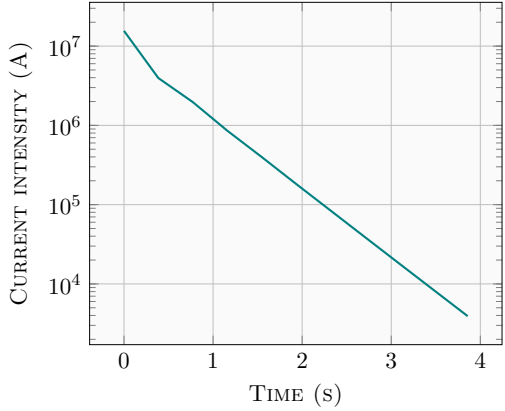

The characteristic time scale $L/R$ or wall time is expressed as
\begin{equation*}
\tau_{\rm vv} = \dfrac{L}{R_t},
\end{equation*}
with $L$ the inductance and the toroidal resistance
\begin{equation*}
R_t = \dfrac{2 \pi R \eta}{\mathcal{A}};
\end{equation*}
 $\mathcal{A}$ being the cross section.
The current decay is such that 
\begin{equation*}
I(t) = I_0 \exp(-\dfrac{t}{\tau_{\rm vv}}).
\end{equation*}
With the slopes computed previously, we can deduce that the resistivity value corresponding to $\tau_{\rm vv} = \SI{500}{\milli\second}$ is: 
\begin{equation}
\eta_{\rm vv} = \SI{1.30288e-6}{\ohm\meter} .
\label{eqn:vvresistivity}
\end{equation}
\noindent With such a resistivity ($S_{\rm vv} = 21219200$) inside the vacuum vessel and $\eta = \SI{1.30288e-3}{\ohm\meter}$ ($S = 21219.2$) elsewhere, we obtain the current evolution depicted in Figure~\ref{fig:current_vvdiffusion_S2e+7_2e+4_time}. 
The slope 
is equal to:
\begin{equation*}
a_3 := \dfrac{\ln(3922.26) - \ln(860919)}{3.858030 - 1.157409} = \SI{-1.99633062361}{\second^{-1}},
\end{equation*} 
which is very close to the expected value $\dfrac{-1}{\tau_{\rm vv}} = \dfrac{-1}{0.5} = \SI{-2}{\second^{-1}}$.

\subsection{Preparing the initial state for the quasi-static force-free model}
\label{sec:initial}

This section discusses the preparation of the initial condition which will be used in all the rest numerical tests.
The quasi-static force-free model targets a post-thermal-quench plasma
in which the plasma beta is negligibly small. We prepare such an
initial force-free state via a free-boundary Grad-Shafranov
solver~\cite{Li-Tang2-SIAMJSC-2021} by zeroing out the plasma beta of
an original full-beta \SI{15}{\mega\ampere} ITER equilibrium, while holding the
poloidal magnetic flux fixed from the original full-beta \SI{15}{\mega\ampere}
free-boundary Grad-Shafranov equilibrium.  This frozen flux boundary
condition simply reflects the fact that on the time scale of the
thermal quench, which is anticipated to be on the order of a
millisecond and hence much shorter than the vacuum vessel wall time,
the vacuum vessel acts like a perfect flux conserver.  This initial
state thus has zero pressure gradient, \SI{15}{\mega\ampere} of toroidal plasma
current, and one x-point at the bottom. The details of the force-free
equilibrium solver can be found in Ref.~\cite{Li-Tang2-SIAMJSC-2021}.
The numerical solution, once transferred to the staggered grid of the
mimetic finite difference solver, has $(\nabla \times \B) \times \B$
that is not exactly zero but close to $10^{-3}$, and that introduces
an error at the first time step of the quasi-static model in
Equations~\eqref{eq:4.566} and~\eqref{eq:4.568} since the initial
velocity is nil. This gross violation of force-free condition creates
difficulties for the quasi-static force-free solver at the Newton
iteration of the first time step.

A workaround for this issue is to call a time-dependent model that
would decrease $(\nabla \times \B) \times \B$ and also compute a
velocity that would restore some balance in Equations~\eqref{eq:4.566}
and~\eqref{eq:4.568}.  This time-dependent model is as follows.
\begin{itemize}
\item In the plasma region:
\begin{align}
  m_i n_0 \frac{\partial {\V}_{i\perp} }{\partial t} - \nu m_i n_0 \nabla^2 \V_{i\perp} - \frac{1}{\mu_0} (\nabla\times{\B})\times{\B} = & ~0 \qquad \text{in} \quad \Omega^{P} , \label{eq:4.597} \\
  - \nabla^2 \Phi + \nabla\cdot\left[ {\V}_{i\perp}\times {\B}  \right] = & ~ 0 \qquad \text{in} \quad \Omega^{P}, \label{eq:4.598} \\
  \boldsymbol{\tau} - \nabla\Phi + {\V}_{i\perp}\times {\B}  = & ~ \mathbf{0} \qquad \text{in} \quad \Omega^{P}, \label{eq:4.599} \\
  \frac{\partial{\B}}{\partial t} + \nabla\times\boldsymbol{\tau} = & ~ \mathbf{0} \qquad \text{in} \quad \Omega^{P}. \label{eq:4.600}
\end{align}
\item In the wall region:
\begin{align}
  {\V}_{i\perp} = & ~ \mathbf{0} \qquad \text{in} \quad \Omega^{W} , \label{eq:4.602} \\
  - \nabla^2 \Phi = & ~ 0 \qquad \text{in} \quad \Omega^{W} \label{eq:4.603} \\
  \boldsymbol{\tau} - \nabla\Phi = & ~ \mathbf{0} \qquad \text{in} \quad \Omega^{W}, \label{eq:4.604} \\
  \frac{\partial{\B}}{\partial t} + \nabla\times\boldsymbol{\tau} = & ~ \mathbf{0} \qquad \text{in} \quad \Omega^{W} . \label{eq:4.605}
\end{align}
\item At the wall/plasma interface:
\begin{alignat}{3}
  &{\V}_{i\perp} = ~ 0 \qquad &&\text{on} \quad \Gamma^{PW} , \label{eq:4.606} \\
  &\Phi \qquad &&\text{continous across} \quad  \Gamma^{PW} ,\label{eq:phicont} \\
  &\boldsymbol{\tau} \times \mathbf{n}  \qquad &&  \text{continous across} \quad \Gamma^{PW}. \label{eq:4.608} \\
  &{\B} \cdot \mathbf{n} \qquad &&  \text{continous across} \quad \Gamma^{PW} , \label{eq:4.607} 
\end{alignat}
\item At the outer rectangular boundary $\partial\Omega$:
\begin{align}
  {\V}_{i\perp} & =  \mathbf{0},\\
  \Phi & =  {0},\\
  \boldsymbol{\tau} & =  \mathbf{0}, \\
 \B & = \B_0.
\end{align}
\end{itemize}
Here the resistivity diffusion is deliberately turned off. 
Note that unlike the quasi-static model, this time-dependent model has an inertia term in~\eqref{eq:4.597},
 and thus it supports the Alfven wave and is subject to the time step constraint due to the wave.
 A physics-based preconditioner for the small flow limit of a similar extended MHD model 
 was proposed in Ref.~\cite{Chacon2008}.
 This preconditioner is extended to our MFD solver for efficiently inverting the linearized system.

Evolving the model produces a magnetic field that has a much smaller force balancing than the original field loaded into the MFD solver. 
In the final step of the initial condition preparation, the resulting velocity $\V_{i\perp,0}$ and magnetic field $\B_0$ from
this time-dependent model 
are then used to update $\boldsymbol{\tau}_0$ the divergence-free
component of the electric field and $\Phi_0$ the electrostatic
potential such that we have
\begin{itemize}
\item in the plasma region:
\begin{align*}
  - \nabla^2 \Phi_0 - \nabla\cdot\left[ - {\V}_{i\perp,0}\times {\B_0} + \frac{\eta}{\mu_0}\left(\nabla\times{\B_0}\right) \right] = & ~ 0 \qquad \text{in} \quad \Omega^{P},  \\
  \boldsymbol{\tau}_0 - \nabla\Phi_0 + {\V}_{i\perp,0}\times {\B_0} - \frac{\eta}{\mu_0}\left(\nabla\times{\B_0}\right) = & ~ \mathbf{0} \qquad \text{in} \quad \Omega^{P}, 
\end{align*}
\item in the wall region:
\begin{align*}
  - \nabla^2 \Phi_0 - \nabla\cdot\left[ \frac{\eta}{\mu_0}\left(\nabla\times{\B_0}\right) \right] = & ~ 0 \qquad \text{in} \quad \Omega^{W}  \\
  \boldsymbol{\tau}_0 - \nabla\Phi_0 - \frac{\eta}{\mu_0}\left(\nabla\times{\B_0}\right) = & ~ \mathbf{0} \qquad \text{in} \quad \Omega^{W},
\end{align*}
\end{itemize}
This last step is necessary to guarantee the initial condition is consistent with the quasi-static model. 
Finally, we set a uniform initial density of $n_0 = \SI{1e+20}{\meter}^{-3}$ and that completes
the initial state $(n_0, {\V}_{i\perp,0}, \Phi_0, \boldsymbol{\tau}_0,
{\B_0} )$ for the quasi-static model.

\subsection{Comparing the different regularizations}  \label{sec:num_regularization}

As noted in section~\ref{sec:regularization}, the quasi-static
force-free model requires regularization that invokes either a
fictitious drag coefficient $\epsilon$ or an artificial viscosity
$\nu.$ Since they enter directly into the force-balance equation that
would otherwise constrain the magnetic field to be force-free, we
anticipate their presence will modify the quasi-static dynamics but
the effect will diminish as $\epsilon$ or $\nu$ gets smaller. This is
a convergence issue related to modeling $\mathbf{B}(\mathbf{x},t)$ as a function of
decreasing $\epsilon$ and $\nu.$ It is important to note that with
smaller $\epsilon$ or $\nu,$ the regularization becomes weaker so the
numerical matrix inversion for the implicit solve of the quasi-static
model becomes more difficult computationally. So a practically useful
regularization scheme should produce good convergence in
$\mathbf{B}(\mathbf{x},t)$ solution for modestly small $\epsilon$ or
$\nu.$ For this convergence check, we will gauge the quality of the
solution for $\mathbf{B}$ through two quantities.  The first is the
vertical position of the magnetic axis $Z_a$ as a function of time
during a VDE.  The second is the total toroidal plasma current in the
chamber, $I_p(t),$ as a function of time during a VDE. Next we first
explain the set up of the test case, and then show how $Z_a(t)$ and
$I_p(t)$ scale with $\epsilon$ and $\nu.$

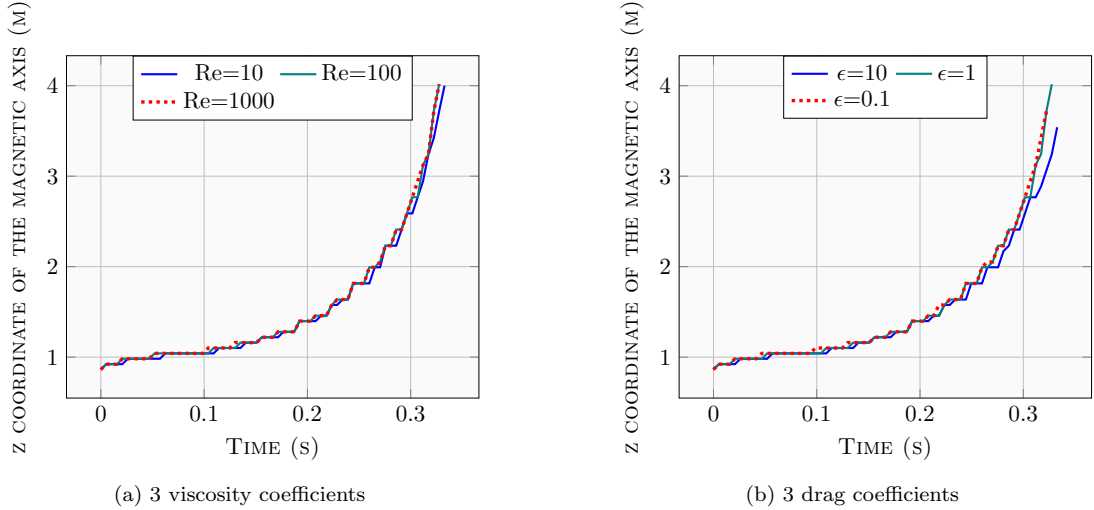
\begin{figure}[!h]
\centering
\begin{subfigure}[b]{0.47\textwidth} 
\centering
\begin{tikzpicture}[scale=0.8,>=latex]
\begin{axis}[xlabel=\textsc{{Time (s)}}, ylabel=\textsc{{z coordinate of the magnetic axis (m)}}, ymajorgrids, xmajorgrids, axis background/.style={fill=gray!4}, legend style={legend columns=2,at={(0.5,1.0)},anchor=north}]
\addplot+[color=blue,line width=1pt,mark=none] coordinates {(0, 0.86275)	(0.0052, 0.92225)	(0.0104, 0.92225)	(0.0156, 0.92225)	(0.0208, 0.92225)	(0.026, 0.98175)	(0.0312, 0.98175)	(0.0364, 0.98175)	(0.0416, 0.98175)	(0.0468, 0.98175)	(0.052, 0.98175)	(0.0572, 0.98175)	(0.0624, 1.04125)	(0.0676, 1.04125)	(0.0728, 1.04125)	(0.078, 1.04125)	(0.0832, 1.04125)	(0.0884, 1.04125)	(0.0936, 1.04125)	(0.0988, 1.04125)	(0.104, 1.04125)	(0.1092, 1.04125)	(0.1144, 1.10075)	(0.1196, 1.10075)	(0.1248, 1.10075)	(0.13, 1.10075)	         (0.1352, 1.10075)	(0.1404, 1.16025)	(0.1456, 1.16025)	(0.1508, 1.16025)	(0.156, 1.21975)	(0.1612, 1.21975)	(0.1664, 1.21975)	(0.1716, 1.21975)	(0.1768, 1.27925)	(0.182, 1.27925)	(0.1872, 1.27925)	(0.1924, 1.39825)	(0.1976, 1.39825)	(0.2028, 1.39825)	(0.208, 1.39825)	(0.2132, 1.45775)	(0.2184, 1.45775)	(0.2236, 1.57675)	(0.2288, 1.57675)   (0.234, 1.63625)	(0.2392, 1.63625)	(0.2444, 1.81475)	(0.2496, 1.81475)	(0.2548, 1.81475)	(0.26, 1.81475)   (0.2652, 1.99325)     (0.2704, 1.99325)     (0.2756, 2.23125)     (0.2808, 2.23125)     (0.286, 2.23125)       (0.2912, 2.40975)   (0.2964, 2.58825)     (0.3016, 2.58825)     (0.3068, 2.76675)     (0.3120, 2.94525)     (0.3172, 3.24275)     (0.3224, 3.42125)   (0.3276, 3.71875)   (0.3328, 4)};
\addplot+[color=teal,line width=1pt,mark=none] coordinates {(0, 0.86275)	(0.0052, 0.92225)	(0.0104, 0.92225)	(0.0156, 0.92225)	(0.0208, 0.98175)	(0.026, 0.98175)	(0.0312, 0.98175)	(0.0364, 0.98175)	(0.0416, 0.98175)	(0.0468, 0.98175)	(0.052, 1.04125)	(0.0572, 1.04125)	(0.0624, 1.04125)	(0.0676, 1.04125)	(0.0728, 1.04125)	(0.078, 1.04125)	(0.0832, 1.04125)	(0.0884, 1.04125)	(0.0936, 1.04125)	(0.0988, 1.04125)	(0.104, 1.04125)	(0.1092, 1.10075)	(0.1144, 1.10075)	(0.1196, 1.10075)	(0.1248, 1.10075)	(0.13, 1.10075)	         (0.1352, 1.16025)	(0.1404, 1.16025)	(0.1456, 1.16025)	(0.1508, 1.16025)	(0.156, 1.21975)	(0.1612, 1.21975)	(0.1664, 1.21975)	(0.1716, 1.27925)	(0.1768, 1.27925)	(0.182, 1.27925)	(0.1872, 1.27925)	(0.1924, 1.39825)	(0.1976, 1.39825)	(0.2028, 1.39825)	(0.208, 1.45775)	(0.2132, 1.45775)	(0.2184, 1.45775)	(0.2236, 1.57675)	(0.2288, 1.63625)   (0.234, 1.63625)	(0.2392, 1.63625)	(0.2444, 1.81475)	(0.2496, 1.81475)	(0.2548, 1.81475)	(0.26, 1.99325)   (0.2652, 1.99325)     (0.2704, 2.05275)     (0.2756, 2.23125)     (0.2808, 2.23125)     (0.286, 2.40975)       (0.2912, 2.40975)   (0.2964, 2.58825)     (0.3016, 2.76675)     (0.3068, 2.76675)     (0.3120, 3.12375)     (0.3172, 3.24275)     (0.3224, 3.71875)   (0.3276, 4.01625)};
\addplot+[color=red,line width=1pt,mark=none,ultra thick,dotted] coordinates {(0, 0.86275)	   (0.0052, 0.92225)	(0.0104, 0.92225)	(0.0156, 0.92225)	(0.0208, 0.98175)	(0.026, 0.98175)	(0.0312, 0.98175)	(0.0364, 0.98175)	(0.0416, 0.98175)	(0.0468, 0.98175)	(0.052, 1.04125)	(0.0572, 1.04125)	(0.0624, 1.04125)	(0.0676, 1.04125)	(0.0728, 1.04125)	(0.078, 1.04125)	(0.0832, 1.04125)	(0.0884, 1.04125)	(0.0936, 1.04125)	(0.0988, 1.04125)	(0.104, 1.10075)	(0.1092, 1.10075)	(0.1144, 1.10075)	(0.1196, 1.10075)	(0.1248, 1.10075)	(0.13, 1.16025)	         (0.1352, 1.16025)	(0.1404, 1.16025)	(0.1456, 1.16025)	(0.1508, 1.16025)	(0.156, 1.21975)	(0.1612, 1.21975)	(0.1664, 1.21975)	(0.1716, 1.27925)	(0.1768, 1.27925)	(0.182, 1.27925)	(0.1872, 1.27925)	(0.1924, 1.39825)	(0.1976, 1.39825)	(0.2028, 1.39825)	(0.208, 1.45775)	(0.2132, 1.45775)	(0.2184, 1.45775)	(0.2236, 1.57675)	(0.2288, 1.63625)   (0.234, 1.63625)	(0.2392, 1.63625)	(0.2444, 1.81475)	(0.2496, 1.81475)	(0.2548, 1.81475)	(0.26, 1.99325)   (0.2652, 1.99325)     (0.2704, 2.05275)     (0.2756, 2.23125)     (0.2808, 2.23125)     (0.286, 2.40975)       (0.2912, 2.40975)   (0.2964, 2.58825)     (0.3016, 2.76675)     (0.3068, 2.94525)     (0.3120, 3.12375)     (0.3172, 3.24275)     (0.3224, 3.71875)   (0.3276, 4.01625)};
\legend{Re=10,Re=100,Re=1000}
\end{axis}
\end{tikzpicture}
\caption{3 viscosity coefficients}
\label{fig:z_magneticaxis_15_15eV_time_Re1-100}
\end{subfigure} %
~
\begin{subfigure}[b]{0.47\textwidth}
\centering
\begin{tikzpicture}[scale=0.8,>=latex]
\begin{axis}[xlabel=\textsc{{Time (s)}}, ylabel=\textsc{{z coordinate of the magnetic axis (m)}}, ymajorgrids, xmajorgrids, axis background/.style={fill=gray!4}, legend style={legend columns=2,at={(0.5,1.0)},anchor=north}]
\addplot+[color=blue,line width=1pt,mark=none] coordinates {(0, 0.86275)	(0.0052, 0.92225)	(0.0104, 0.92225)	(0.0156, 0.92225)	(0.0208, 0.92225)	(0.026, 0.98175)	(0.0312, 0.98175)	(0.0364, 0.98175)	(0.0416, 0.98175)	(0.0468, 0.98175)	(0.052, 0.98175)	(0.0572, 1.04125)	(0.0624, 1.04125)	(0.0676, 1.04125)	(0.0728, 1.04125)	(0.078, 1.04125)	(0.0832, 1.04125)	(0.0884, 1.04125)	(0.0936, 1.04125)	(0.0988, 1.04125)	(0.104, 1.04125)	(0.1092, 1.04125)	(0.1144, 1.10075)	(0.1196, 1.10075)	(0.1248, 1.10075)	(0.13, 1.10075)	         (0.1352, 1.10075)	(0.1404, 1.16025)	(0.1456, 1.16025)	(0.1508, 1.16025)	(0.156, 1.21975)	(0.1612, 1.21975)	(0.1664, 1.21975)	(0.1716, 1.21975)	(0.1768, 1.27925)	(0.182, 1.27925)	(0.1872, 1.27925)	(0.1924, 1.39825)	(0.1976, 1.39825)	(0.2028, 1.39825)	(0.208, 1.39825)	(0.2132, 1.45775)	(0.2184, 1.45775)	(0.2236, 1.57675)	(0.2288, 1.57675)    (0.234, 1.63625)	(0.2392, 1.63625)	(0.2444, 1.63625)      (0.2496, 1.81475)	(0.2548, 1.81475)	(0.26, 1.81475)	       (0.2652, 1.99325)	(0.2704, 1.99325)	(0.2756, 1.99325)      (0.2808, 2.17175)    (0.2860, 2.23125)      (0.2912, 2.40975)   (0.2964, 2.40975)   (0.3016, 2.58825)     (0.3068, 2.76675)     (0.3120, 2.76675)     (0.3172, 2.88575)      (0.3224, 3.06425)   (0.3276, 3.24275)   (0.3328, 3.54025)};
\addplot+[color=teal,line width=1pt,mark=none] coordinates {(0, 0.86275)	(0.0052, 0.92225)	(0.0104, 0.92225)	(0.0156, 0.92225)	(0.0208, 0.98175)	(0.026, 0.98175)	(0.0312, 0.98175)	(0.0364, 0.98175)	(0.0416, 0.98175)	(0.0468, 0.98175)	(0.052, 1.04125)	(0.0572, 1.04125)	(0.0624, 1.04125)	(0.0676, 1.04125)	(0.0728, 1.04125)	(0.078, 1.04125)	(0.0832, 1.04125)	(0.0884, 1.04125)	(0.0936, 1.04125)	(0.0988, 1.04125)	(0.104, 1.04125)	(0.1092, 1.10075)	(0.1144, 1.10075)	(0.1196, 1.10075)	(0.1248, 1.10075)	(0.13, 1.10075)	(0.1352, 1.16025)	(0.1404, 1.16025)	(0.1456, 1.16025)	(0.1508, 1.16025)	(0.156, 1.21975)	(0.1612, 1.21975)	(0.1664, 1.21975)	(0.1716, 1.27925)	(0.1768, 1.27925)	(0.182, 1.27925)	(0.1872, 1.27925)	(0.1924, 1.39825)	(0.1976, 1.39825)	(0.2028, 1.39825)	(0.208, 1.45775)	(0.2132, 1.45775)	(0.2184, 1.45775)	(0.2236, 1.57675)	(0.2288, 1.63625)   (0.234, 1.63625)	(0.2392, 1.63625)	(0.2444, 1.81475)	(0.2496, 1.81475)	(0.2548, 1.81475)	(0.26, 1.99325)   (0.2652, 1.99325)     (0.2704, 2.05275)     (0.2756, 2.23125)     (0.2808, 2.23125)     (0.286, 2.40975)      (0.2912, 2.40975)   (0.2964, 2.58825)     (0.3016, 2.76675)     (0.3068, 2.76675)     (0.3120, 3.12375)     (0.3172, 3.24275)    (0.3224, 3.71875)  (0.3276, 4.01625)};
\addplot+[color=red,line width=1pt,mark=none,ultra thick,dotted] coordinates {(0, 0.86275)	(0.0052, 0.92225)	(0.0104, 0.92225)	(0.0156, 0.92225)	(0.0208, 0.98175)	(0.026, 0.98175)	(0.0312, 0.98175)	(0.0364, 0.98175)	(0.0416, 0.98175)	(0.0468, 1.04125)	(0.052, 1.04125)	(0.0572, 1.04125)	(0.0624, 1.04125)	(0.0676, 1.04125)	(0.0728, 1.04125)	(0.078, 1.04125)	(0.0832, 1.04125)	(0.0884, 1.04125)	(0.0936, 1.04125)	(0.0988, 1.10075)	(0.104, 1.10075)	(0.1092, 1.10075)	(0.1144, 1.10075)	(0.1196, 1.10075)	(0.1248, 1.10075)	(0.13, 1.16025)	(0.1352, 1.16025)	(0.1404, 1.16025)	(0.1456, 1.16025)	(0.1508, 1.16025)	(0.156, 1.21975)	(0.1612, 1.21975)	(0.1664, 1.21975)	(0.1716, 1.27925)	(0.1768, 1.27925)	(0.182, 1.27925)	(0.1872, 1.27925)	(0.1924, 1.39825)	(0.1976, 1.39825)	(0.2028, 1.39825)	(0.208, 1.45775)	(0.2132, 1.45775)	(0.2184, 1.57675)	(0.2236, 1.57675)	(0.2288, 1.63625)   (0.234, 1.63625)	(0.2392, 1.63625)	(0.2444, 1.81475)	(0.2496, 1.81475)	(0.2548, 1.81475)	(0.26, 1.99325)   (0.2652, 2.05275)     (0.2704, 2.05275)     (0.2756, 2.23125)     (0.2808, 2.23125)     (0.286, 2.40975)      (0.2912, 2.40975)   (0.2964, 2.58825)     (0.3016, 2.76675)     (0.3068, 2.94525)     (0.3120, 3.12375)     (0.3172, 3.42125)     (0.3224, 3.71875) };
\legend{$\epsilon$=10,$\epsilon$=1,$\epsilon$=0.1}
\end{axis}
\end{tikzpicture}
\caption{3 drag coefficients}
\label{fig:z_magneticaxis_15_15eV_time_eps+1-1}
\end{subfigure} 
\caption{Evolution of the z coordinate of the magnetic axis over time for different viscosity and drag coefficients.}
\label{fig:z_magneticaxis_15_15eV_time}
\end{figure}

\begin{figure}[!h]
\centering
\begin{subfigure}[b]{0.5\textwidth} 
\centering
\begin{tikzpicture}[scale=0.8,>=latex]
\begin{axis}[scaled y ticks=base 10:-6, ytick scale label code/.code={}, xlabel=\textsc{{Time (s)}}, ylabel=\textsc{{Current intensity (MA)}}, ymajorgrids, xmajorgrids, axis background/.style={fill=gray!4}, legend style={legend columns=2,at={(0.4,0.05)},anchor=south}]
\addplot+[color=blue,line width=1pt,mark=star] coordinates {(0, 1.48934e+07) (0.026, 1.52149e+07) (0.052, 1.50161e+07) (0.078, 1.45168e+07) (0.104 , 1.38643e+07) (0.13, 1.31421e+07) (0.156, 1.23954e+07) (0.182, 1.16501e+07) (0.208, 1.09213e+07) (0.234, 1.02167e+07) (0.26, 9.53784e+06) (0.286, 8.87986e+06) (0.312, 8.22953e+06) (0.3328, 7.69845e+06)};
\addplot+[color=teal,line width=1pt,mark=triangle] coordinates {(0, 1.48934e+07) (0.026, 1.52405e+07) (0.052, 1.5032e+07) (0.078, 1.45262e+07) (0.104 , 1.38697e+07) (0.13, 1.31466e+07) (0.156, 1.24002e+07) (0.182, 1.16557e+07) (0.208, 1.09273e+07) (0.234, 1.02225e+07) (0.26, 9.54294e+06) (0.286, 8.88372e+06) (0.312, 8.23127e+06) (0.3276, 7.82844e+06)};
\addplot+[color=red,line width=1pt,mark=none,ultra thick,dotted] coordinates {(0, 1.48934e+07) (0.026, 1.52495e+07) (0.052, 1.50363e+07) (0.078, 1.45283e+07) (0.104 , 1.38704e+07) (0.13, 1.31472e+07) (0.156, 1.24011e+07) (0.182, 1.16571e+07) (0.208, 1.09291e+07) (0.234, 1.02241e+07) (0.26, 9.54422e+06) (0.286, 8.88477e+06) (0.312, 8.23172e+06) (0.3276, 7.82664e+06)};
\legend{Re=10,Re=100,Re=1000}
\end{axis}
\end{tikzpicture}
\caption{3 viscosity coefficients}
\label{fig:current_variousvisc_time}
\end{subfigure}%
~
\begin{subfigure}[b]{0.5\textwidth} 
\centering
\begin{tikzpicture}[scale=0.8,>=latex]
\begin{axis}[scaled y ticks=base 10:-6, ytick scale label code/.code={}, xlabel=\textsc{{Time (s)}}, ylabel=\textsc{{Current intensity (MA)}}, ymajorgrids, xmajorgrids, axis background/.style={fill=gray!4}, legend style={legend columns=2,at={(0.4,0.05)},anchor=south}]
\addplot+[color=blue,line width=1pt,mark=triangle] coordinates {(0, 1.48934e+07) (0.026, 1.5222e+07) (0.052, 1.50214e+07) (0.078, 1.45207e+07) (0.104 , 1.38657e+07) (0.13, 1.3141e+07) (0.156, 1.23914e+07) (0.182, 1.16431e+07) (0.208, 1.09114e+07) (0.234, 1.0204e+07) (0.26, 9.52287e+06) (0.286, 8.86403e+06) (0.312, 8.21639e+06) (0.3328, 7.69489e+06)};
\addplot+[color=teal,line width=1pt,mark=triangle] coordinates {(0, 1.48934e+07) (0.026, 1.52405e+07) (0.052, 1.5032e+07) (0.078, 1.45262e+07) (0.104 , 1.38697e+07) (0.13, 1.31466e+07) (0.156, 1.24002e+07) (0.182, 1.16557e+07) (0.208, 1.09273e+07) (0.234, 1.02225e+07) (0.26, 9.54294e+06) (0.286, 8.88372e+06) (0.312, 8.23127e+06)  (0.3276, 7.82844e+06)};
\addplot+[color=red,line width=1pt,mark=none,ultra thick,dotted] coordinates {(0, 1.48934e+07) (0.026, 1.52739e+07) (0.052, 1.50496e+07) (0.078, 1.45334e+07) (0.104 , 1.38713e+07) (0.13, 1.31469e+07) (0.156, 1.24011e+07) (0.182, 1.16581e+07) (0.208, 1.09313e+07) (0.234, 1.02267e+07) (0.26, 9.5458e+06) (0.286, 8.88566e+06) (0.312, 8.22992e+06) (0.3224, 7.95837e+06)};
\legend{$\epsilon$=10,$\epsilon$=1,$\epsilon$=0.1}
\end{axis}
\end{tikzpicture}
\caption{3 drag coefficients}
\label{fig:current_variousdrag_time}
\end{subfigure}
\caption{Evolution of the current intensity inside the plasma chamber over time with different artificial viscosity and fictitious drag coefficients.}
\label{fig:current_time}
\end{figure}
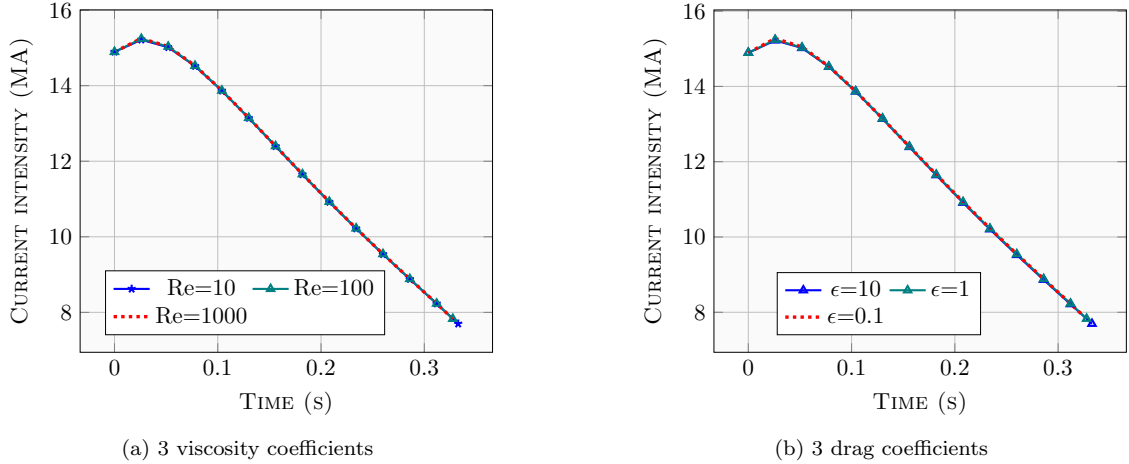

We consider a 5-layer configuration for the resistivity as shown in
Figure~\ref{fig:5layer} with the following resistivity values:
\begin{itemize}
    \item $\eta = \SI{9.66e-6}{\ohm\meter}$ in the plasma chamber
      corresponding to values 1 and 2 for the levelset function in
      Figure~\ref{fig:5layer};
    \item $\eta = \SI{4.4e-2}{\ohm\meter}$ in the blanket module
      corresponding to a value of 0 for the levelset function in
      Figure~\ref{fig:5layer};
    \item $\eta = \SI{1.30288e-6}{\ohm\meter}$ in the vacuum vessel
      corresponding to a value of -1 for the levelset function in
      Figure~\ref{fig:5layer};
    \item $\eta = \SI{1.30288e-3}{\ohm\meter}$ in the region
      outside the wall corresponding to a value of -2 for the levelset
      function in Figure~\ref{fig:5layer}.
\end{itemize}
and compare the results of the quasi-static models with fictitious
drag term~\eqref{eq:drag-regularize} and with fictitious viscous
term~\eqref{eq:viscous-regularize} using different viscosity and drag
coefficients. We plot the evolution in time of the $z$ coordinate of the
magnetic axis and the current intensity in
Figures~\ref{fig:z_magneticaxis_15_15eV_time}
and~\ref{fig:current_time}. Note that after normalization of
Equation~\eqref{eq:viscous-regularize}, $\mu_0 \nu$ gets replaced by
the viscosity coefficient ${1}/{Re}$.

We observe in Figures~\ref{fig:z_magneticaxis_15_15eV_time}
and~\ref{fig:current_time} that there is no significant change in the
curves for the current intensity and magnetic axis coordinate when we
vary the viscosity/drag coefficients, which could indicate
that the regularized model has already converged with such values and
there is no need to further decrease the regularization term as that
would induce more computational effort for the solver but with little
and insignificant impact on the relevant physical indicators or
properties.


 A subtler effect of regularization, as explained in
Section~\ref{sec:regularization}, is that it sets a radial electric
field $\varphi(\psi),$ which is not constrained by the quasi-static
force-free model. This is thus an artificial radial electric field,
which interestingly enough, does not modify the magnetic field
evolution directly, since the curl of an electrostatic field vanishes
in the Faraday's law.  In Fig.~\ref{fig:EP_avg_psi_isovolumes}, we numerically compute
$\varphi(\psi)$ from $\Phi(R,Z)$ by performing an average of $\Phi$ along
the field lines which are contours of poloidal magnetic flux $\psi.$
This $\varphi(\Psi)$ is then plotted as a function of $\psi.$ Here we
shall classify two classes of magnetic field lines or constant $\psi$
surfaces (contour lines in the poloidal cross section of an
axisymmetric configuration).  Where $\psi$ forms closed contours in
the plasma domain, one has closed flux surfaces.  There are regions in
which $\psi$ contours or magnetic field lines intercept the chamber
wall, these are known as the open flux or magnetic field line region.
Since the chamber wall is a conductor and $\Phi$ is set to zero there,
so $\varphi$ on the open flux region is close to zero, in the order of
$10^{-10}$ to $10^{-8}$ (normalized unit) range.  The radial
electrostatic potential $\varphi(\psi)$ can take on a much larger value
inside the closed flux region, resulting in a sizable radial electric
field across the separatrix (if it exists) or the last closed flux
surface that scrapes off the chamber wall.

 By comparing Fig.~\ref{fig:EP_avg_psi_isovolumes_visc} and
Fig.~\ref{fig:EP_avg_psi_isovolumes_drag}, one can see that both the
fictitious drag $(\epsilon)$ and artificial viscosity $(\nu)$ produce
$\varphi(\psi)$ of comparable amplitude. In the case of artificial
viscosity, the $Re=100$ and $Re=1000$ cases appear to converge to the
same $\varphi(\psi).$ The fictitious drag coefficients of $\epsilon=10$
and $\epsilon=0.1$ evidently yield $\varphi(\psi)$ with sizable
difference in the closed flux region. 
That is also the case after 25 time steps in Figure~\ref{fig:EP_avg_psi_isovolumes_25dt}.
It should be emphasized again
that the $\varphi(\psi)$ so produced is an artificial radial electric
field, entirely the result of the regularization scheme. The
quasi-static MHD model itself does not physically constrain
$\varphi(\psi).$

 At a later time, all the fictitious drag and artificial viscosity coefficients produce almost identical $\varphi(\psi)$ amplitudes. This could be observed in  Figure~\ref{fig:EP_avg_psi_isovolumes_50dt} that shows the curves of electrostatic potential average in function of the poloidal magnetic flux function after 50 time steps, which corresponds to the time when the magnetic axis moves to mid-way (in vertical position) between the eventual impact point at the first wall and the initial state.
The same observation holds after 56 time steps in Figure~\ref{fig:EP_avg_psi_isovolumes_56dt}.

\pgfplotstableread{data_to_plot/dataRe10.txt}\tableone
\pgfplotstablecreatecol[create col/expr={-log10(-\thisrow{y} * 55000000)}]{log}\tableone
\pgfplotstablecreatecol[create col/expr={5*(\thisrow{x})}]{nonnormx}\tableone

\pgfplotstableread{data_to_plot/dataRe100.txt}\tabletwo
\pgfplotstablecreatecol[create col/expr={-log10(-\thisrow{y} * 55000000)}]{log}\tabletwo
\pgfplotstablecreatecol[create col/expr={5*(\thisrow{x})}]{nonnormx}\tabletwo

\pgfplotstableread{data_to_plot/dataRe1000.txt}\tablethree
\pgfplotstablecreatecol[create col/expr={-log10(-\thisrow{y} * 55000000)}]{log}\tablethree
\pgfplotstablecreatecol[create col/expr={5*(\thisrow{x})}]{nonnormx}\tablethree

\pgfplotstableread{data_to_plot/datadampV+1.txt}\tablefour
\pgfplotstablecreatecol[create col/expr={1/log10(-\thisrow{y})}]{log}\tablefour
\pgfplotstablecreatecol[create col/expr={5*(\thisrow{x})}]{nonnormx}\tablefour

\pgfplotstableread{data_to_plot/datadampV-1.txt}\tablesix
\pgfplotstablecreatecol[create col/expr={\thisrow{y} > 0 ? -1/log10(\thisrow{y}) : 1/log10(-\thisrow{y})}]{log}\tablesix 
\pgfplotstablecreatecol[create col/expr={5*(\thisrow{x})}]{nonnormx}\tablesix

\pgfplotstableread{data_to_plot/datadampV0.txt}\tablefive
\pgfplotstablecreatecol[create col/expr={\thisrow{y} > 0 ? -1/log10(\thisrow{y}) : 1/log10(-\thisrow{y})}]{log}\tablefive 
\pgfplotstablecreatecol[create col/expr={5*(\thisrow{x})}]{nonnormx}\tablefive

\pgfplotstableread{data_to_plot/datadampV-2.txt}\tablesixbis
\pgfplotstablecreatecol[create col/expr={\thisrow{y} > 0 ? -1/log10(\thisrow{y}) : 1/log10(-\thisrow{y})}]{log}\tablesixbis 
\pgfplotstablecreatecol[create col/expr={5*(\thisrow{x})}]{nonnormx}\tablesixbis

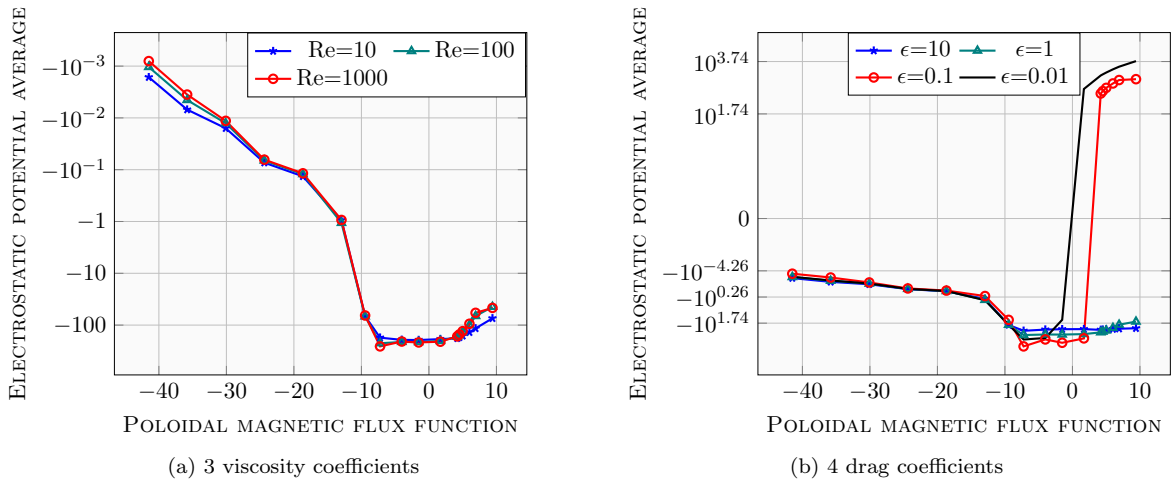
\begin{figure}[!h]
\centering
\begin{subfigure}[b]{0.47\textwidth} 
\begin{tikzpicture}[scale=0.80,>=latex]
\begin{axis}[xlabel=\textsc{{Poloidal magnetic flux function}}, ylabel=\textsc{{Electrostatic potential average}} , ytick={4,3,2,1,0,-1,-2}, yticklabels={$-10^{-4}$,$-10^{-3}$,$-10^{-2}$,$-10^{-1}$,$-1$,$-10$,$-100$}, ymajorgrids, xmajorgrids, axis background/.style={fill=gray!4}, legend style={legend columns=2,at={(0.66,1)},anchor=north}]
\addplot+[color=blue,line width=1pt,mark=star] table[x=nonnormx, y=log] \tableone;
\addplot+[color=teal,line width=1pt,mark=triangle] table[x=nonnormx, y=log] \tabletwo;
\addplot+[color=red,line width=1pt,mark=o] table[x=nonnormx, y=log] \tablethree;
\legend{Re=10,Re=100,Re=1000}
\end{axis}
\end{tikzpicture}
\caption{3 viscosity coefficients}
\label{fig:EP_avg_psi_isovolumes_visc}
\end{subfigure}%
~
\begin{subfigure}[b]{0.47\textwidth} 
\centering
\begin{tikzpicture}[scale=0.80,>=latex]
\begin{axis}[xlabel=\textsc{{Poloidal magnetic flux function}}, ylabel=\textsc{{Electrostatic potential average}} , ytick={-0.0833,-0.125,-0.1666,-0.25,0,0.1666,0.25}, yticklabels={$-10^{-4.26}$,$-10^{0.26}$,$-10^{1.74}$,$-10^{3.74}$,0,$10^{1.74}$,$10^{3.74}$}, ymajorgrids, xmajorgrids, axis background/.style={fill=gray!4}, legend style={legend columns=2,at={(0.5,1)},anchor=north}]
\addplot+[color=blue,line width=1pt,mark=star] table[x=nonnormx, y=log] \tablefour;
\addplot+[color=teal,line width=1pt,mark=triangle] table[x=nonnormx, y=log] \tablefive;
\addplot+[color=red,line width=1pt,mark=o] table[x=nonnormx, y=log] \tablesix;
\addplot+[color=black,line width=1pt,mark=none] table[x=nonnormx, y=log] \tablesixbis;
\legend{$\epsilon$=10,$\epsilon$=1,$\epsilon$=0.1,$\epsilon$=0.01}
\end{axis}
\end{tikzpicture}
\caption{4 drag coefficients}
\label{fig:EP_avg_psi_isovolumes_drag}
\end{subfigure}
\caption{Evolution of the electrostatic potential average over poloidal magnetic flux function with different artificial viscosity and fictitious drag coefficients at the first time step.}
\label{fig:EP_avg_psi_isovolumes}
\end{figure}


\pgfplotstableread{data_to_plot/datadampV+1_25dt.txt}\tablethirteen
\pgfplotstablecreatecol[create col/expr={\thisrow{y} > 0 ? -1/log10(\thisrow{y}) : 1/log10(-\thisrow{y})}]{log}\tablethirteen
\pgfplotstablecreatecol[create col/expr={5*(\thisrow{x})}]{nonnormx}\tablethirteen

\pgfplotstableread{data_to_plot/datadampV0_25dt.txt}\tablefourteen
\pgfplotstablecreatecol[create col/expr={\thisrow{y} > 0 ? -1/log10(\thisrow{y}) : 1/log10(-\thisrow{y})}]{log}\tablefourteen
\pgfplotstablecreatecol[create col/expr={5*(\thisrow{x})}]{nonnormx}\tablefourteen

\pgfplotstableread{data_to_plot/datadampV-1_25dt.txt}\tablefifteen
\pgfplotstablecreatecol[create col/expr={\thisrow{y} > 0 ? -1/log10(\thisrow{y}) : 1/log10(-\thisrow{y})}]{log}\tablefifteen
\pgfplotstablecreatecol[create col/expr={5*(\thisrow{x})}]{nonnormx}\tablefifteen

\pgfplotstableread{data_to_plot/dataRe10_25dt.txt}\tablesixteen
\pgfplotstablecreatecol[create col/expr={\thisrow{y} > 0 ? -1/log10(\thisrow{y}) : 1/log10(-\thisrow{y})}]{log}\tablesixteen
\pgfplotstablecreatecol[create col/expr={5*(\thisrow{x})}]{nonnormx}\tablesixteen

\pgfplotstableread{data_to_plot/dataRe100_25dt.txt}\tableseventeen
\pgfplotstablecreatecol[create col/expr={\thisrow{y} > 0 ? -1/log10(\thisrow{y}) : 1/log10(-\thisrow{y})}]{log}\tableseventeen
\pgfplotstablecreatecol[create col/expr={5*(\thisrow{x})}]{nonnormx}\tableseventeen

\pgfplotstableread{data_to_plot/dataRe1000_25dt.txt}\tableeighteen
\pgfplotstablecreatecol[create col/expr={\thisrow{y} > 0 ? -1/log10(\thisrow{y}) : 1/log10(-\thisrow{y})}]{log}\tableeighteen
\pgfplotstablecreatecol[create col/expr={5*(\thisrow{x})}]{nonnormx}\tableeighteen

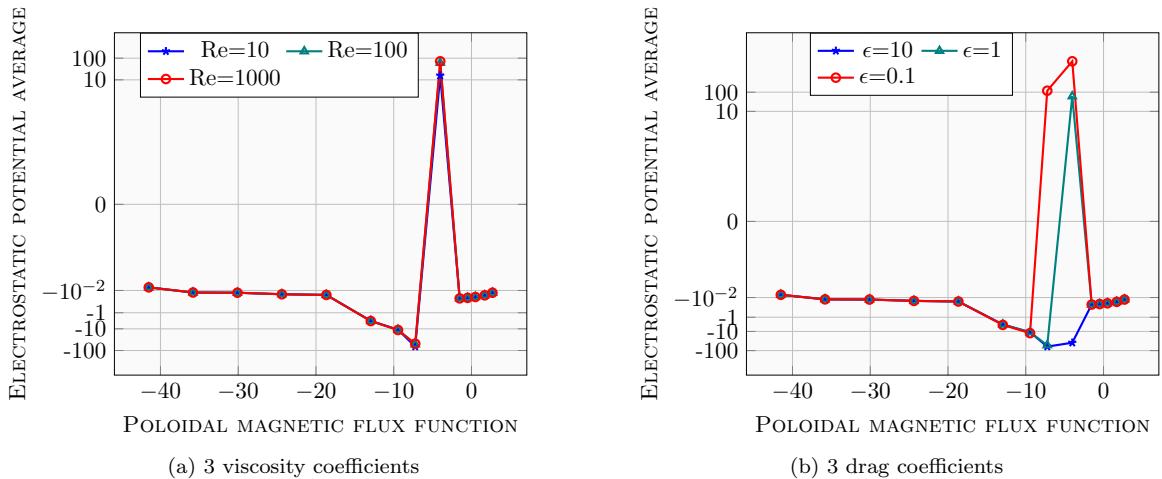
\begin{figure}[!h]
\centering
\begin{subfigure}[b]{0.47\textwidth} 
\begin{tikzpicture}[scale=0.80,>=latex]
\begin{axis}[xlabel=\textsc{{Poloidal magnetic flux function}}, ylabel=\textsc{{Electrostatic potential average}} , ytick={-0.17420500732,-0.14835996906,-0.12919291255,-0.10266,0, 0.14835996906,0.17420500732}, yticklabels={-100,-10,-1,$-10^{-2}$,0,10,100}, ymajorgrids, xmajorgrids, axis background/.style={fill=gray!4}, legend style={legend columns=2,at={(0.4,1)},anchor=north}]
\addplot+[color=blue,line width=1pt,mark=star] table[x=nonnormx, y=log] \tablesixteen;
\addplot+[color=teal,line width=1pt,mark=triangle] table[x=nonnormx, y=log] \tableseventeen;
\addplot+[color=red,line width=1pt,mark=o] table[x=nonnormx, y=log] \tableeighteen;
\legend{Re=10,Re=100,Re=1000}
\end{axis}
\end{tikzpicture}
\caption{3 viscosity coefficients}
\label{fig:EP_avg_psi_isovolumes_visc_25dt}
\end{subfigure}%
~
\begin{subfigure}[b]{0.47\textwidth} 
\centering
\begin{tikzpicture}[scale=0.80,>=latex]
\begin{axis}[xlabel=\textsc{{Poloidal magnetic flux function}}, ylabel=\textsc{{Electrostatic potential average}} , ytick={-0.17420500732,-0.14835996906,-0.12919291255,-0.10266,0, 0.14835996906,0.17420500732}, yticklabels={-100,-10,-1,$-10^{-2}$,0,10,100}, ymajorgrids, xmajorgrids, axis background/.style={fill=gray!4}, legend style={legend columns=2,at={(0.4,1)},anchor=north}]
\addplot+[color=blue,line width=1pt,mark=star] table[x=nonnormx, y=log] \tablethirteen;
\addplot+[color=teal,line width=1pt,mark=triangle] table[x=nonnormx, y=log] \tablefourteen;
\addplot+[color=red,line width=1pt,mark=o] table[x=nonnormx, y=log] \tablefifteen;
\legend{$\epsilon$=10,$\epsilon$=1,$\epsilon$=0.1}
\end{axis}
\end{tikzpicture}
\caption{3 drag coefficients}
\label{fig:EP_avg_psi_isovolumes_drag_25dt}
\end{subfigure}
\caption{Evolution of the electrostatic potential average over poloidal magnetic flux function with different artificial viscosity and fictitious drag coefficients after 25 time steps.}
\label{fig:EP_avg_psi_isovolumes_25dt}
\end{figure}


\pgfplotstableread{data_to_plot/datadampV+1_50dt.txt}\tableten
\pgfplotstablecreatecol[create col/expr={-log10(-\thisrow{y} * 55000000)}]{log}\tableten
\pgfplotstablecreatecol[create col/expr={5*(\thisrow{x})}]{nonnormx}\tableten

\pgfplotstableread{data_to_plot/datadampV0_50dt.txt}\tableeleven
\pgfplotstablecreatecol[create col/expr={-log10(-\thisrow{y} * 55000000)}]{log}\tableeleven
\pgfplotstablecreatecol[create col/expr={5*(\thisrow{x})}]{nonnormx}\tableeleven

\pgfplotstableread{data_to_plot/datadampV-1_50dt.txt}\tabletwelve
\pgfplotstablecreatecol[create col/expr={-log10(-\thisrow{y} * 55000000)}]{log}\tabletwelve
\pgfplotstablecreatecol[create col/expr={5*(\thisrow{x})}]{nonnormx}\tabletwelve

\pgfplotstableread{data_to_plot/dataRe10_50dt.txt}\tableseven
\pgfplotstablecreatecol[create col/expr={-log10(-\thisrow{y} * 55000000)}]{log}\tableseven
\pgfplotstablecreatecol[create col/expr={5*(\thisrow{x})}]{nonnormx}\tableseven

\pgfplotstableread{data_to_plot/dataRe100_50dt.txt}\tableeight
\pgfplotstablecreatecol[create col/expr={-log10(-\thisrow{y} * 55000000)}]{log}\tableeight
\pgfplotstablecreatecol[create col/expr={5*(\thisrow{x})}]{nonnormx}\tableeight

\pgfplotstableread{data_to_plot/dataRe1000_50dt.txt}\tablenine
\pgfplotstablecreatecol[create col/expr={-log10(-\thisrow{y} * 55000000)}]{log}\tablenine
\pgfplotstablecreatecol[create col/expr={5*(\thisrow{x})}]{nonnormx}\tablenine

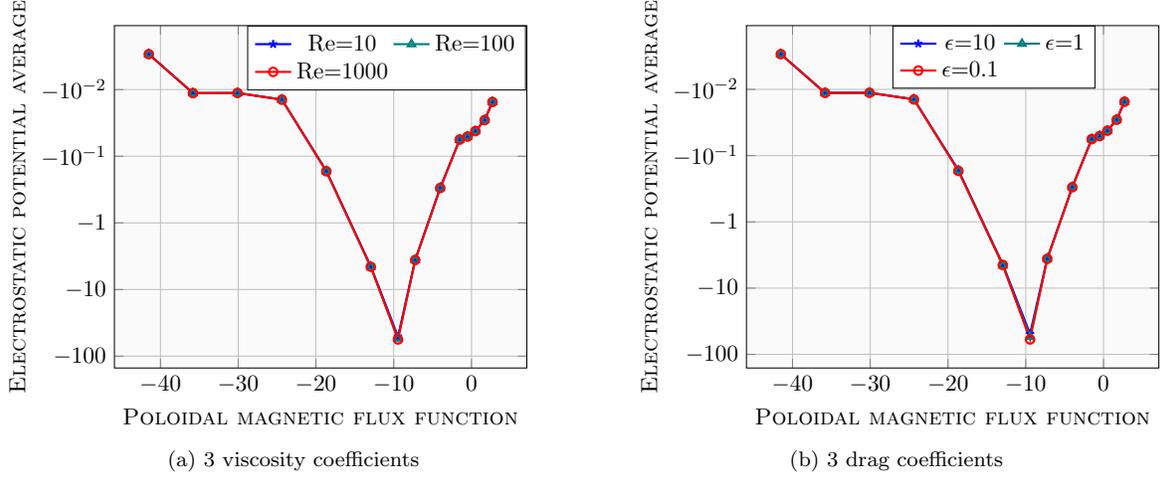
\begin{figure}[!h]
\centering
\begin{subfigure}[b]{0.47\textwidth} 
\begin{tikzpicture}[scale=0.80,>=latex]
\begin{axis}[xlabel=\textsc{{Poloidal magnetic flux function}}, ylabel=\textsc{{Electrostatic potential average}} , ytick={4,3,2,1,0,-1,-2}, yticklabels={$-10^{-4}$,$-10^{-3}$,$-10^{-2}$,$-10^{-1}$,$-1$,$-10$,$-100$}, ymajorgrids, xmajorgrids, axis background/.style={fill=gray!4}, legend style={legend columns=2,at={(0.66,1)},anchor=north}]
\addplot+[color=blue,line width=1pt,mark=star] table[x=nonnormx, y=log] \tableseven;
\addplot+[color=teal,line width=1pt,mark=triangle] table[x=nonnormx, y=log] \tableeight;
\addplot+[color=red,line width=1pt,mark=o] table[x=nonnormx, y=log] \tablenine;
\legend{Re=10,Re=100,Re=1000}
\end{axis}
\end{tikzpicture}
\caption{3 viscosity coefficients}
\label{fig:EP_avg_psi_isovolumes_visc_50dt}
\end{subfigure}%
~
\begin{subfigure}[b]{0.47\textwidth} 
\centering
\begin{tikzpicture}[scale=0.80,>=latex]
\begin{axis}[xlabel=\textsc{{Poloidal magnetic flux function}}, ylabel=\textsc{{Electrostatic potential average}} , ytick={4,3,2,1,0,-1,-2}, yticklabels={$-10^{-4}$,$-10^{-3}$,$-10^{-2}$,$-10^{-1}$,$-1$,$-10$,$-100$}, ymajorgrids, xmajorgrids, axis background/.style={fill=gray!4}, legend style={legend columns=2,at={(0.6,1)},anchor=north}]
\addplot+[color=blue,line width=1pt,mark=star] table[x=nonnormx, y=log] \tableten;
\addplot+[color=teal,line width=1pt,mark=triangle] table[x=nonnormx, y=log] \tableeleven;
\addplot+[color=red,line width=1pt,mark=o] table[x=nonnormx, y=log] \tabletwelve;
\legend{$\epsilon$=10,$\epsilon$=1,$\epsilon$=0.1}
\end{axis}
\end{tikzpicture}
\caption{3 drag coefficients}
\label{fig:EP_avg_psi_isovolumes_drag_50dt}
\end{subfigure}
\caption{Evolution of the electrostatic potential average over poloidal magnetic flux function with different artificial viscosity and fictitious drag coefficients after 50 time steps.}
\label{fig:EP_avg_psi_isovolumes_50dt}
\end{figure}


\pgfplotstableread{data_to_plot/datadampV+1_56dt.txt}\tablenineteen
\pgfplotstablecreatecol[create col/expr={-log10(-\thisrow{y} * 55000000)}]{log}\tablenineteen
\pgfplotstablecreatecol[create col/expr={5*(\thisrow{x})}]{nonnormx}\tablenineteen

\pgfplotstableread{data_to_plot/datadampV0_56dt.txt}\tabletwenty
\pgfplotstablecreatecol[create col/expr={-log10(-\thisrow{y} * 55000000)}]{log}\tabletwenty
\pgfplotstablecreatecol[create col/expr={5*(\thisrow{x})}]{nonnormx}\tabletwenty

\pgfplotstableread{data_to_plot/datadampV-1_56dt.txt}\tabletwentyone
\pgfplotstablecreatecol[create col/expr={-log10(-\thisrow{y} * 55000000)}]{log}\tabletwentyone
\pgfplotstablecreatecol[create col/expr={5*(\thisrow{x})}]{nonnormx}\tabletwentyone

\pgfplotstableread{data_to_plot/dataRe10_56dt.txt}\tabletwentytwo
\pgfplotstablecreatecol[create col/expr={-log10(-\thisrow{y} * 55000000)}]{log}\tabletwentytwo
\pgfplotstablecreatecol[create col/expr={5*(\thisrow{x})}]{nonnormx}\tabletwentytwo

\pgfplotstableread{data_to_plot/dataRe100_56dt.txt}\tabletwentythree
\pgfplotstablecreatecol[create col/expr={-log10(-\thisrow{y} * 55000000)}]{log}\tabletwentythree
\pgfplotstablecreatecol[create col/expr={5*(\thisrow{x})}]{nonnormx}\tabletwentythree

\pgfplotstableread{data_to_plot/dataRe1000_56dt.txt}\tabletwentyfour
\pgfplotstablecreatecol[create col/expr={-log10(-\thisrow{y} * 55000000)}]{log}\tabletwentyfour
\pgfplotstablecreatecol[create col/expr={5*(\thisrow{x})}]{nonnormx}\tabletwentyfour

\begin{figure}[!h]
\centering
\begin{subfigure}[b]{0.47\textwidth} 
\begin{tikzpicture}[scale=0.80,>=latex]
\begin{axis}[xlabel=\textsc{{Poloidal magnetic flux function}}, ylabel=\textsc{{Electrostatic potential average}} , ytick={4,3,2,1,0,-1,-2}, yticklabels={$-10^{-4}$,$-10^{-3}$,$-10^{-2}$,$-10^{-1}$,$-1$,$-10$,$-100$}, ymajorgrids, xmajorgrids, axis background/.style={fill=gray!4}, legend style={legend columns=2,at={(0.66,1)},anchor=north}]
\addplot+[color=blue,line width=1pt,mark=star] table[x=nonnormx, y=log] \tabletwentytwo;
\addplot+[color=teal,line width=1pt,mark=triangle] table[x=nonnormx, y=log] \tabletwentythree;
\addplot+[color=red,line width=1pt,mark=o] table[x=nonnormx, y=log] \tabletwentyfour;
\legend{Re=10,Re=100,Re=1000}
\end{axis}
\end{tikzpicture}
\caption{3 viscosity coefficients}
\label{fig:EP_avg_psi_isovolumes_visc_56dt}
\end{subfigure}%
~
\begin{subfigure}[b]{0.47\textwidth} 
\centering
\begin{tikzpicture}[scale=0.80,>=latex]
\begin{axis}[xlabel=\textsc{{Poloidal magnetic flux function}}, ylabel=\textsc{{Electrostatic potential average}} , ytick={4,3,2,1,0,-1,-2}, yticklabels={$-10^{-4}$,$-10^{-3}$,$-10^{-2}$,$-10^{-1}$,$-1$,$-10$,$-100$}, ymajorgrids, xmajorgrids, axis background/.style={fill=gray!4}, legend style={legend columns=2,at={(0.6,1)},anchor=north}]
\addplot+[color=blue,line width=1pt,mark=star] table[x=nonnormx, y=log] \tablenineteen;
\addplot+[color=teal,line width=1pt,mark=triangle] table[x=nonnormx, y=log] \tabletwenty;
\addplot+[color=red,line width=1pt,mark=o] table[x=nonnormx, y=log] \tabletwentyone;
\legend{$\epsilon$=10,$\epsilon$=1,$\epsilon$=0.1}
\end{axis}
\end{tikzpicture}
\caption{3 drag coefficients}
\label{fig:EP_avg_psi_isovolumes_drag_56dt}
\end{subfigure}
\caption{Evolution of the electrostatic potential average over poloidal magnetic flux function with different artificial viscosity and fictitious drag coefficients after 56 time steps.}
\label{fig:EP_avg_psi_isovolumes_56dt}
\end{figure}
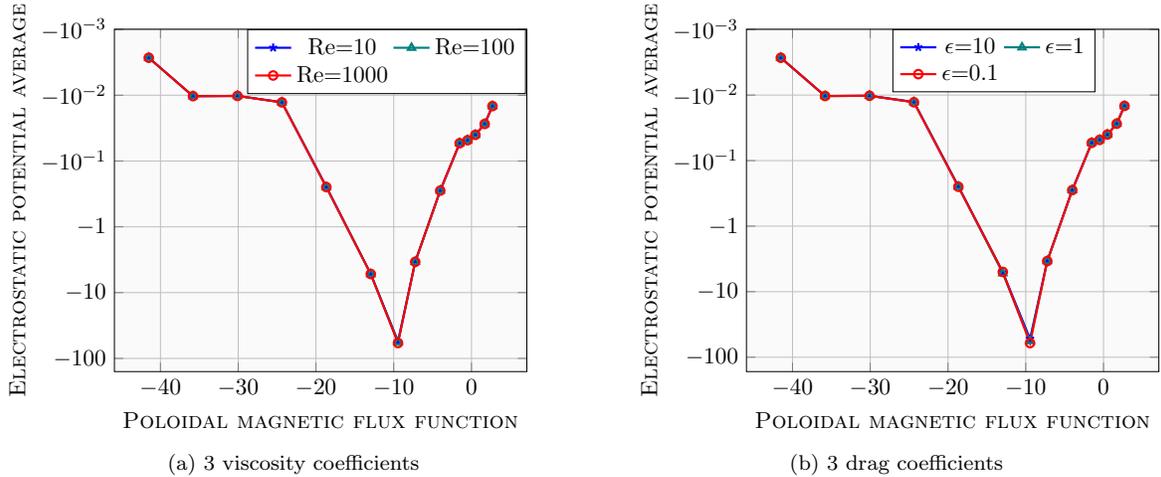

\begin{figure}[!h]
\centering
\begin{subfigure}[b]{0.32\textwidth} 
\begin{center}
\includegraphics[width=\textwidth]{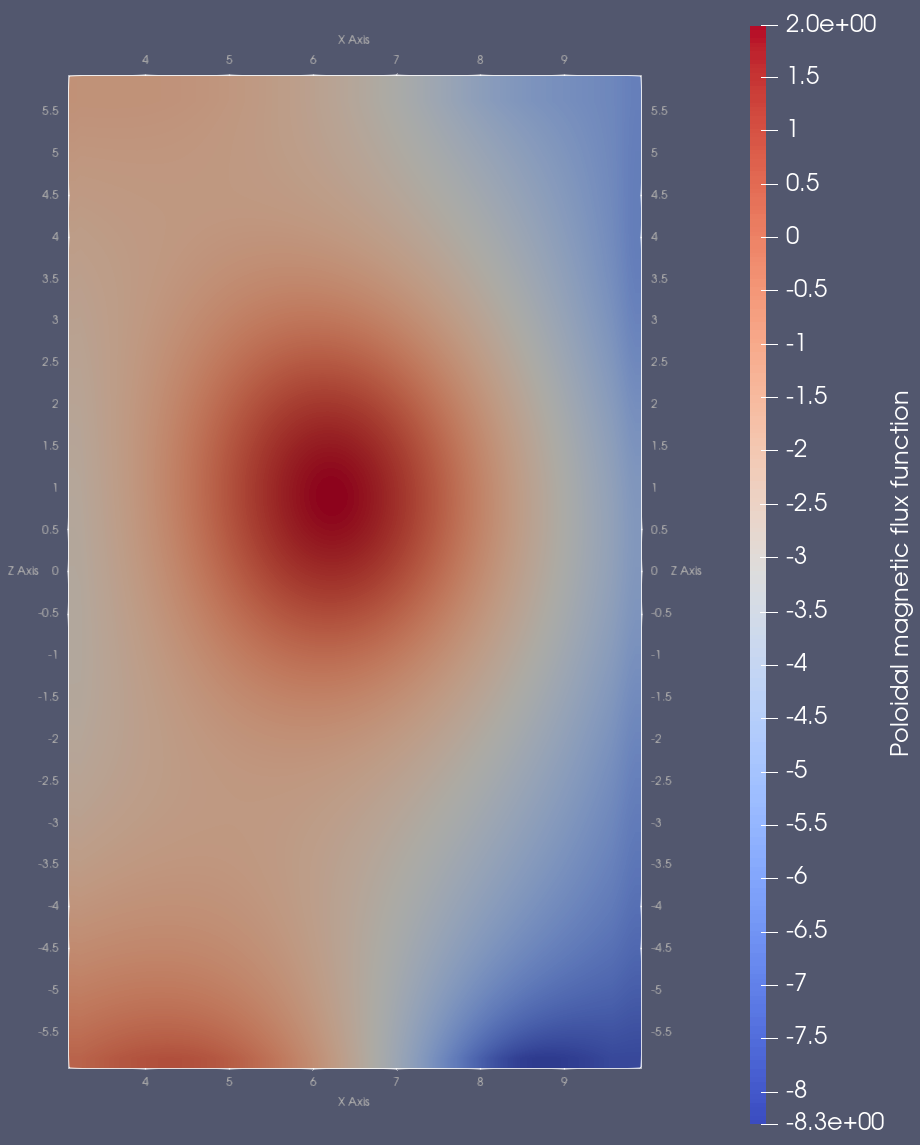}
\caption{${\mathrm Re}=10$}
\label{fig:psi_drag_10}
\end{center}
\end{subfigure}%
~
\begin{subfigure}[b]{0.32\textwidth} 
\begin{center}
\includegraphics[width=\textwidth]{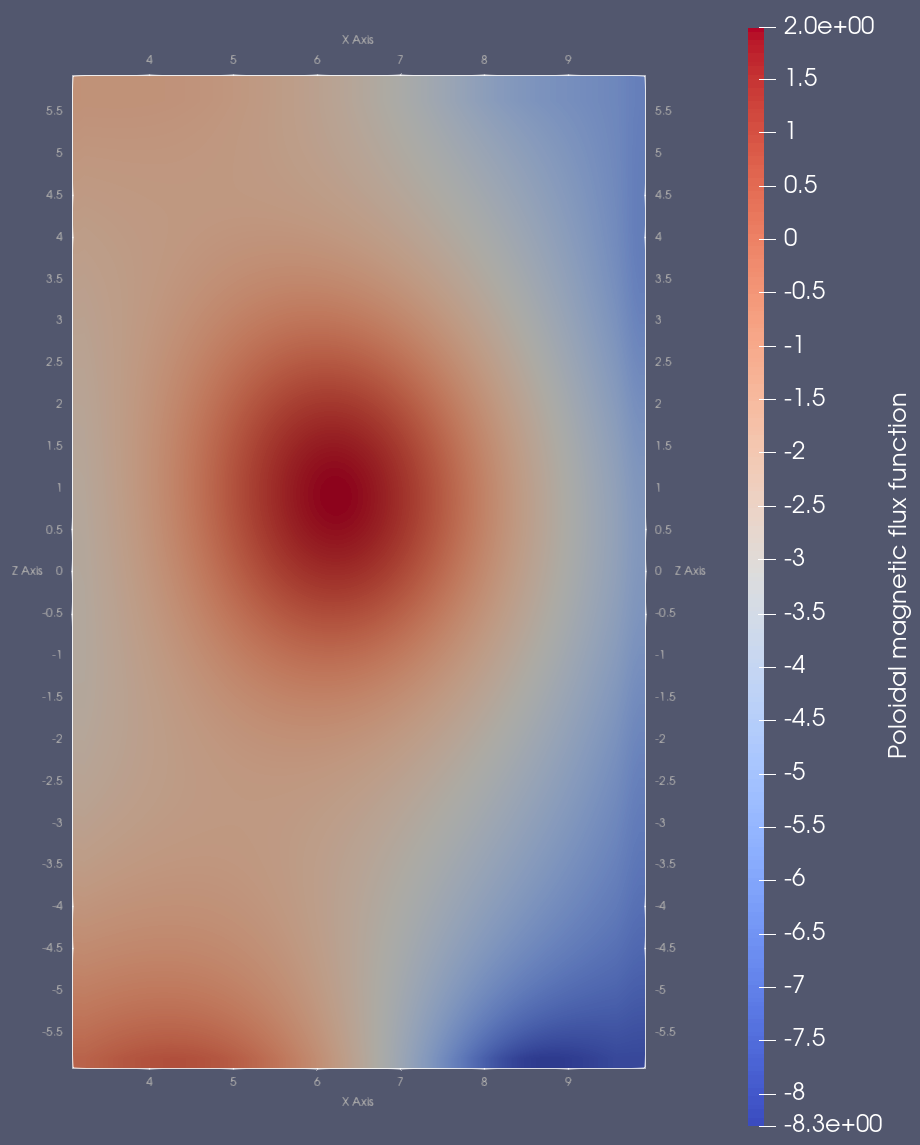}
\caption{${\mathrm Re}=100$}
\label{fig:psi_drag_100}
\end{center}
\end{subfigure}%
~
\begin{subfigure}[b]{0.32\textwidth} 
\begin{center}
\includegraphics[width=\textwidth]{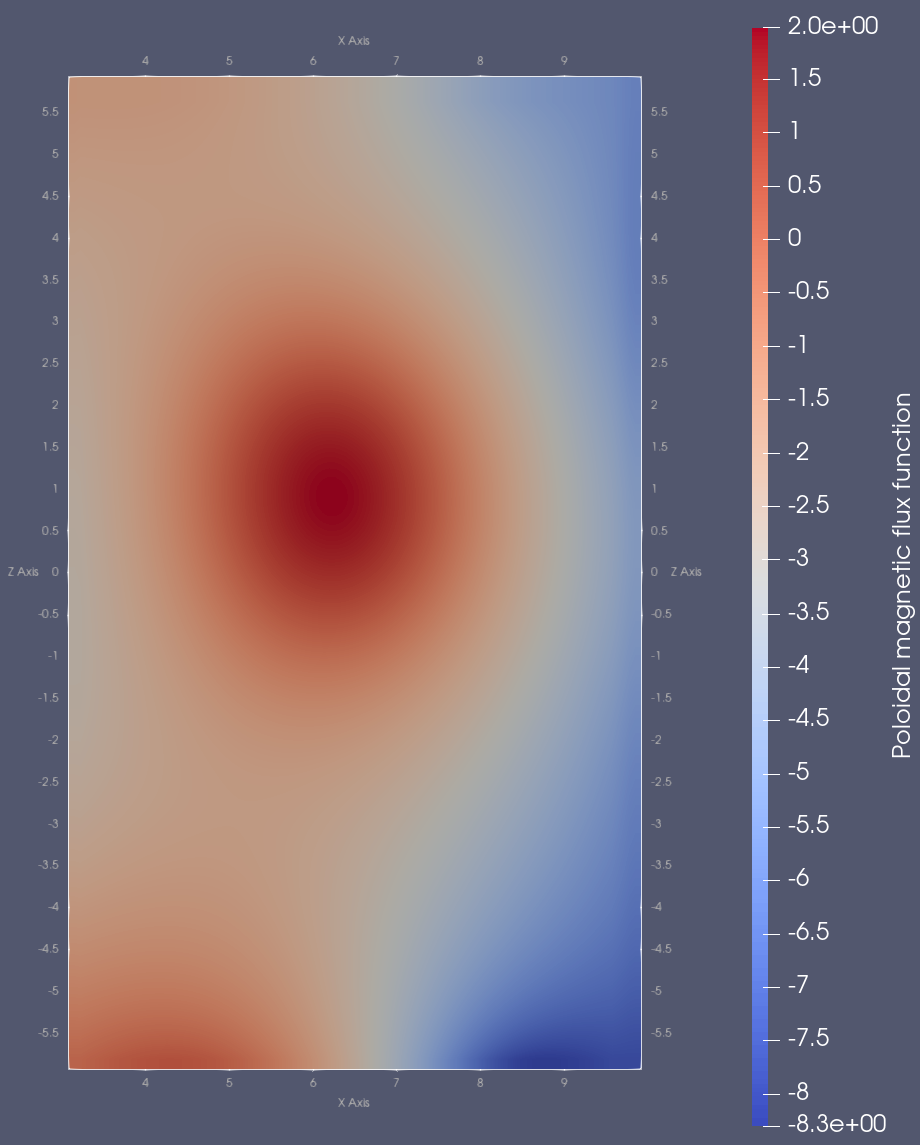}
\caption{${\mathrm Re}=1000$}
\label{fig:psi_drag_1000}
\end{center}
\end{subfigure}
\caption{The poloidal magnetic flux function with different artificial viscosity coefficients at the first time step}
\label{fig:psi_1st_step}
\end{figure}

\begin{figure}[!h]
\centering
\begin{subfigure}[b]{0.32\textwidth} 
\begin{center}
\includegraphics[width=\textwidth]{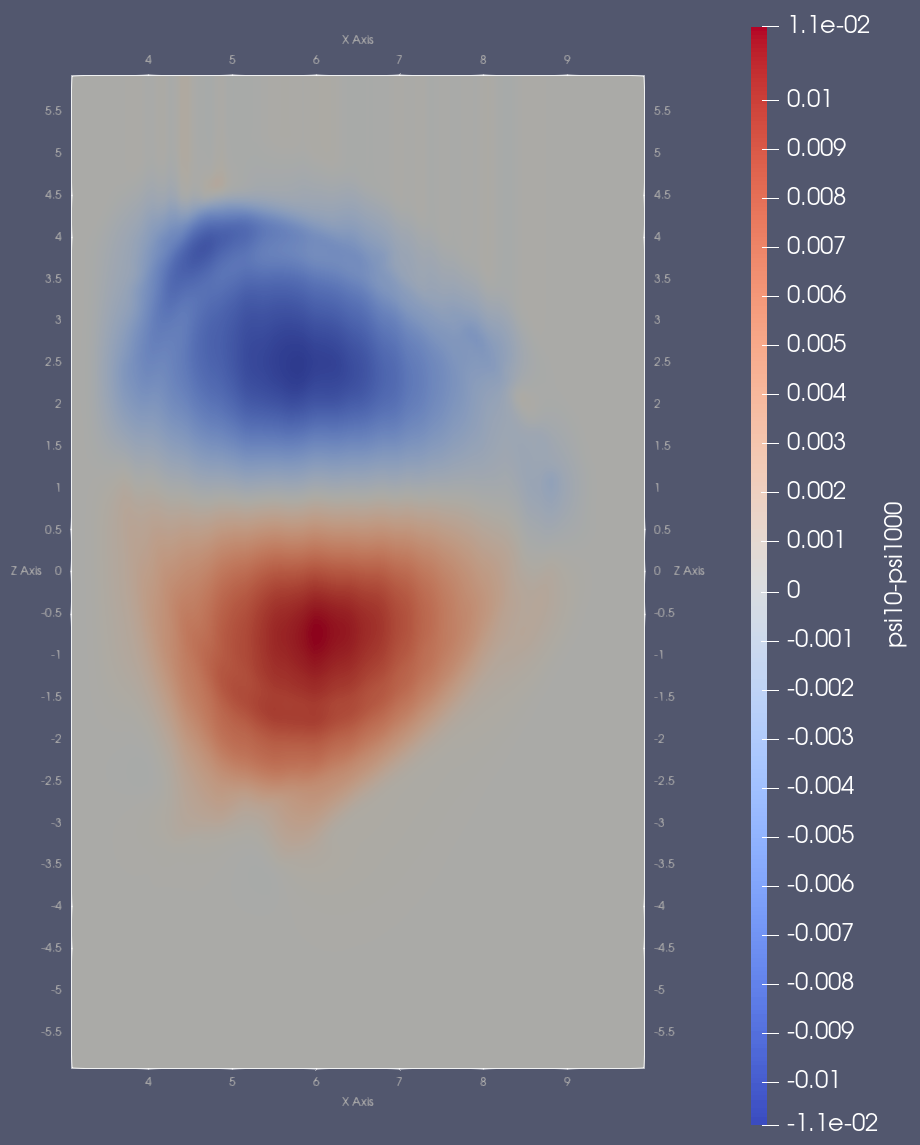}
\caption{$\Psi({\mathrm Re}=10) - \Psi({\mathrm Re}=1000)$}
\label{fig:Deltapsi_drag_10}
\end{center}
\end{subfigure}%
~
\begin{subfigure}[b]{0.32\textwidth} 
\begin{center}
\includegraphics[width=\textwidth]{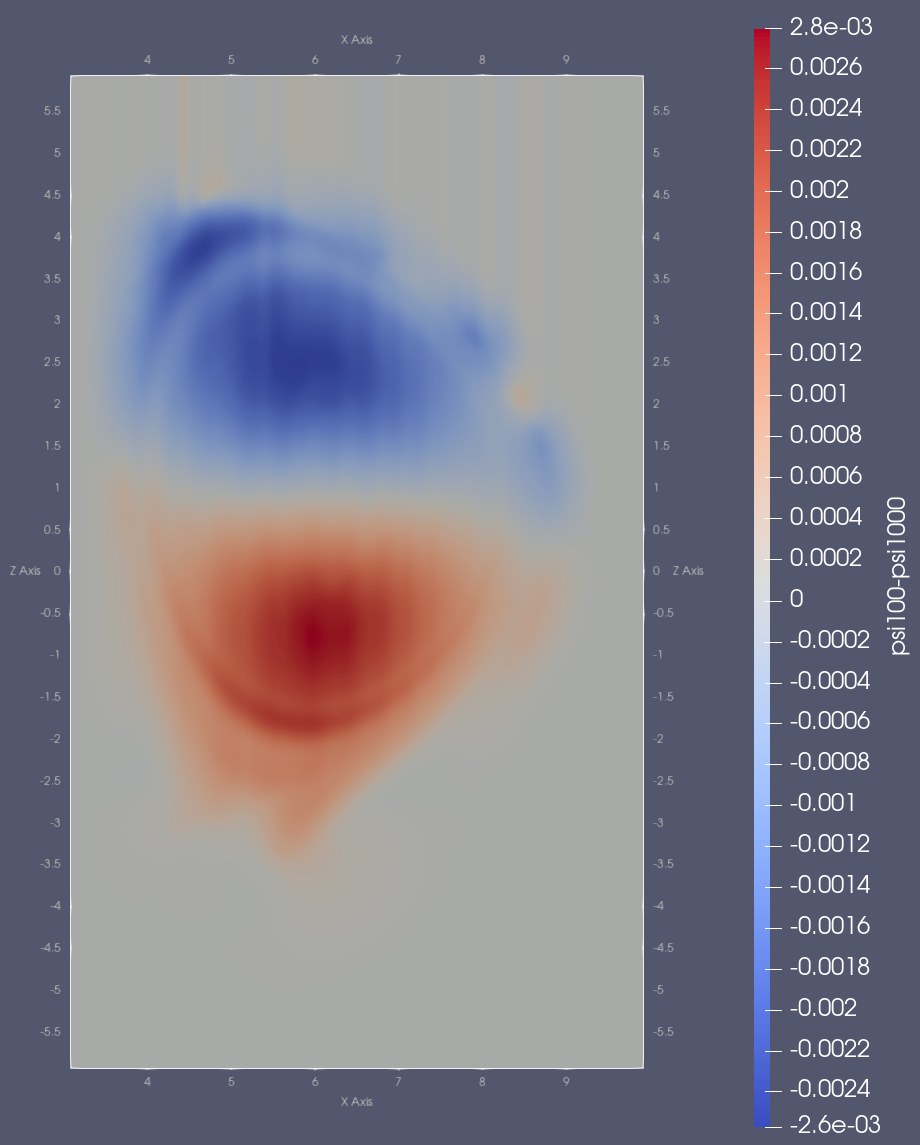}
\caption{$\Psi({\mathrm Re}=100) - \Psi({\mathrm Re}=1000)$}
\label{fig:Deltapsi_drag_100}
\end{center}
\end{subfigure}%
~
\begin{subfigure}[b]{0.32\textwidth} 
\begin{center}
\includegraphics[width=\textwidth]{psi_Re1000.png}
\caption{$\Psi({\mathrm Re}=1000)$}
\label{fig:psi_drag_1000_bis}
\end{center}
\end{subfigure}
\caption{The poloidal magnetic flux function for ${\mathrm Re} = 1000$ and the differences $\Psi({\mathrm Re}=100) - \Psi({\mathrm Re}=1000)$ and $\Psi({\mathrm Re}=10) - \Psi({\mathrm Re}=1000)$ at the first time step.}
\label{fig:deltapsi_1st_step}
\end{figure}

\begin{figure}[!h]
\centering
\begin{subfigure}[b]{0.32\textwidth} 
\begin{center}
\includegraphics[width=\textwidth]{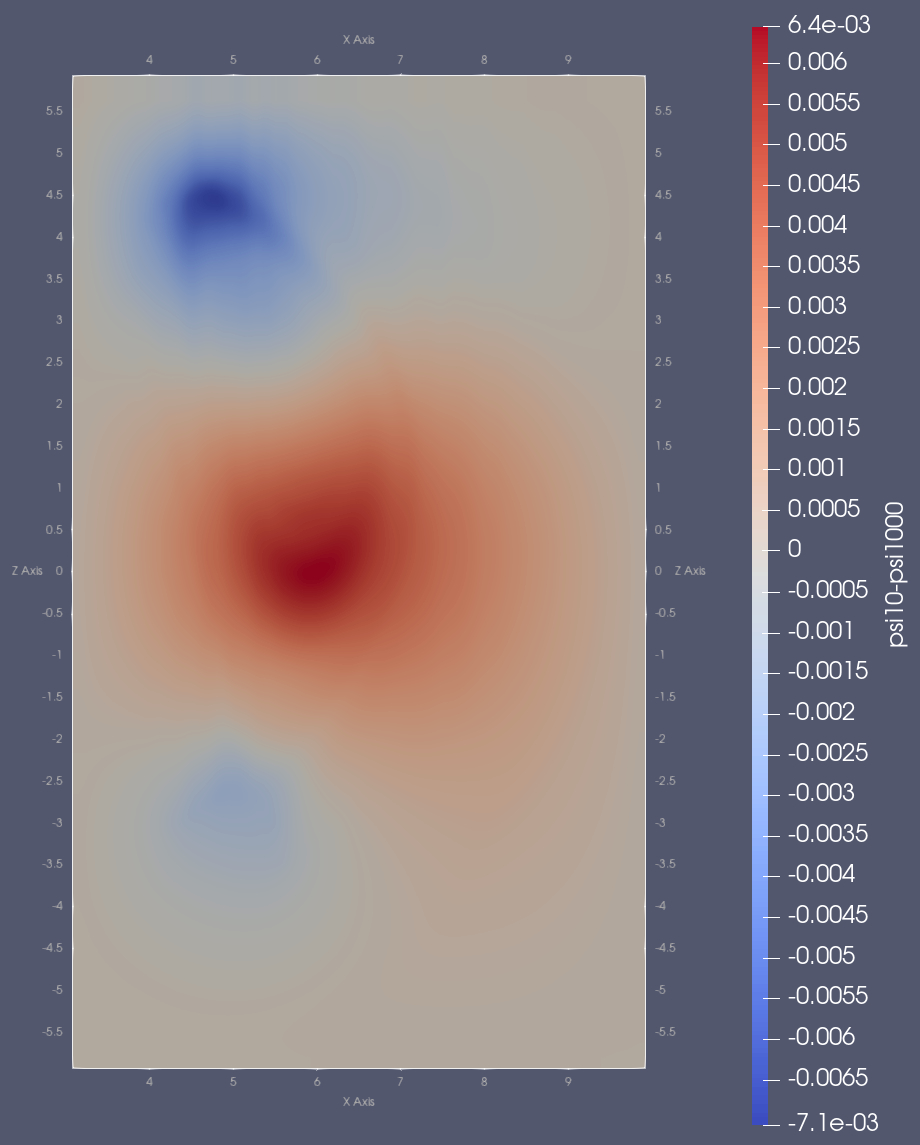}
\caption{$\Psi({\mathrm Re}=10) - \Psi({\mathrm Re}=1000)$}
\label{fig:Deltapsi_drag_10_t50}
\end{center}
\end{subfigure}%
~
\begin{subfigure}[b]{0.32\textwidth} 
\begin{center}
\includegraphics[width=\textwidth]{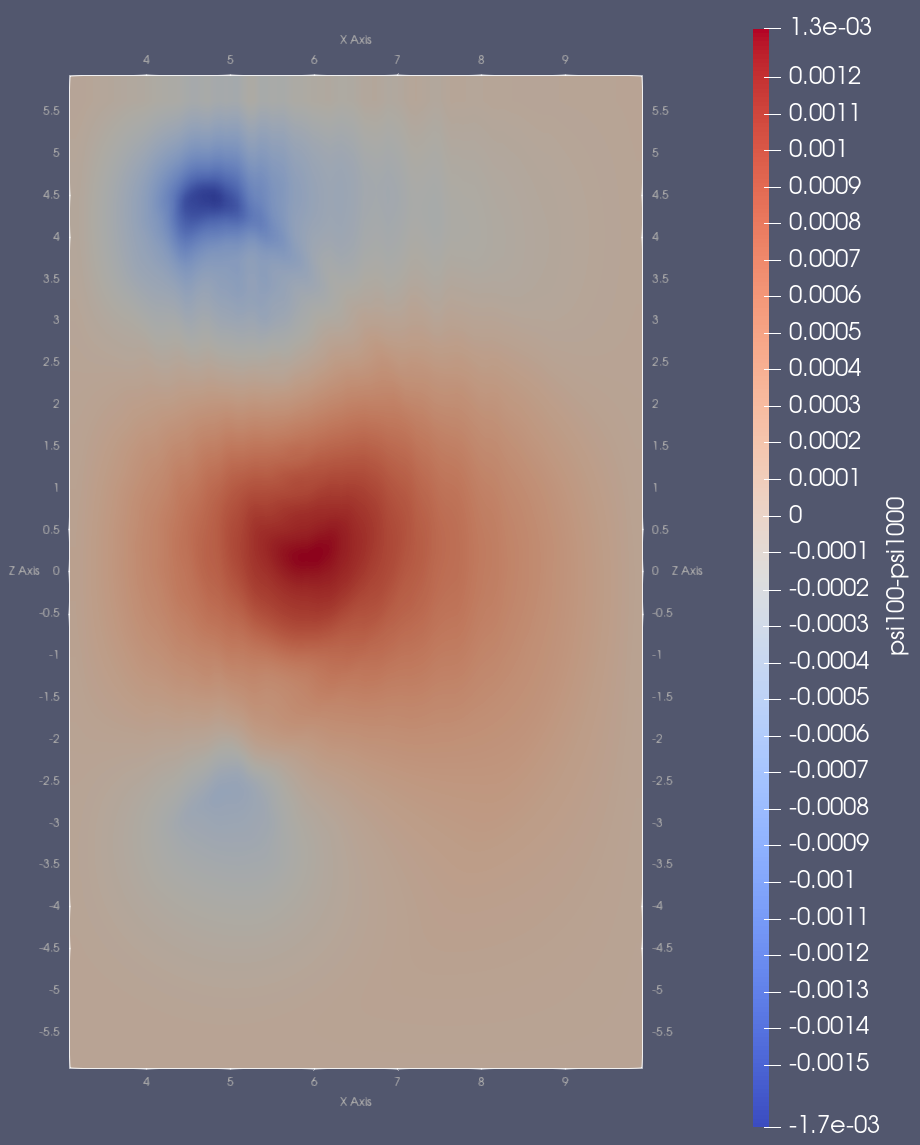}
\caption{$\Psi({\mathrm Re}=100) - \Psi({\mathrm Re}=1000)$}
\label{fig:Deltapsi_drag_100_t50}
\end{center}
\end{subfigure}%
~
\begin{subfigure}[b]{0.32\textwidth} 
\begin{center}
\includegraphics[width=\textwidth]{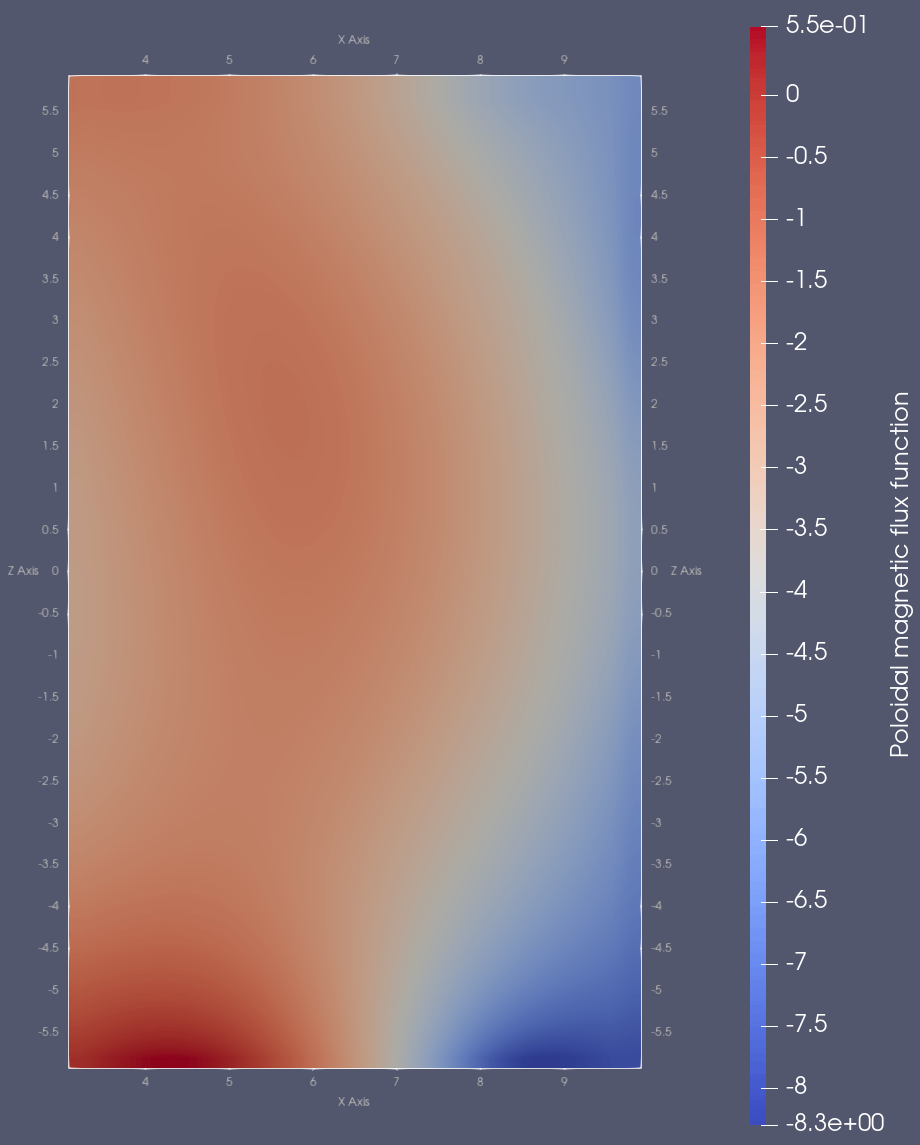}
\caption{$\Psi({\mathrm Re}=1000)$}
\label{fig:psi_drag_1000_t50}
\end{center}
\end{subfigure}
\caption{The poloidal magnetic flux function for ${\mathrm Re} = 1000$ and the differences $\Psi({\mathrm Re}=100) - \Psi({\mathrm Re}=1000)$ and $\Psi({\mathrm Re}=10) - \Psi({\mathrm Re}=1000)$ after 50 time steps.}
\label{fig:deltapsi_50th_step}
\end{figure}

\begin{figure}[!h]
\centering
\begin{subfigure}[b]{0.32\textwidth} 
\begin{center}
\includegraphics[width=\textwidth]{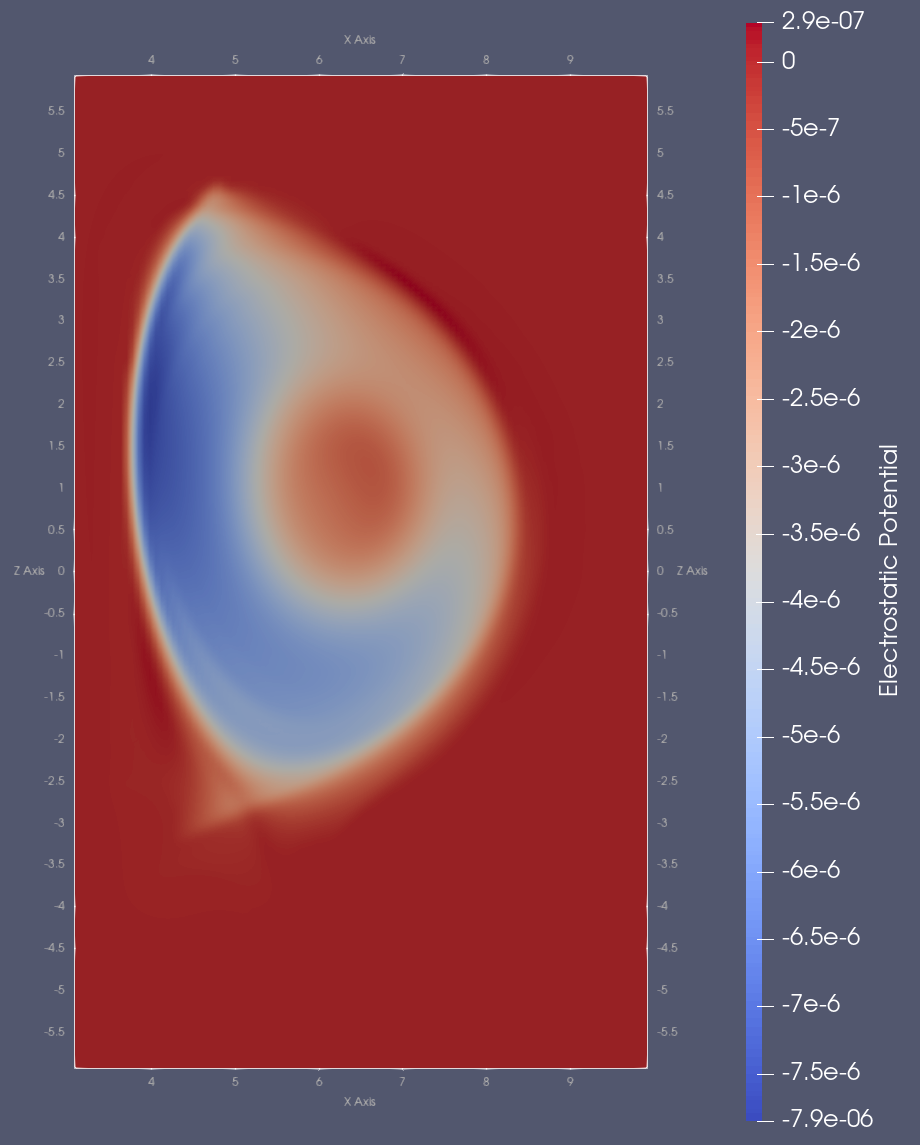}
\caption{${\mathrm Re}=10$}
\label{fig:EP_drag_10}
\end{center}
\end{subfigure}%
~
\begin{subfigure}[b]{0.32\textwidth} 
\begin{center}
\includegraphics[width=\textwidth]{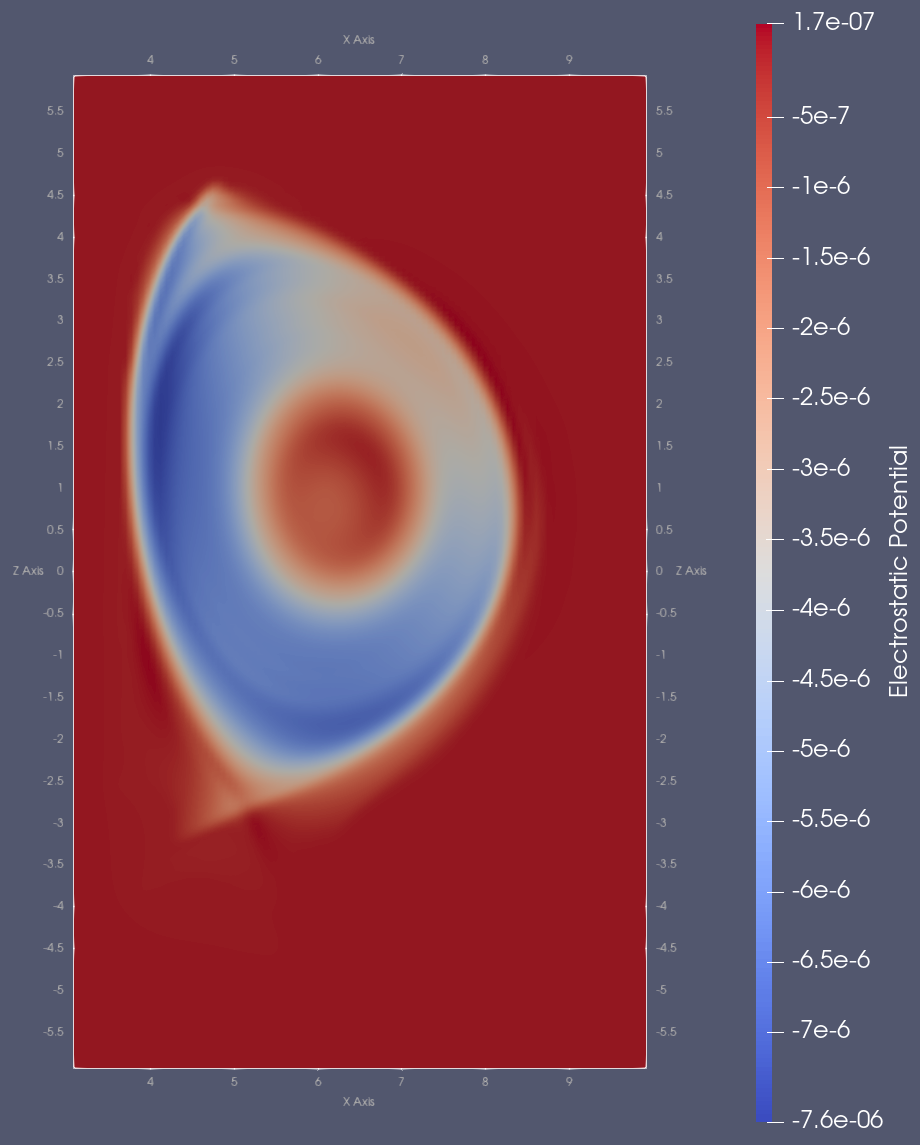}
\caption{${\mathrm Re}=100$}
\label{fig:EP_drag_100}
\end{center}
\end{subfigure}%
~
\begin{subfigure}[b]{0.32\textwidth} 
\begin{center}
\includegraphics[width=\textwidth]{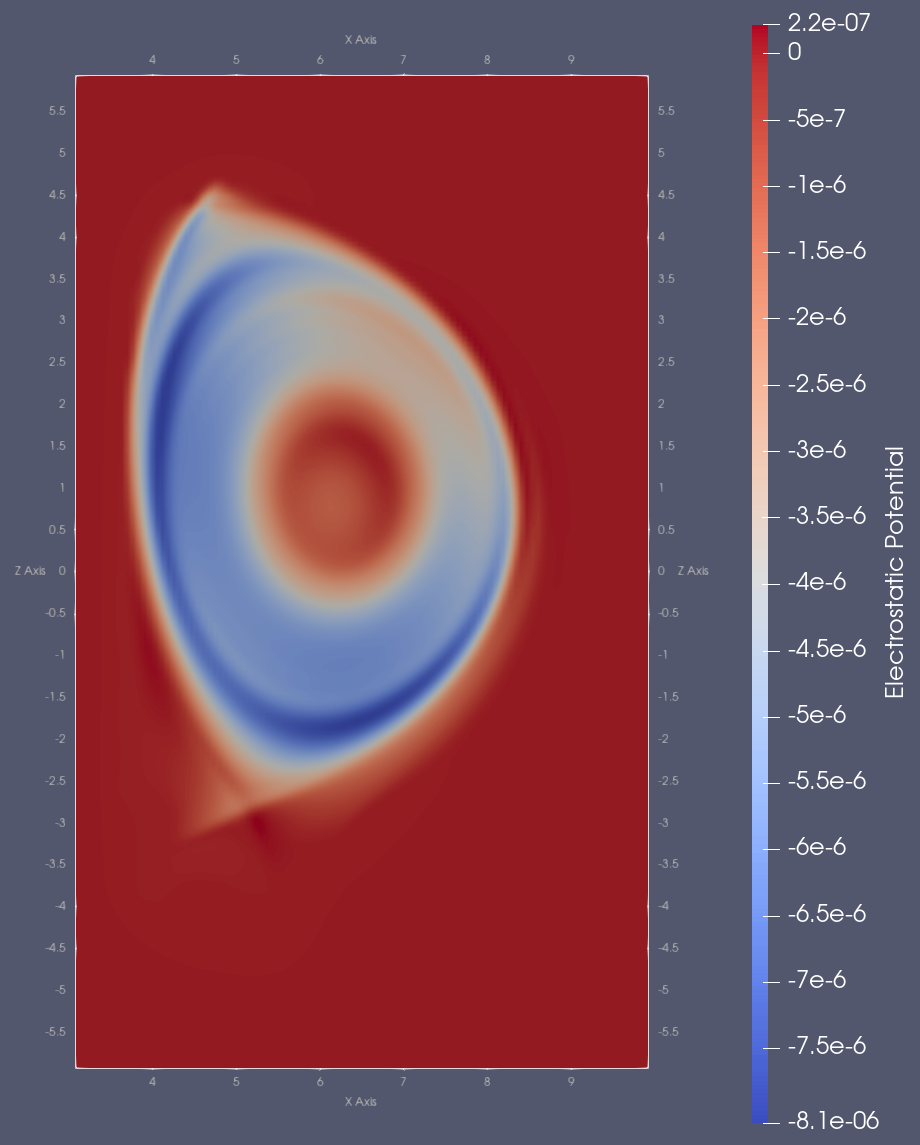}
\caption{${\mathrm Re}=1000$}
\label{fig:EP_drag_1000}
\end{center}
\end{subfigure}
\caption{The electrostatic potential with different artificial viscosity coefficients at the first time step.}
\label{fig:EP_1st_step}
\end{figure}
\begin{figure}[!h]
\centering
\begin{subfigure}[b]{0.32\textwidth} 
\begin{center}
\includegraphics[width=\textwidth]{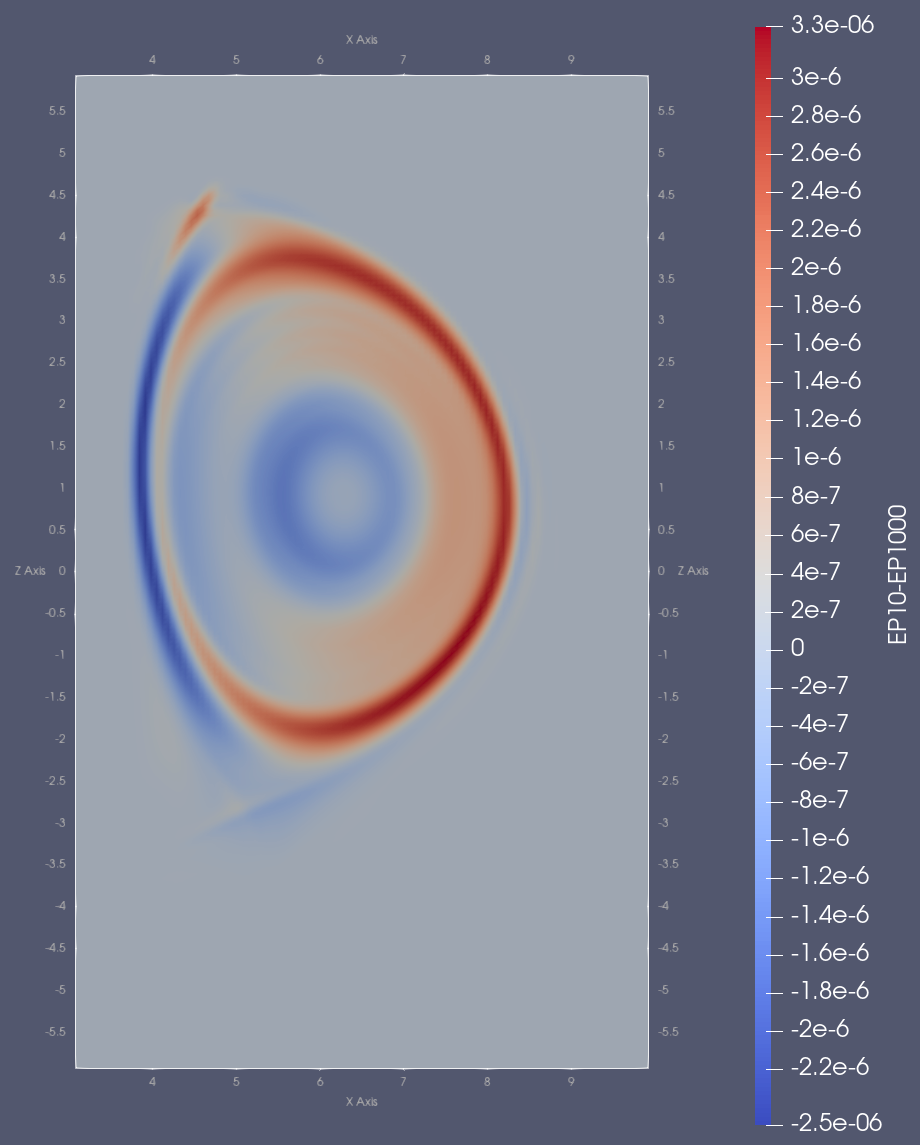}
\caption{$\Phi({\mathrm Re}=10) - \Phi({\mathrm Re}=1000)$}
\label{fig:DeltaEP_drag_10}
\end{center}
\end{subfigure}%
~
\begin{subfigure}[b]{0.32\textwidth} 
\begin{center}
\includegraphics[width=\textwidth]{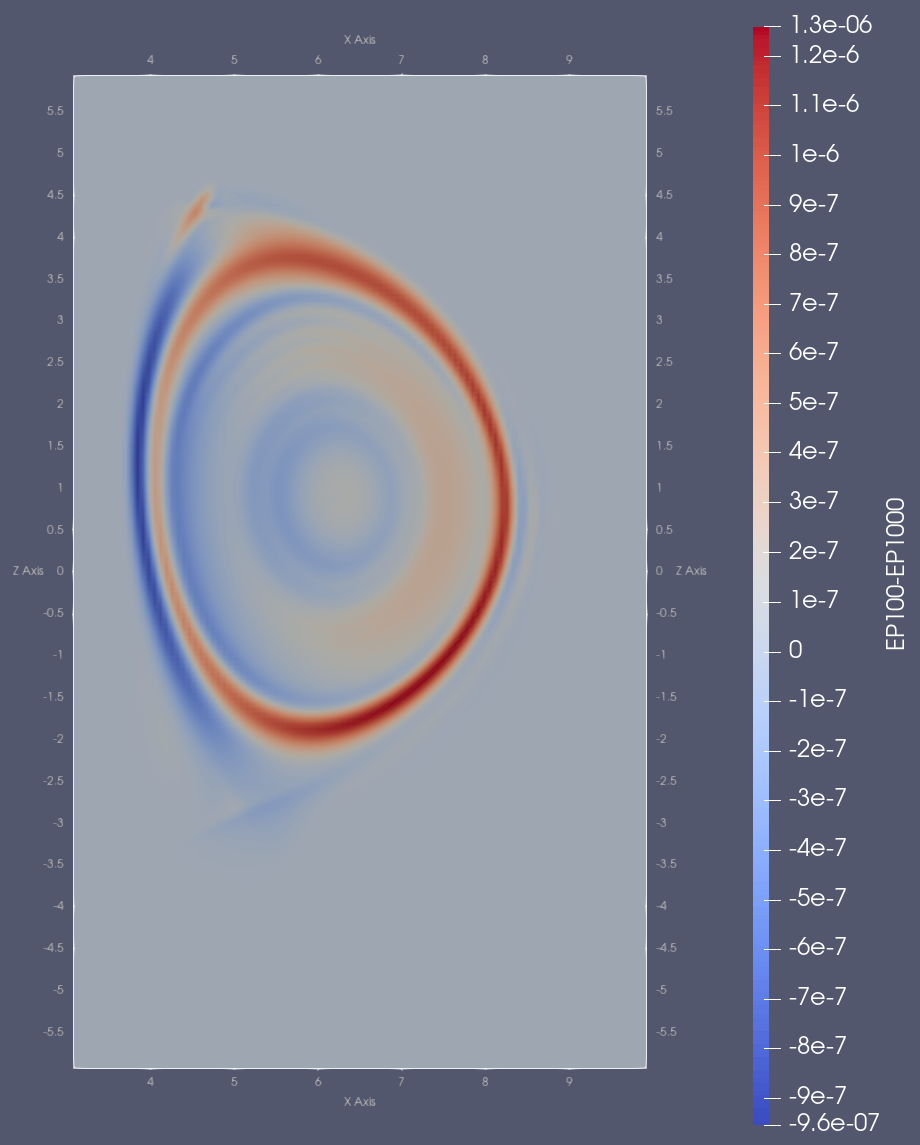}
\caption{$\Phi({\mathrm Re}=100) - \Phi({\mathrm Re}=1000)$}
\label{fig:DeltaEP_drag_100}
\end{center}
\end{subfigure}%
~
\begin{subfigure}[b]{0.32\textwidth} 
\begin{center}
\includegraphics[width=\textwidth]{EP_Re1000.png}
\caption{$\Phi({\mathrm Re}=1000)$}
\label{fig:EP_drag_1000_bis}
\end{center}
\end{subfigure}
\caption{The electrostatic potential for ${\mathrm Re} = 1000$ and the differences $\Phi({\mathrm Re}=100) - \Phi({\mathrm Re}=1000)$ and $\Phi({\mathrm Re}=10) - \Phi({\mathrm Re}=1000)$ at the first time step.}
\label{fig:DeltaEP_1st_step}
\end{figure}

\begin{figure}[!h]
\centering
\begin{subfigure}[b]{0.32\textwidth} 
\begin{center}
\includegraphics[width=\textwidth]{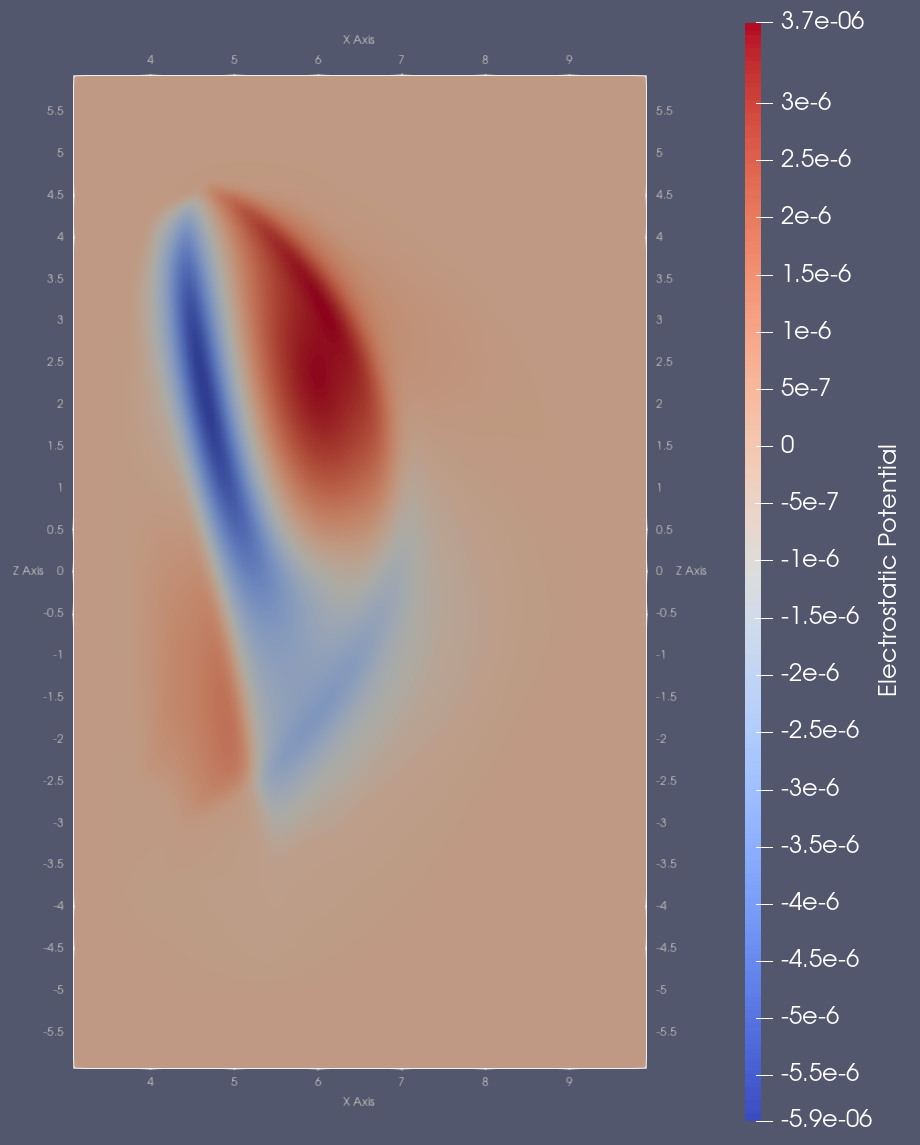}
\caption{${\mathrm Re}=10$}
\label{fig:EP_drag_10_t50}
\end{center}
\end{subfigure}%
~
\begin{subfigure}[b]{0.32\textwidth} 
\begin{center}
\includegraphics[width=\textwidth]{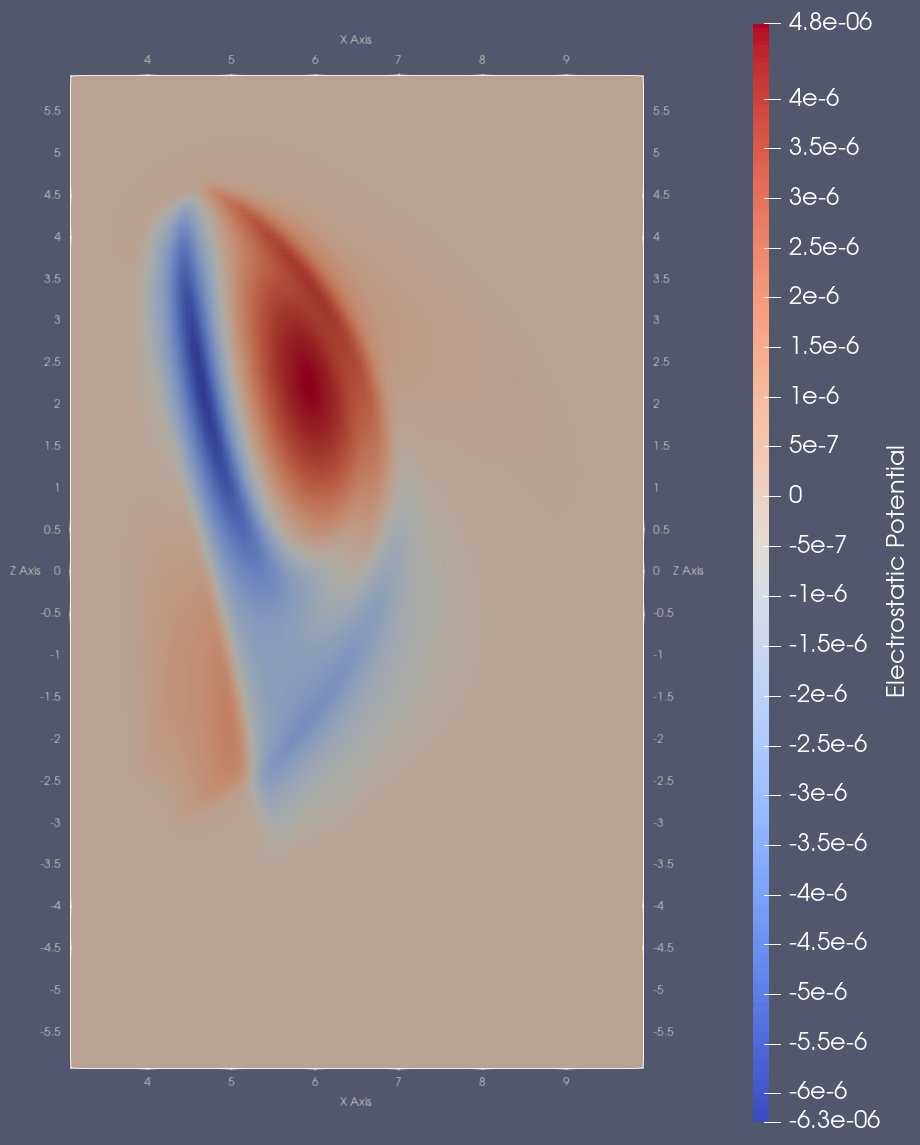}
\caption{${\mathrm Re}=100$}
\label{fig:EP_drag_100_t50}
\end{center}
\end{subfigure}%
~
\begin{subfigure}[b]{0.32\textwidth} 
\begin{center}
\includegraphics[width=\textwidth]{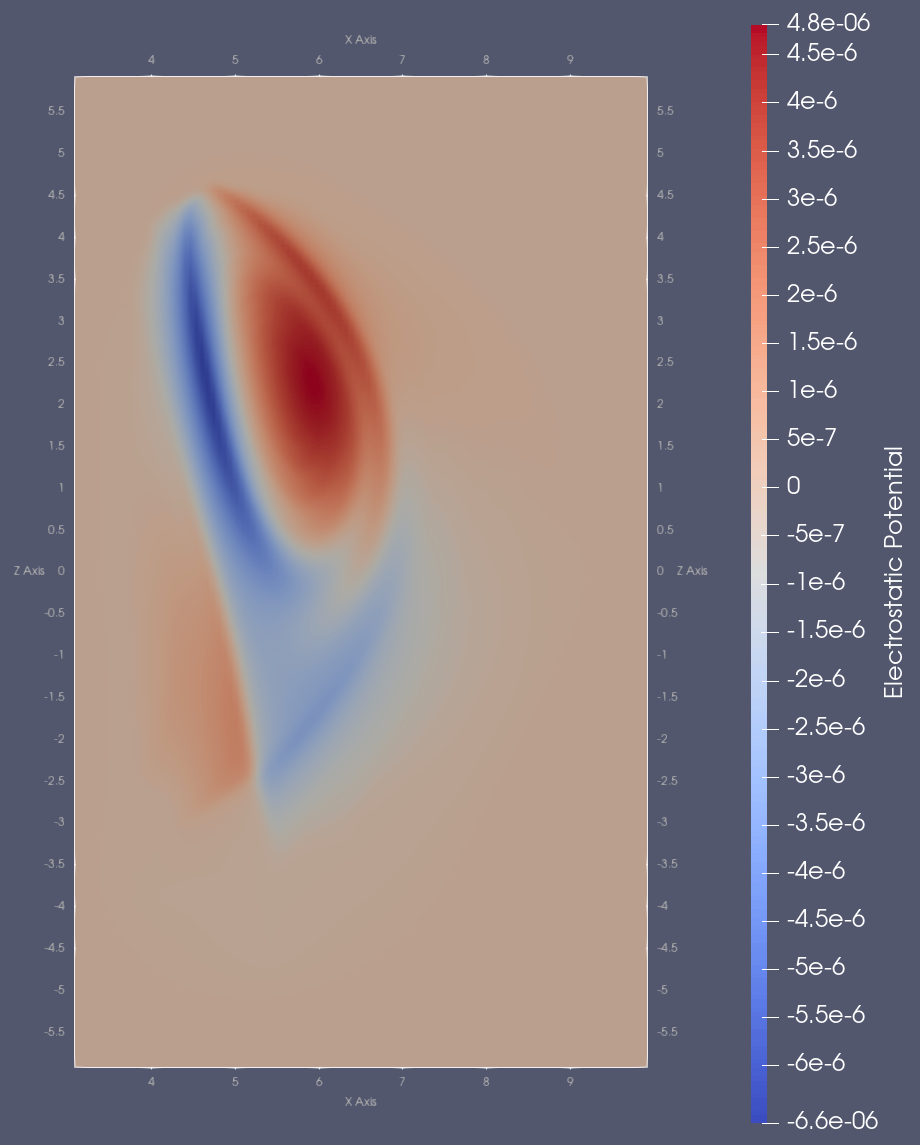}
\caption{${\mathrm Re}=1000$}
\label{fig:EP_drag_1000_t50}
\end{center}
\end{subfigure}
\caption{The electrostatic potential with different artificial viscosity coefficients after 50 time steps.}
\label{fig:EP_50th_step}
\end{figure}

\begin{figure}[!h]
\centering
\begin{subfigure}[b]{0.32\textwidth} 
\begin{center}
\includegraphics[width=\textwidth]{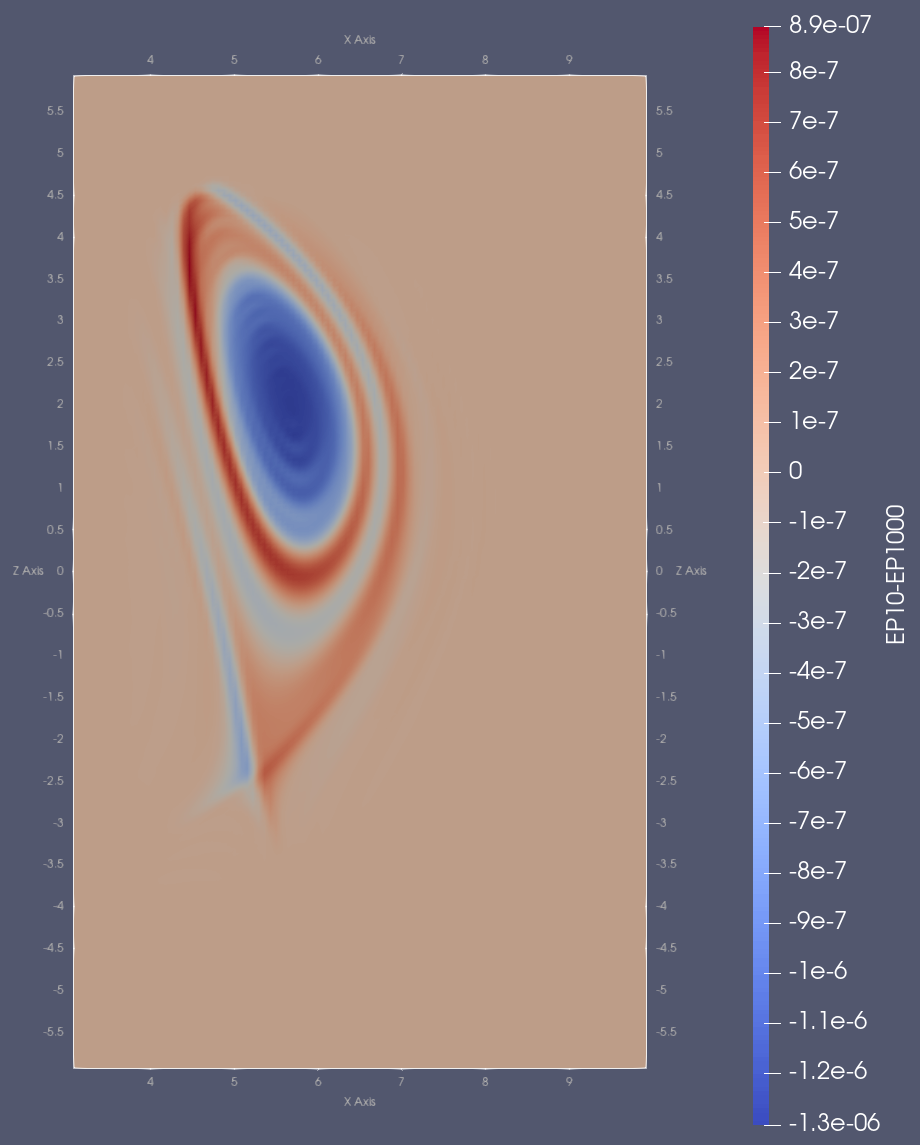}
\caption{$\Phi({\mathrm Re}=10) - \Phi({\mathrm Re}=1000)$}
\label{fig:DeltaEP_drag_10_t50}
\end{center}
\end{subfigure}%
~
\begin{subfigure}[b]{0.32\textwidth} 
\begin{center}
\includegraphics[width=\textwidth]{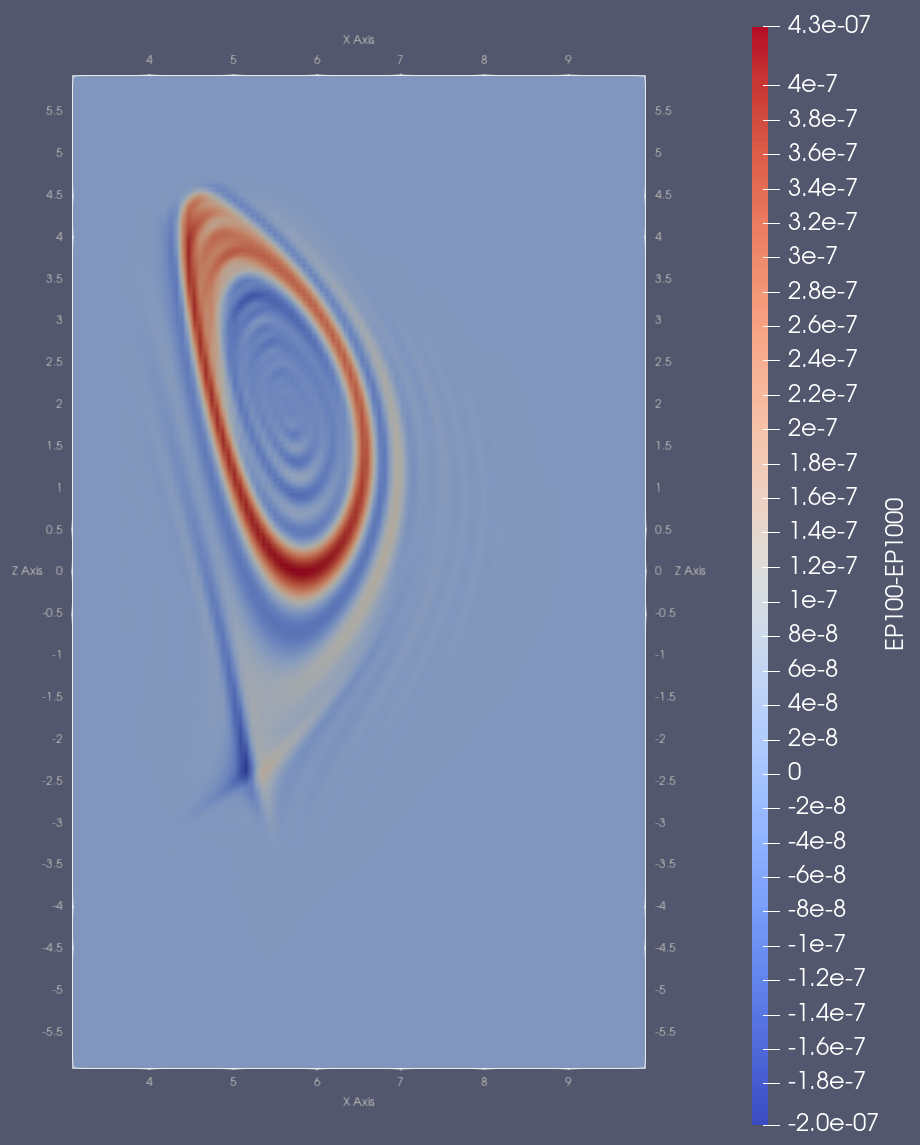}
\caption{$\Phi({\mathrm Re}=100) - \Phi({\mathrm Re}=1000)$}
\label{fig:DeltaEP_drag_100_t50}
\end{center}
\end{subfigure}%
~
\begin{subfigure}[b]{0.32\textwidth} 
\begin{center}
\includegraphics[width=\textwidth]{EP_Re1000_t50.png}
\caption{$\Phi({\mathrm Re}=1000)$}
\label{fig:EP_drag_1000_t50_bis}
\end{center}
\end{subfigure}
\caption{The electrostatic potential for ${\mathrm Re} = 1000$ and the differences $\Phi({\mathrm Re}=100) - \Phi({\mathrm Re}=1000)$ and $\Phi({\mathrm Re}=10) - \Phi({\mathrm Re}=1000)$ after 50 time steps.}
\label{fig:DeltaEP_50th_step}
\end{figure}

 Figures~\ref{fig:psi_1st_step}--\ref{fig:deltapsi_50th_step} and~\ref{fig:EP_1st_step}--\ref{fig:DeltaEP_50th_step} show the poloidal magnetic flux function and the electrostatic potential respectively with different artificial viscosity coefficients at the first time step and after 50 time steps. The former figures show that the poloidal magnetic flux function is identical for the three viscosity coefficients whereas the latter figures show that there are differences in the electrostatic potential but these tend to be rather minimal. 

\subsection{Full cold ITER VDE simulation}
\label{sec:vde}
In this section, we present the results of a full ITER VDE
simulation. We distinguish two settings for the blanket module:
isotropic resistivity and anisotropic resistivity.  In the istropic
case, a higher resistivity inhits both toroidal and poloidal current
flowing inside the blanket.
The wall current mostly resides in the vacuum vessel, driven by inductively
by the inductive electric fild.
For the anisotropic case, the poloidal
resistivity is sufficiently reduced that ploidal current can flow
inside the blanket but the toroidal current is still inhibited by a
higher toroidal resistivity. The poloidal current in the blanket now
provides the pathways for halo current that enters the blanket from
the plasma, poloidally tranverses the blanket, and then exit into the
vacuum vessel, and vice versa.

\subsubsection{Isotropic setting}
\label{sec:vde_isotropic}

{
\begin{figure}[ht]
\centering
\begin{subfigure}[b]{0.32\textwidth} 
\begin{center}
\includegraphics[width=\textwidth]{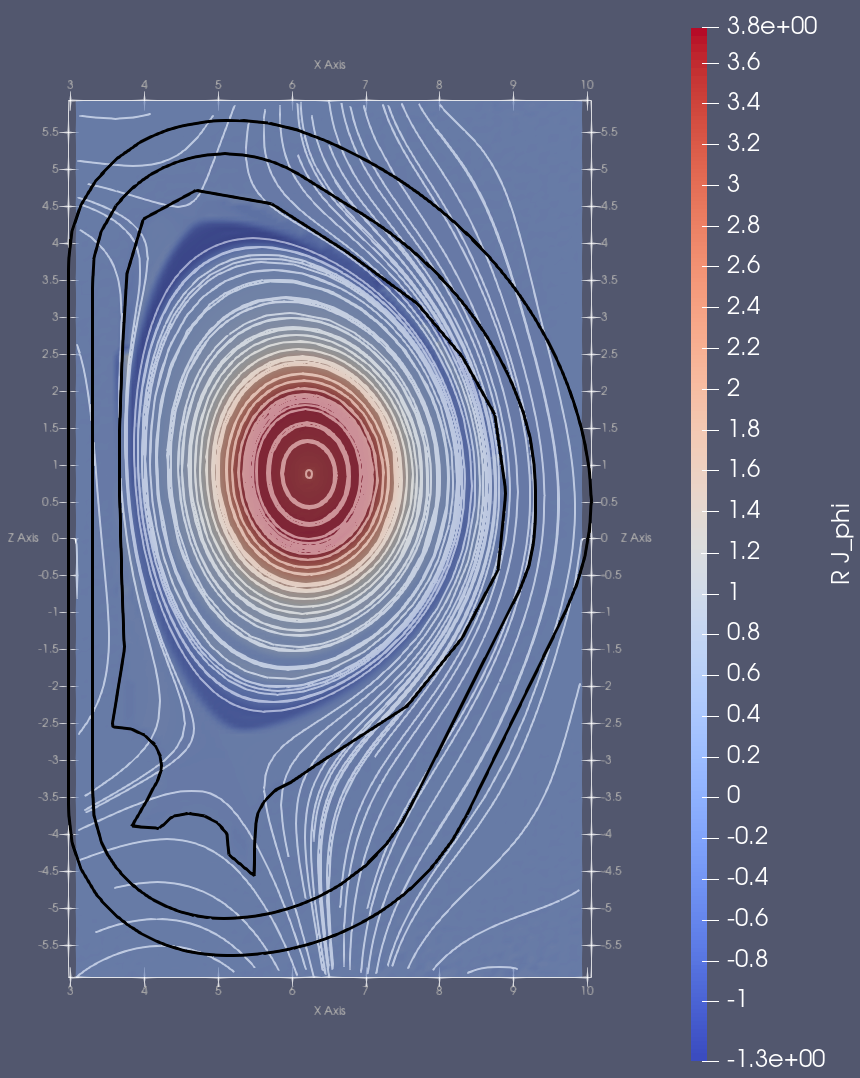}
\caption{$t=\SI{0}{\second}$}
\label{fig:VDE_t0_2}
\end{center}
\end{subfigure}%
~
\begin{subfigure}[b]{0.32\textwidth} 
\begin{center}
\includegraphics[width=\textwidth]{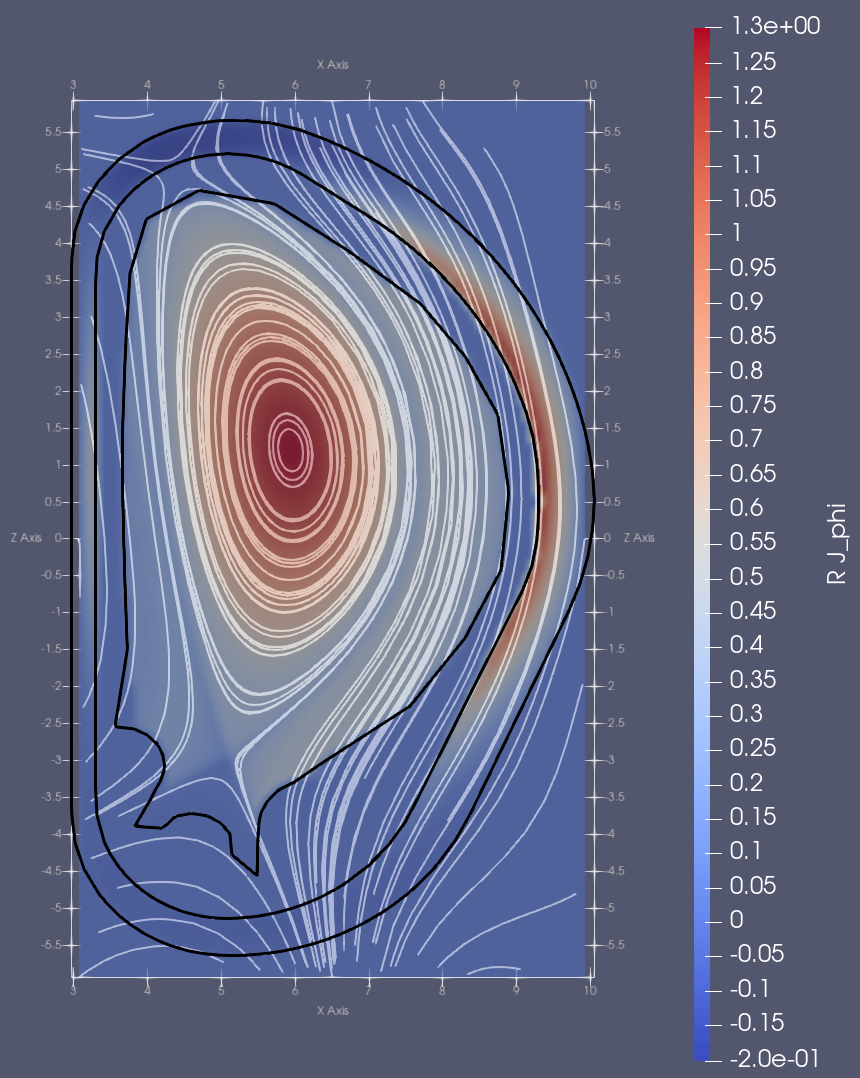}
\caption{$t=\SI{156}{\milli\second}$}
\label{fig:VDE_30dt_2}
\end{center}
\end{subfigure}%
~
\begin{subfigure}[b]{0.32\textwidth} 
\begin{center}
\includegraphics[width=\textwidth]{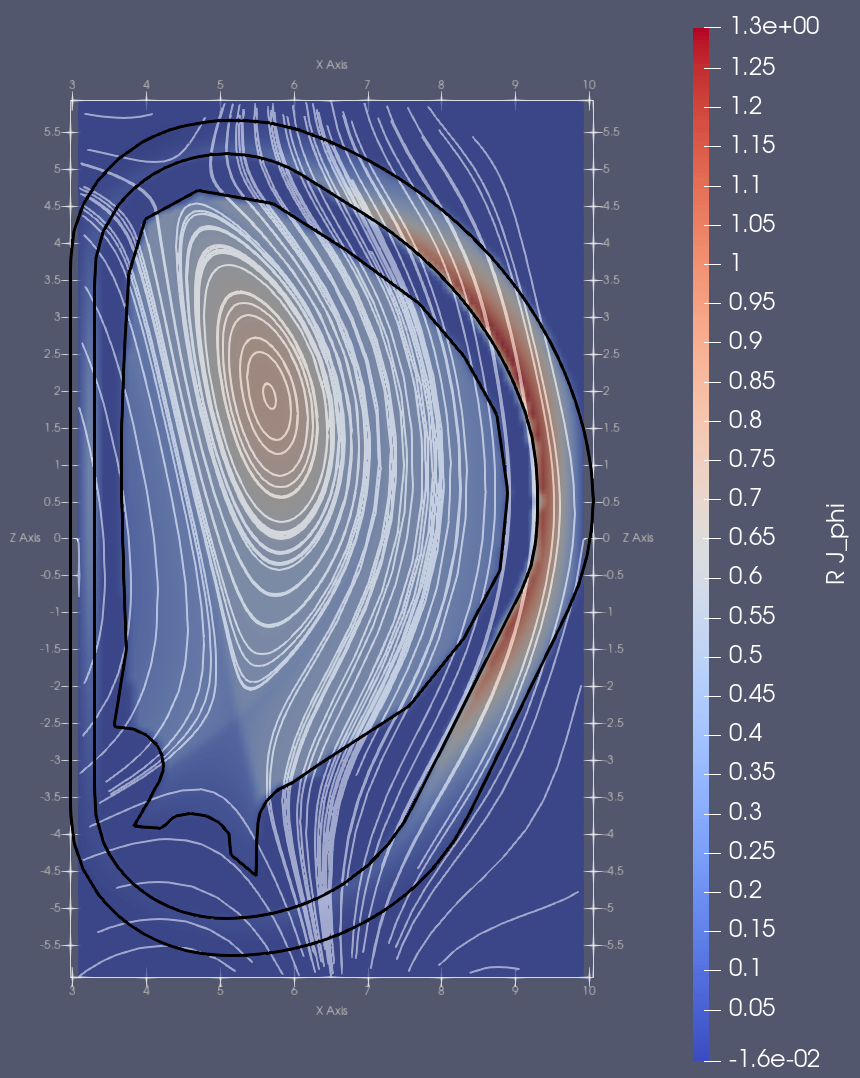}
\caption{$t=\SI{260}{\milli\second}$}
\label{fig:VDE_50dt_2}
\end{center}
\end{subfigure}
\begin{subfigure}[b]{0.32\textwidth} 
\begin{center}
\includegraphics[width=\textwidth]{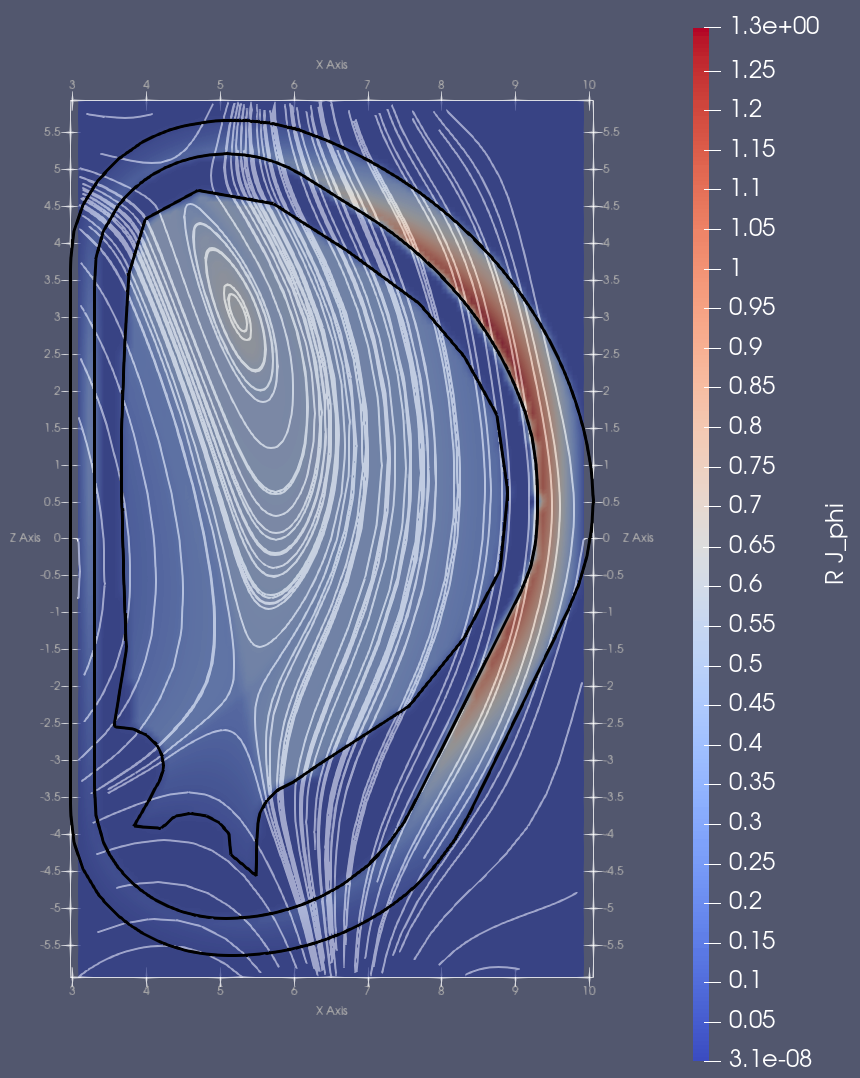}
\caption{$t=\SI{312}{\milli\second}$}
\label{fig:VDE_60dt_2}
\end{center}
\end{subfigure}%
~
\begin{subfigure}[b]{0.32\textwidth} 
\begin{center}
\includegraphics[width=\textwidth]{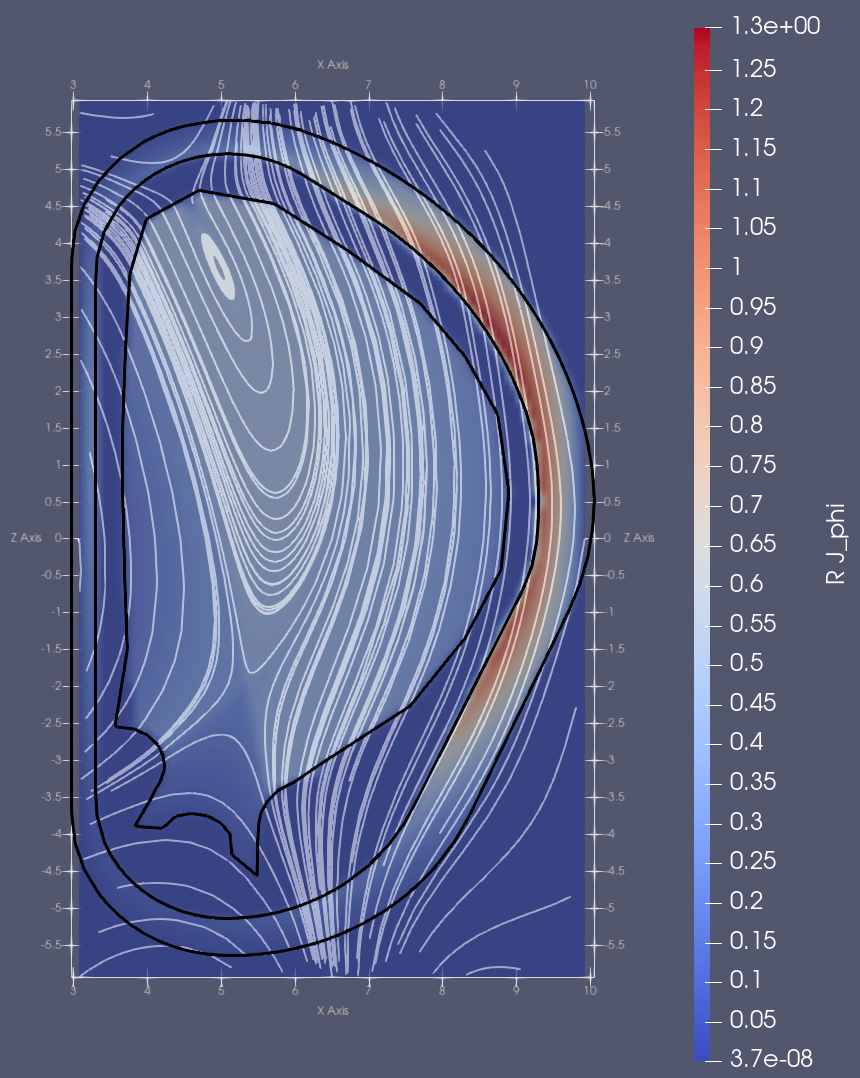}
\caption{$t=\SI{322.4}{\milli\second}$}
\label{fig:VDE_62dt_2}
\end{center}
\end{subfigure}%
~
\begin{subfigure}[b]{0.32\textwidth} 
\begin{center}
\includegraphics[width=\textwidth]{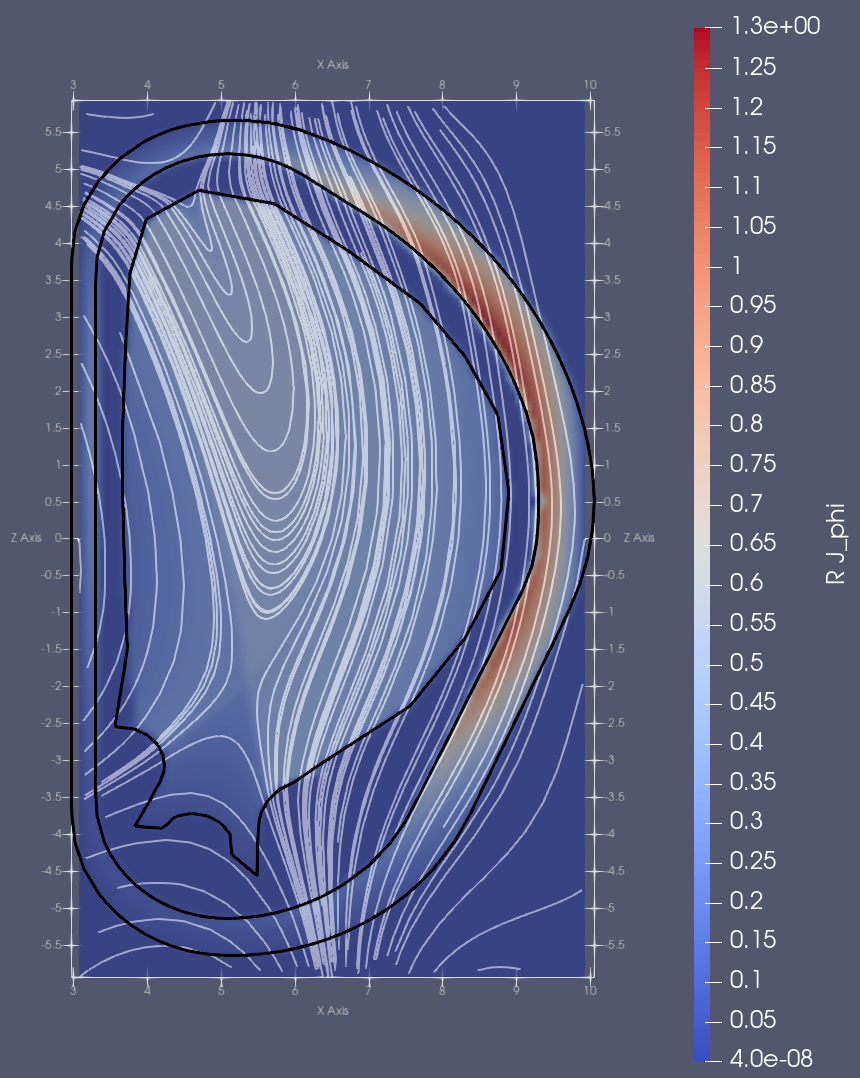}
\caption{$t=\SI{327.6}{\milli\second}$}
\label{fig:VDE_63dt_2}
\end{center}
\end{subfigure}
\caption{The toroidal current $R (\nabla \times \B)_\phi$ and streamlines of the poloidal magnetic field over time.}
\label{fig:VDE_time_2}
\end{figure}
}

First, we consider the resistivity setting described in
Section~\ref{sec:regularization} and a fictitious viscosity
regularization term with $Re = 1000$. The VDE phenomenon is
illustrated in Figure~\ref{fig:VDE_time_2} where we notice that
vertical motion of the plasma column is correlated with the resistive
dissipation of the plasma current. The toroidal current,
$j_\phi:=(\nabla \times \B)_\phi$, in the blanket is mininal because
of the much higher material resistivity there.  In contrast, much
toroidal current can be inductively driven in the vacuum vessel, which
has a much lower resistivity that gives rise to the 500~ms wall time
for the ITER vacuum vessel.  The magnetic field lines are also
projected to the poloidal plane in Figure~\ref{fig:VDE_time_2}.  It is
noted that simultaneously the magnetic axis (the o-point of the
streamlines) progressively moves upward until it hits the rigid wall
and the toroidal current in the plasma chamber gets almost entirely
dissipated. The induced toroidal current in the vacuum vessel persists
till the end of the simulation ($t=\SI{327.6}{\milli\second}$), because the ITER vacuum
vessel has a wall time that is longer ($\SI{500}{\milli\second}$).

\pgfplotstableread{data_to_plot/dataRe1000_divB.txt}\Bdivtable
\pgfplotstablecreatecol[create col/expr={\thisrow{y} * 5)}]{nonnormB}\Bdivtable
\pgfplotstablecreatecol[create col/expr={0.0052*(\thisrow{x})}]{realtime}\Bdivtable

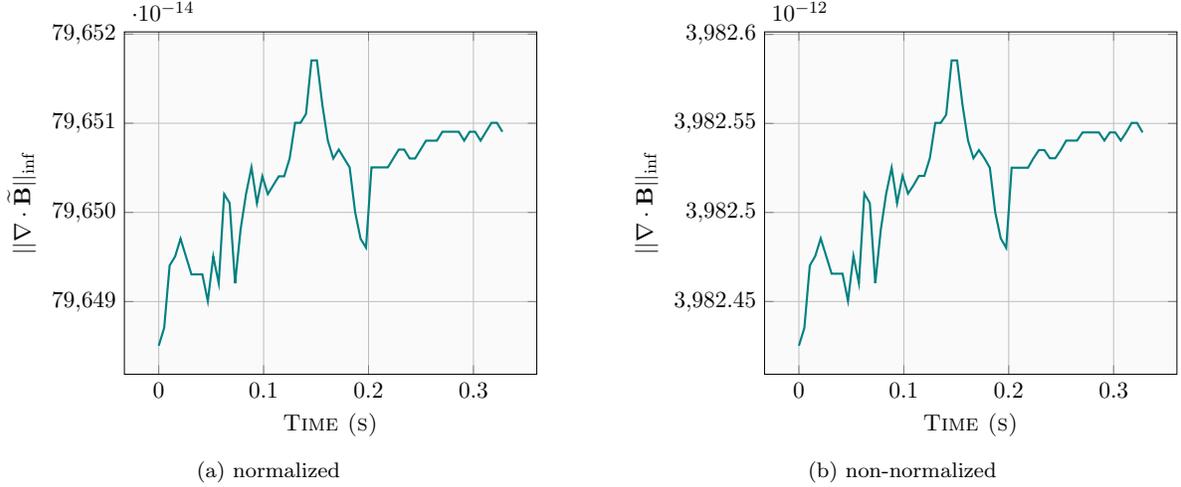
\begin{figure}[ht]
\centering
\begin{subfigure}[b]{0.5\textwidth} 
\begin{center}
\begin{tikzpicture}[scale=0.8,>=latex]
\begin{axis}[xlabel=\textsc{{Time (s)}}, scaled y ticks=base 10:14,
    y tick label style={
        /pgf/number format/fixed,
        /pgf/number format/precision=1
    },
     ylabel=\textsc{{$||\nabla \cdot \widetilde{\B}||_{\inf}$}} , ymajorgrids, xmajorgrids, axis background/.style={fill=gray!4}]
\addplot+[color=teal,line width=1pt,mark=none] table[x=realtime,y] \Bdivtable;
\end{axis}
\end{tikzpicture}
\caption{normalized}
\label{fig:normBdiv_evolution}
\end{center}
\end{subfigure}%
~
\begin{subfigure}[b]{0.5\textwidth} 
\begin{center}
\begin{tikzpicture}[scale=0.8,>=latex]
\begin{axis}[xlabel=\textsc{{Time (s)}}, scaled y ticks=base 10:12, ytick scale label code/.code={$10^{-12}$}, ylabel=\textsc{{$||\nabla \cdot \B||_{\inf}$}} , ymajorgrids, xmajorgrids, axis background/.style={fill=gray!4}]
\addplot+[color=teal,line width=1pt,mark=none] table[x=realtime,y=nonnormB] \Bdivtable;
\end{axis}
\end{tikzpicture}
\caption{non-normalized}
\label{fig:nonnormBdiv_evolution}
\end{center}
\end{subfigure}
\caption{Evolution of the infinity norm of the divergence of the magnetic field over time.}
\label{fig:Bdiv_evolution}
\end{figure}

This example is further used to investigate the divergence-free
property of the magnetic field of the MFD solver.
Figure~\ref{fig:Bdiv_evolution} depicts the evolution of the infinity
norm of the divergence of the magnetic field over
time. Figure~\ref{fig:nonnormBdiv_evolution} shows the non-normalized
quantity whereas Figure~\ref{fig:normBdiv_evolution} corresponds to
the normalized one.  Taking Figure~\ref{fig:normBdiv_evolution} as an
example, the initial divergence is not exactly zero but rather of the
order of $10^{-10}$.  Note that the initial condition is prepared
through several steps as described in Section~\ref{sec:initial}, which
makes it difficult to get an exactly discrete divergence-free field,
especially corresponding to the mimetic divergence operator.
Nevertheless, as time evolves, the divergence fluctuates between two
successive time steps by $\pm 10^{-15}$, which confirms the
divergence-free property of the magnetic field as stated in
Section~\ref{sec:mfd_discretization}.  Here the fluctuation is due to
an iterative nonlinear solver being used.  Nevertheless, such a small
error is more than enough for the tests considered here for avoiding
any potential numerical artifacts due to the divergence error.

{
\begin{figure}[ht]
\centering
\begin{subfigure}[b]{0.32\textwidth} 
\begin{center}
\includegraphics[width=\textwidth]{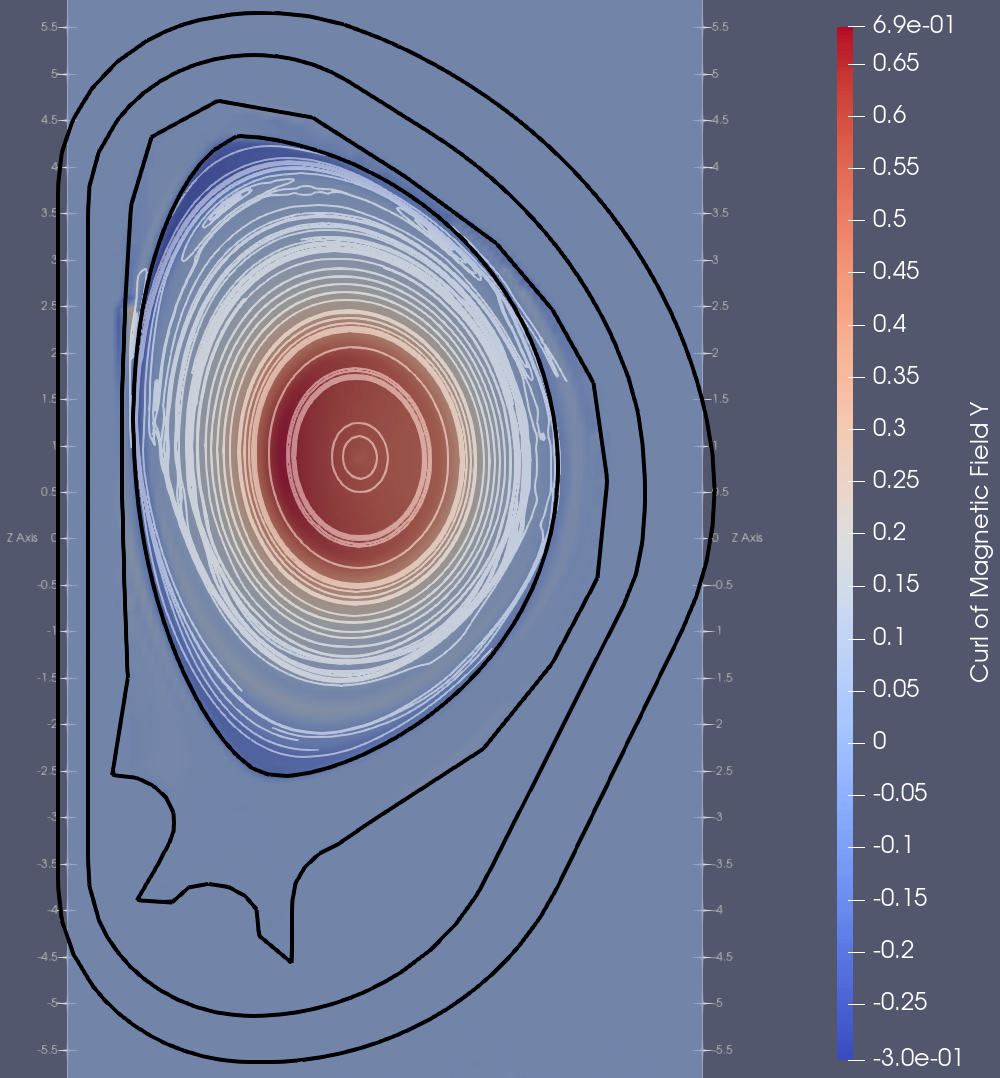}
\caption{$t=\SI{0}{\second}$}
\label{fig:VDE_halo_t0}
\end{center}
\end{subfigure}%
~
\begin{subfigure}[b]{0.32\textwidth} 
\begin{center}
\includegraphics[width=\textwidth]{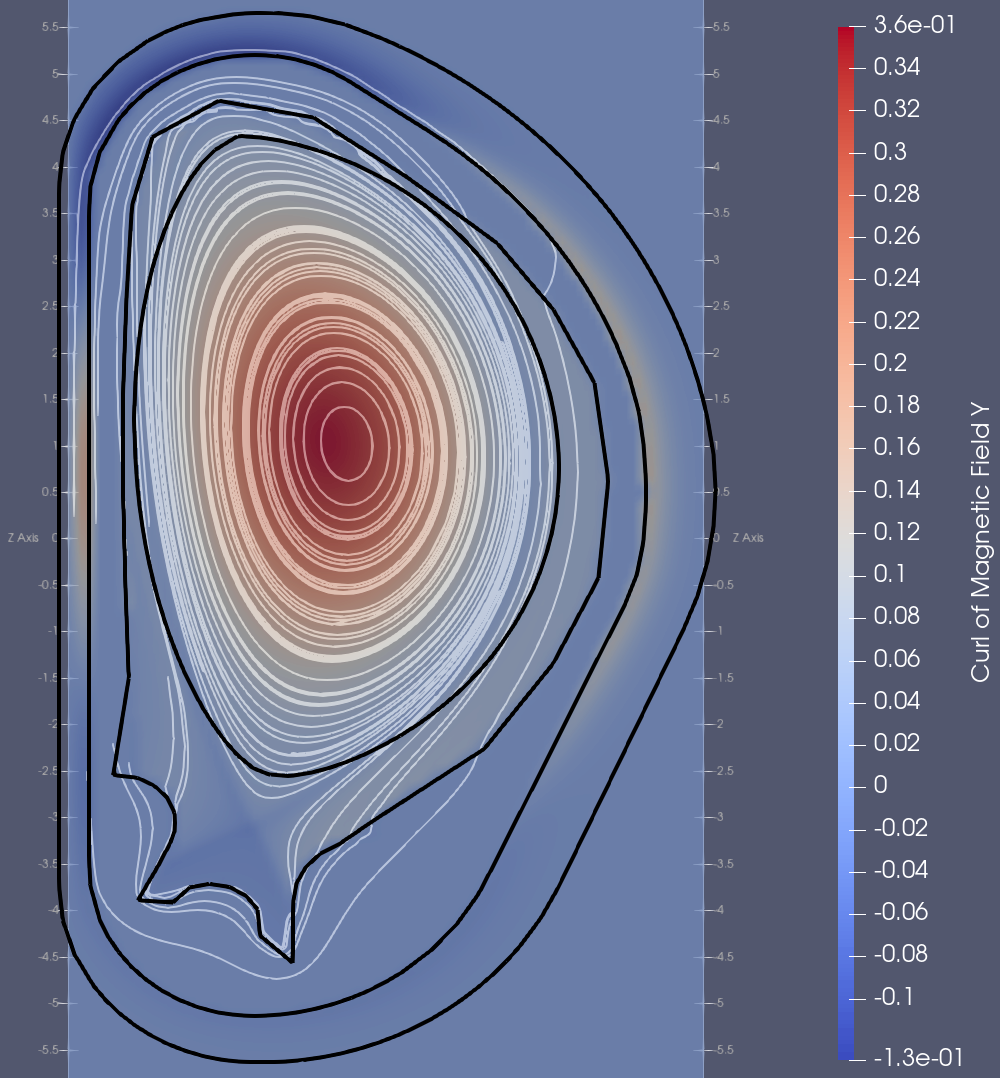}
\caption{$t=\SI{44.19}{\milli\second}$}
\label{fig:VDE_halo_15dt}
\end{center}
\end{subfigure}%
~
\begin{subfigure}[b]{0.32\textwidth} 
\begin{center}
\includegraphics[width=\textwidth]{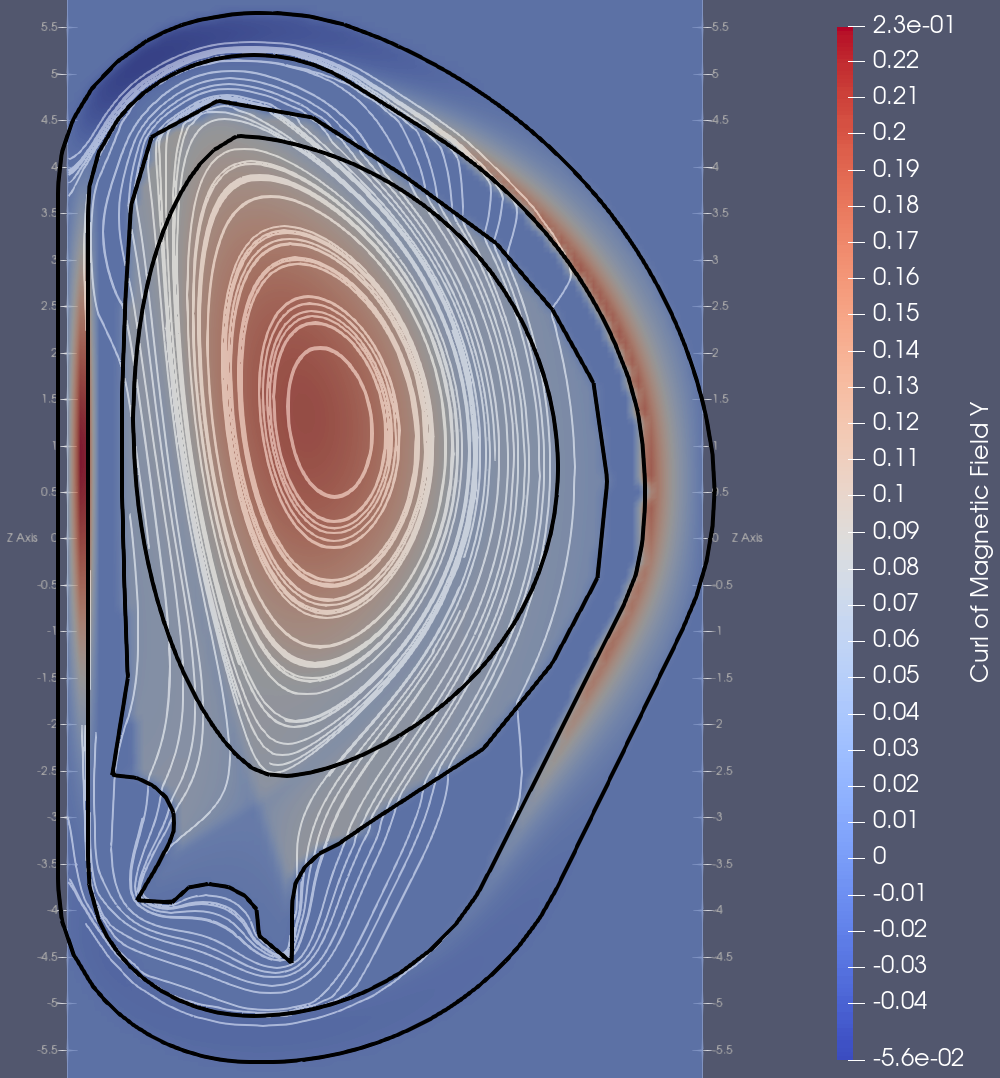}
\caption{$t=\SI{103.11}{\milli\second}$}
\label{fig:VDE_halo_35dt}
\end{center}
\end{subfigure}
\begin{subfigure}[b]{0.32\textwidth} 
\begin{center}
\includegraphics[width=\textwidth]{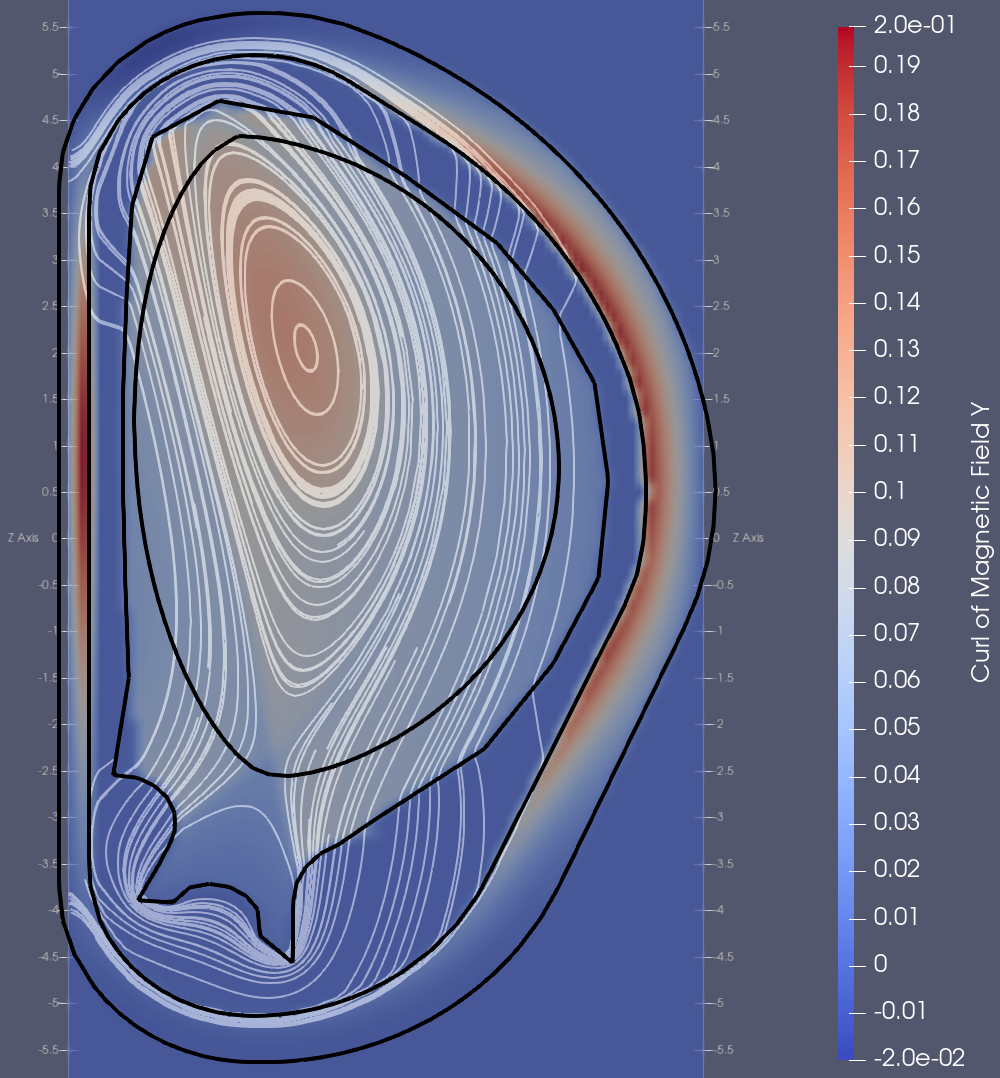}
\caption{$t=\SI{156.14}{\milli\second}$}
\label{fig:VDE_halo_53dt}
\end{center}
\end{subfigure}%
~
\begin{subfigure}[b]{0.32\textwidth} 
\begin{center}
\includegraphics[width=\textwidth]{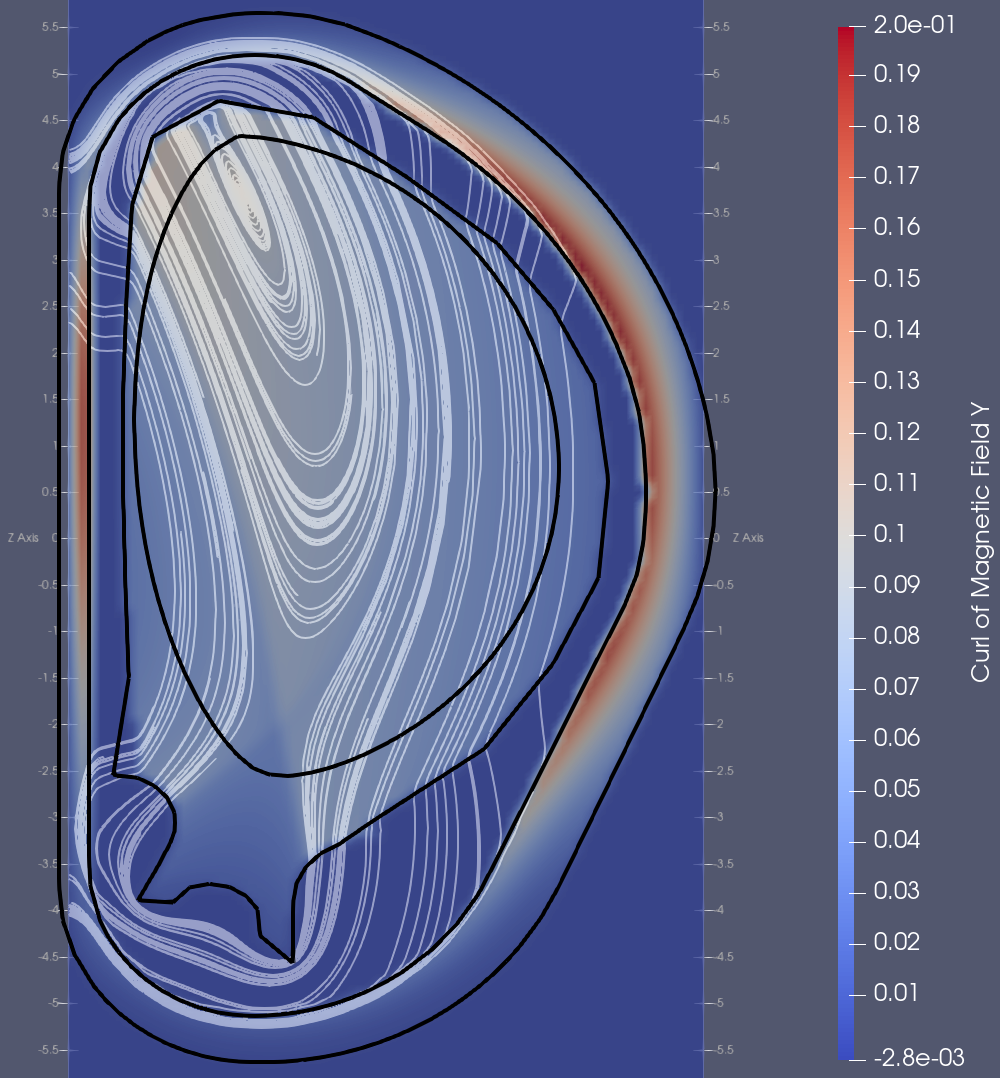}
\caption{$t=\SI{185.60}{\milli\second}$}
\label{fig:VDE_halo_63dt}
\end{center}
\end{subfigure}%
~
\begin{subfigure}[b]{0.32\textwidth} 
\begin{center}
\includegraphics[width=\textwidth]{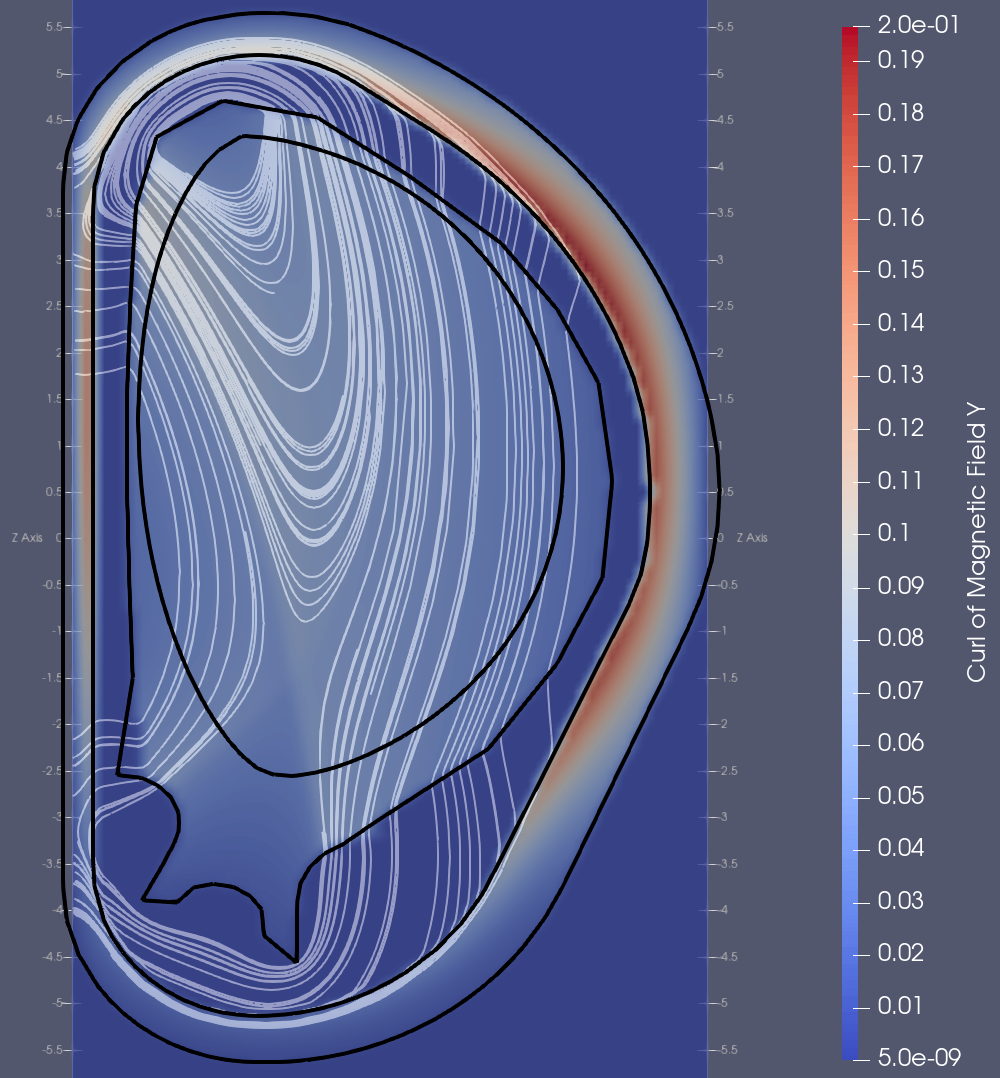}
\caption{$t=\SI{220.96}{\milli\second}$}
\label{fig:VDE_halo_75dt}
\end{center}
\end{subfigure}
\caption{The toroidal current $(\nabla \times \B)_\phi$ and streamlines of the poloidal current $\left(\nabla\times{\B}\right)_{\rm pol}$ over time. At t=$\SI{103.11}{\milli\second}$, a halo current (on the top left region in subfigure~\ref{fig:VDE_halo_35dt}) is observed to form.}
\label{fig:VDE_halocurrent_time}
\end{figure}
}

\subsubsection{Anisotropic setting}
\label{sec:vde_isotropic}
Lastly, we provide the results of a full ITER VDE simulation for the
case of anisotropic resistivity at the first wall and blanket module
(see Formulas~\eqref{eqn:eq34} and~\eqref{eqn:eq35} in
Section~\ref{sec:wall_model}) and a fictitious viscosity
regularization term with $Re = 100$.  The resistivity values tested
are as follows:
\begin{itemize}
    \item $\eta = \SI{1.71e-5}{\ohm\meter}$ in the plasma chamber (that corresponds to an electron temperature of $T_e = 10{\rm eV}$); 
    \item $\eta_t = \SI{4.4e-2}{\ohm\cdot\meter}$ in the blanket module;
    \item $\eta_p = \SI{1.71e-5}{\ohm\cdot\meter}$ in the blanket module;
    \item $\eta = \SI{1.3e-6}{\ohm\meter}$ in the vacuum vessel;
    \item $\eta = \SI{1.3e-3}{\ohm\meter}$ in the region outside the wall.
\end{itemize}

{
\begin{figure}[ht]
\centering
\begin{subfigure}[b]{0.32\textwidth} 
\begin{center}
\includegraphics[width=\textwidth]{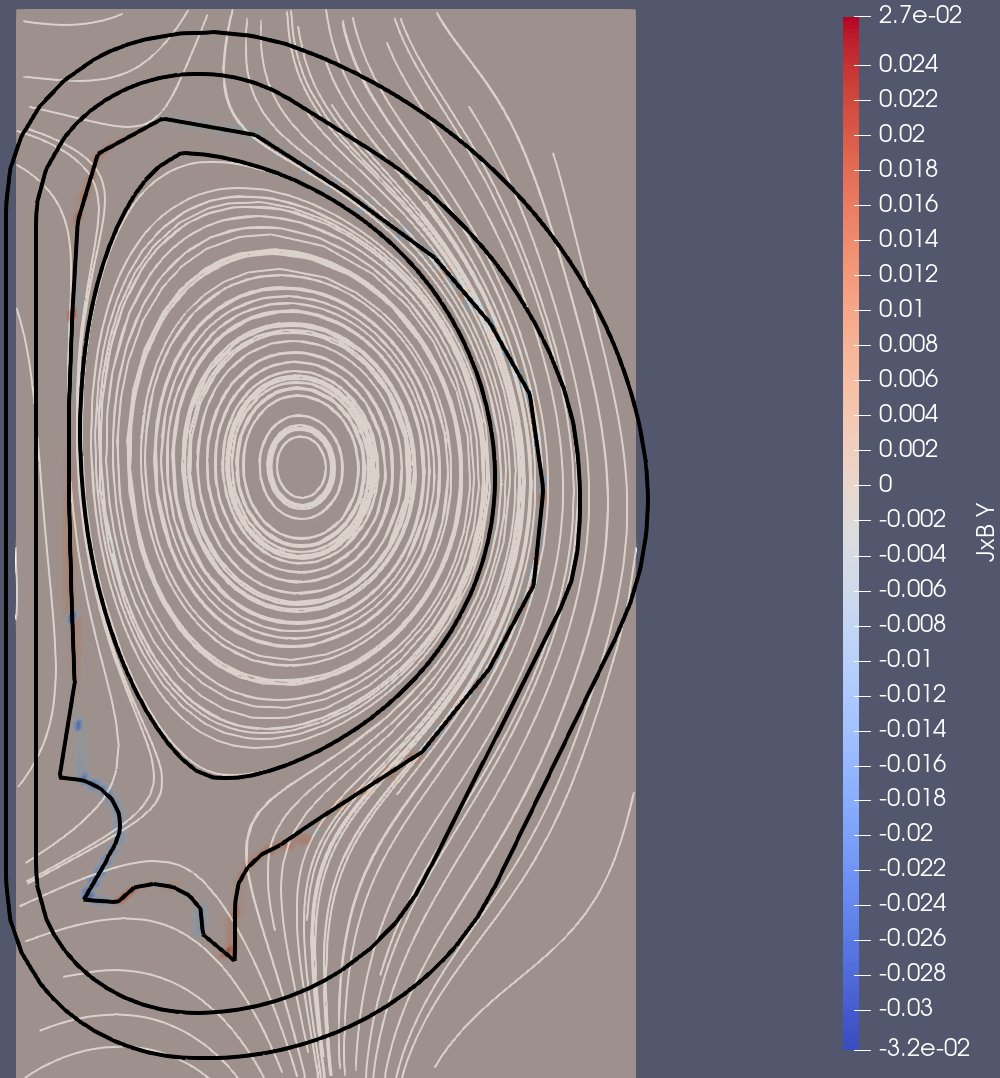}
\caption{$t=\SI{0}{\second}$}
\label{fig:jxB_phi_halo_t0}
\end{center}
\end{subfigure}%
~
\begin{subfigure}[b]{0.32\textwidth} 
\begin{center}
\includegraphics[width=\textwidth]{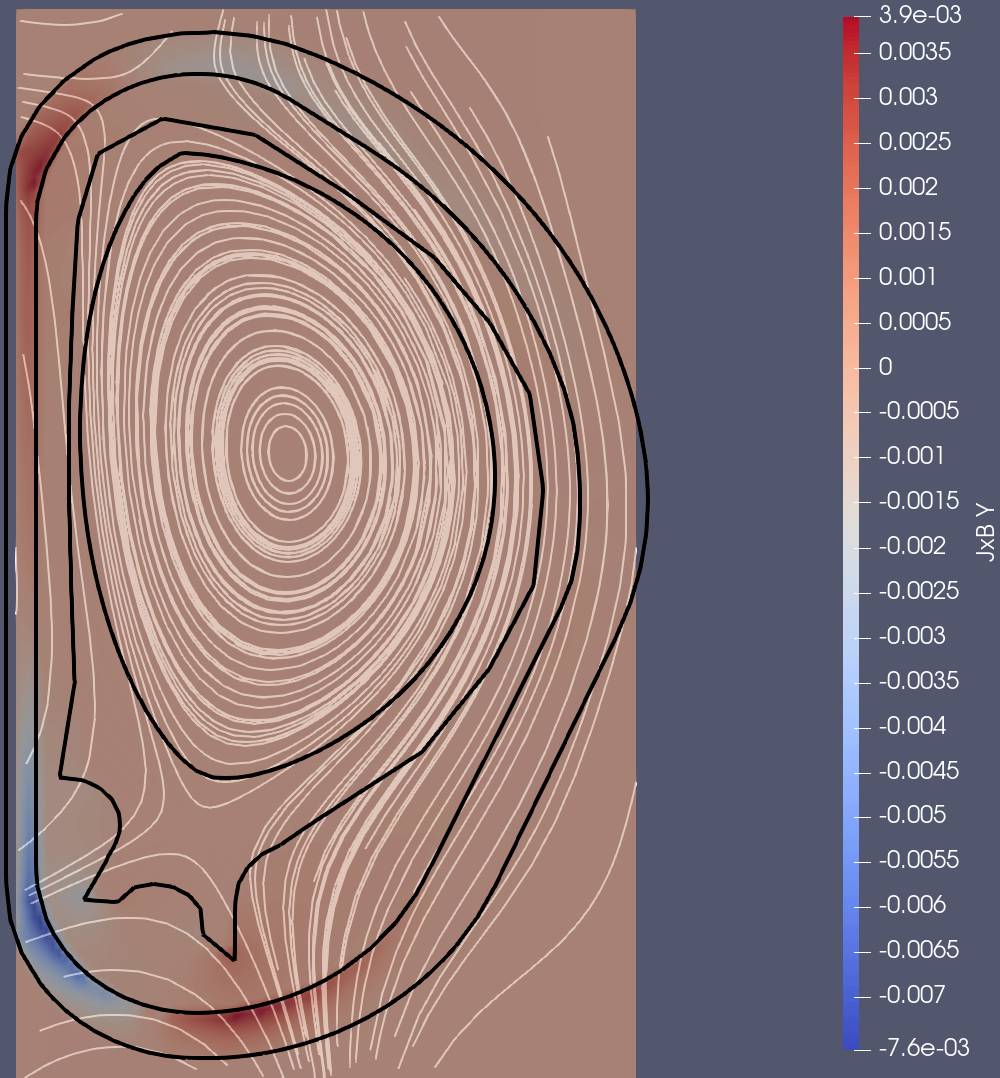}
\caption{$t=\SI{44.19}{\milli\second}$}
\label{fig:jxB_phi_halo_15dt}
\end{center}
\end{subfigure}%
~
\begin{subfigure}[b]{0.32\textwidth} 
\begin{center}
\includegraphics[width=\textwidth]{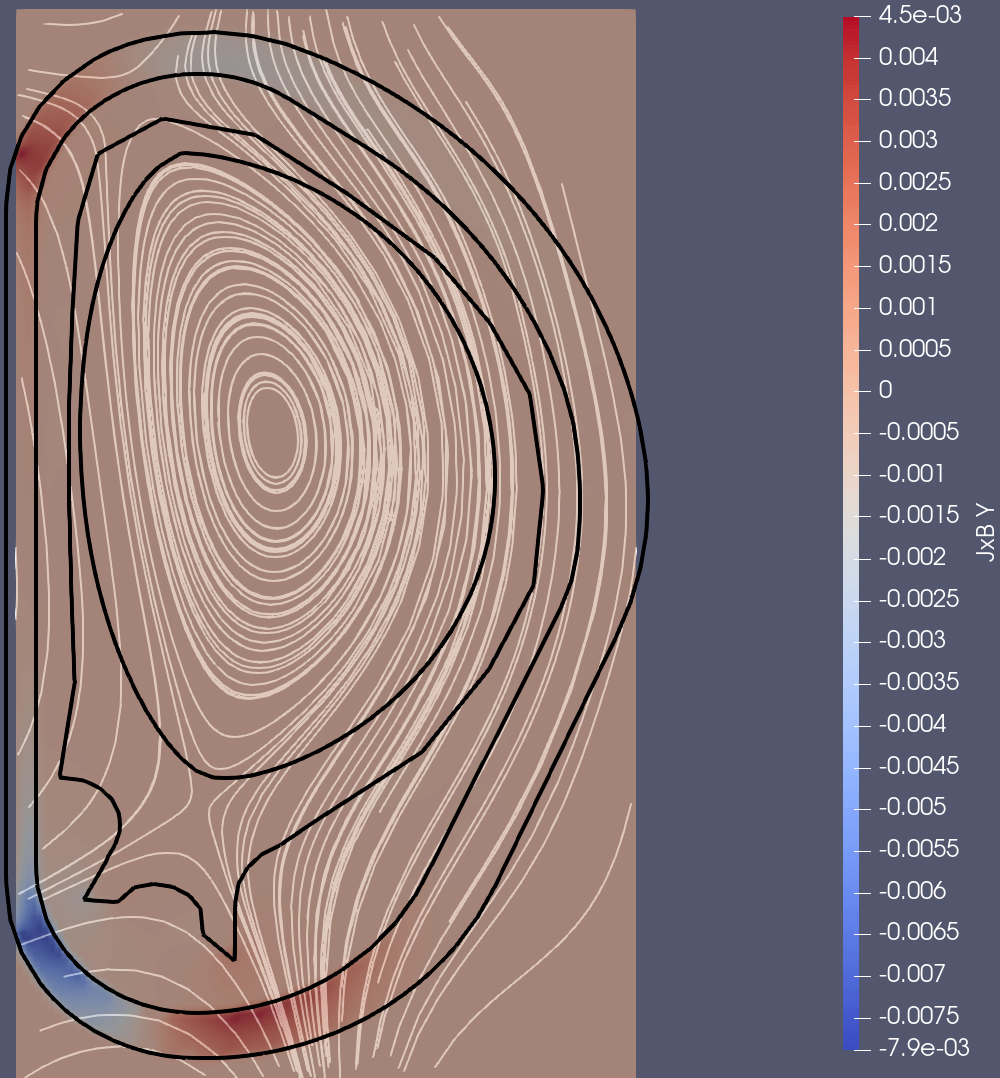}
\caption{$t=\SI{103.11}{\milli\second}$}
\label{fig:jxB_phi_halo_35dt}
\end{center}
\end{subfigure}
\begin{subfigure}[b]{0.32\textwidth} 
\begin{center}
\includegraphics[width=\textwidth]{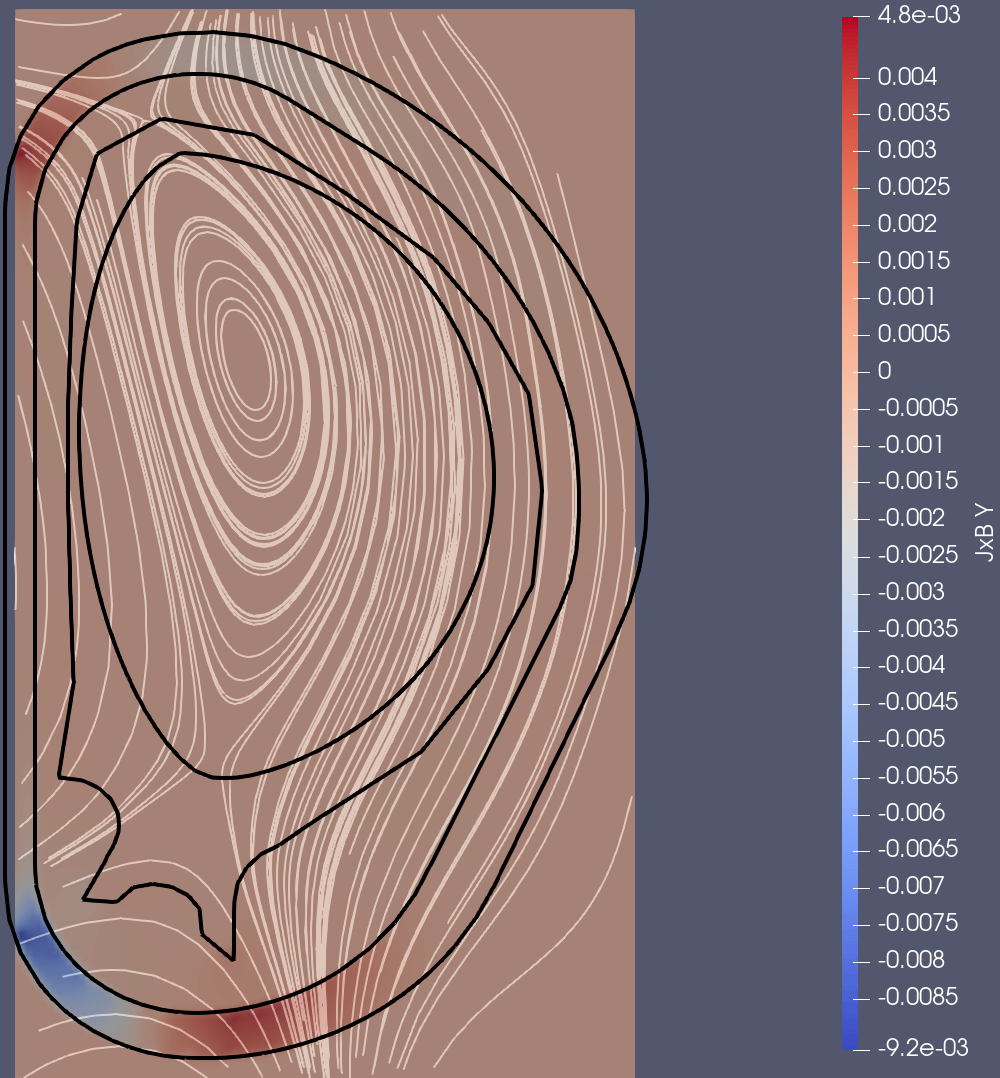}
\caption{$t=\SI{156.14}{\milli\second}$}
\label{fig:jxB_phi_halo_53dt}
\end{center}
\end{subfigure}%
~
\begin{subfigure}[b]{0.32\textwidth} 
\begin{center}
\includegraphics[width=\textwidth]{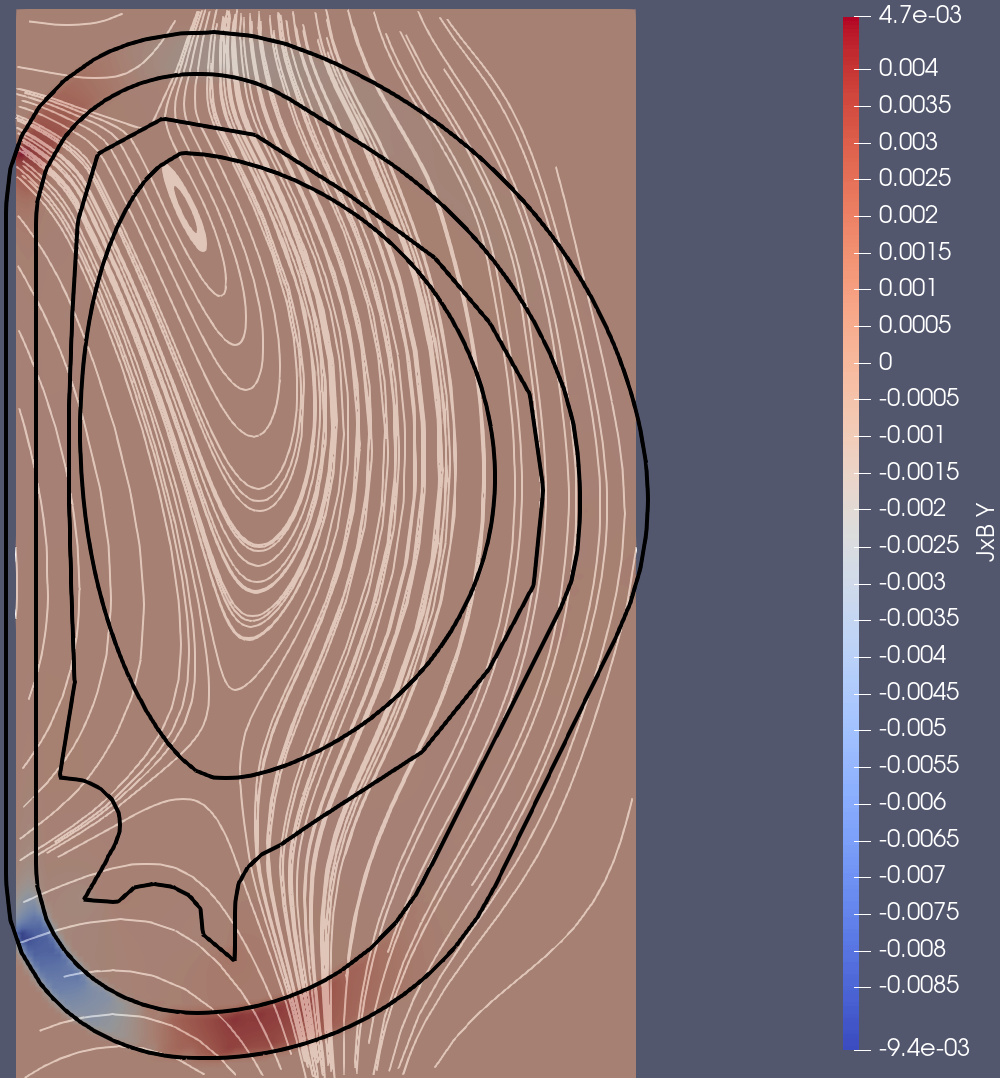}
\caption{$t=\SI{185.60}{\milli\second}$}
\label{fig:jxB_phi_halo_63dt}
\end{center}
\end{subfigure}%
~
\begin{subfigure}[b]{0.32\textwidth} 
\begin{center}
\includegraphics[width=\textwidth]{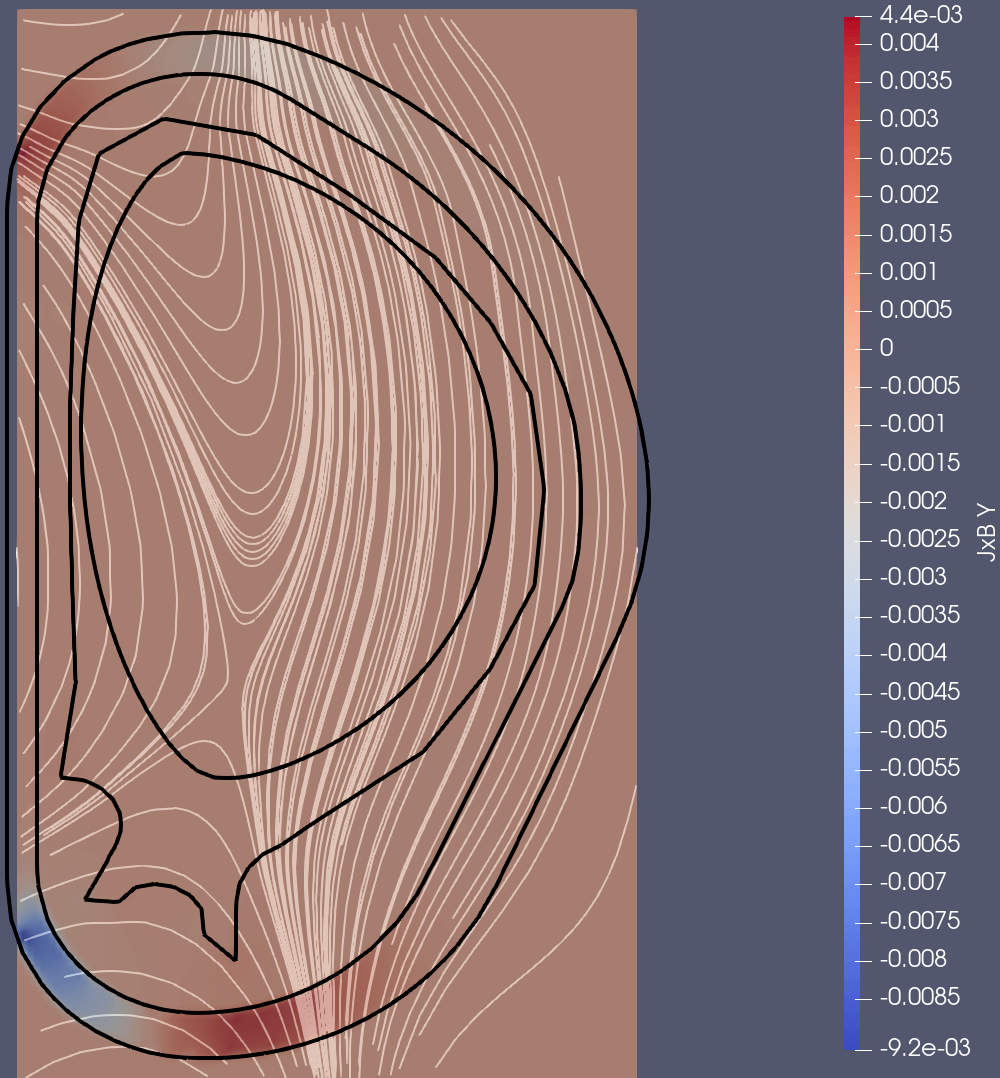}
\caption{$t=\SI{220.96}{\milli\second}$}
\label{fig:jxB_phi_halo_75dt}
\end{center}
\end{subfigure}
\caption{The toroidal component of the Lorentz Force $((\nabla \times \B) \times \B) _\phi$ and streamlines of the poloidal magnetic field over time.}
\label{fig:jxB_phi_halocurrent_time}
\end{figure}
}

The so called ``halo currents'' are illustrated in Figure~\ref{fig:VDE_halocurrent_time} where the toroidal current is given by the colormap and the poloidal current is presented  as streamlines.
We observe that the toroidal current is eliminated and the halo current flows only poloidally in the blanket to enter the vacuum vessel, where the electrical current can have a strong toroidal component.

{
\begin{figure}[ht]
\centering
\begin{subfigure}[b]{0.32\textwidth} 
\begin{center}
\includegraphics[width=\textwidth]{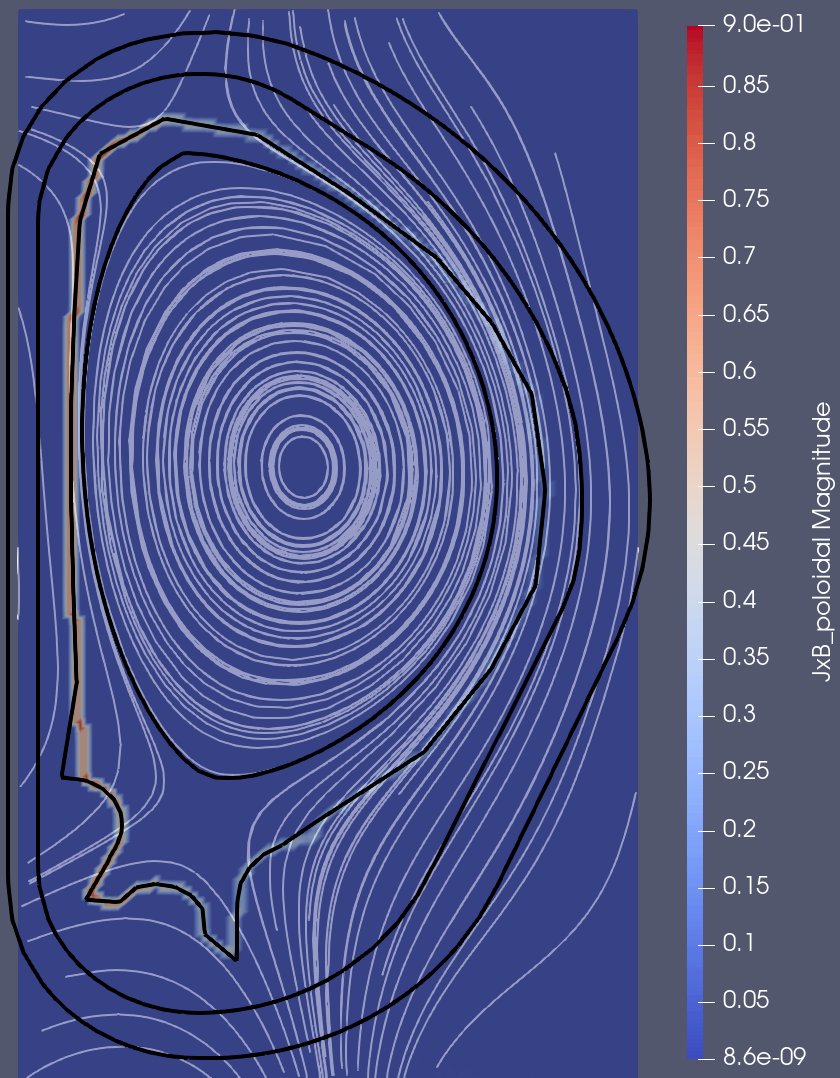}
\caption{$t=\SI{0}{\second}$}
\label{fig:jxB_poloidal_halo_t0}
\end{center}
\end{subfigure}%
~
\begin{subfigure}[b]{0.32\textwidth} 
\begin{center}
\includegraphics[width=\textwidth]{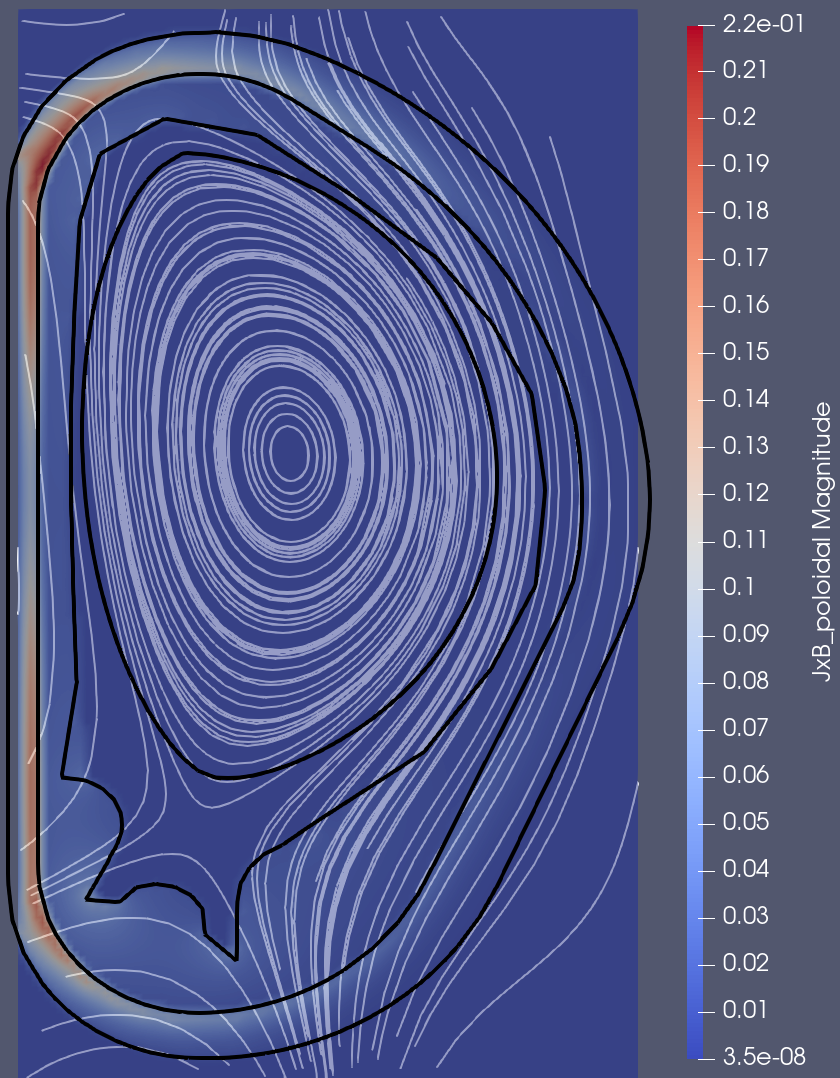}
\caption{$t=\SI{44.19}{\milli\second}$}
\label{fig:jxB_poloidal_halo_15dt}
\end{center}
\end{subfigure}%
~
\begin{subfigure}[b]{0.32\textwidth} 
\begin{center}
\includegraphics[width=\textwidth]{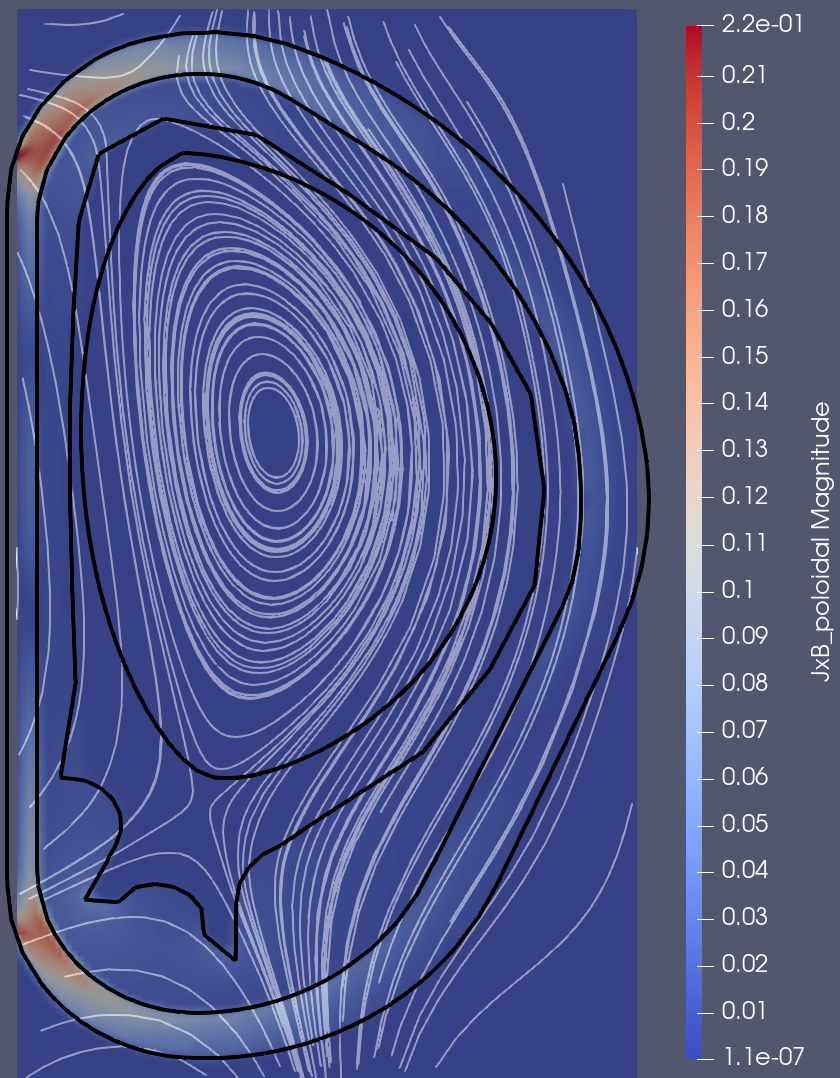}
\caption{$t=\SI{103.11}{\milli\second}$}
\label{fig:jxB_poloidal_halo_35dt}
\end{center}
\end{subfigure}
\begin{subfigure}[b]{0.32\textwidth} 
\begin{center}
\includegraphics[width=\textwidth]{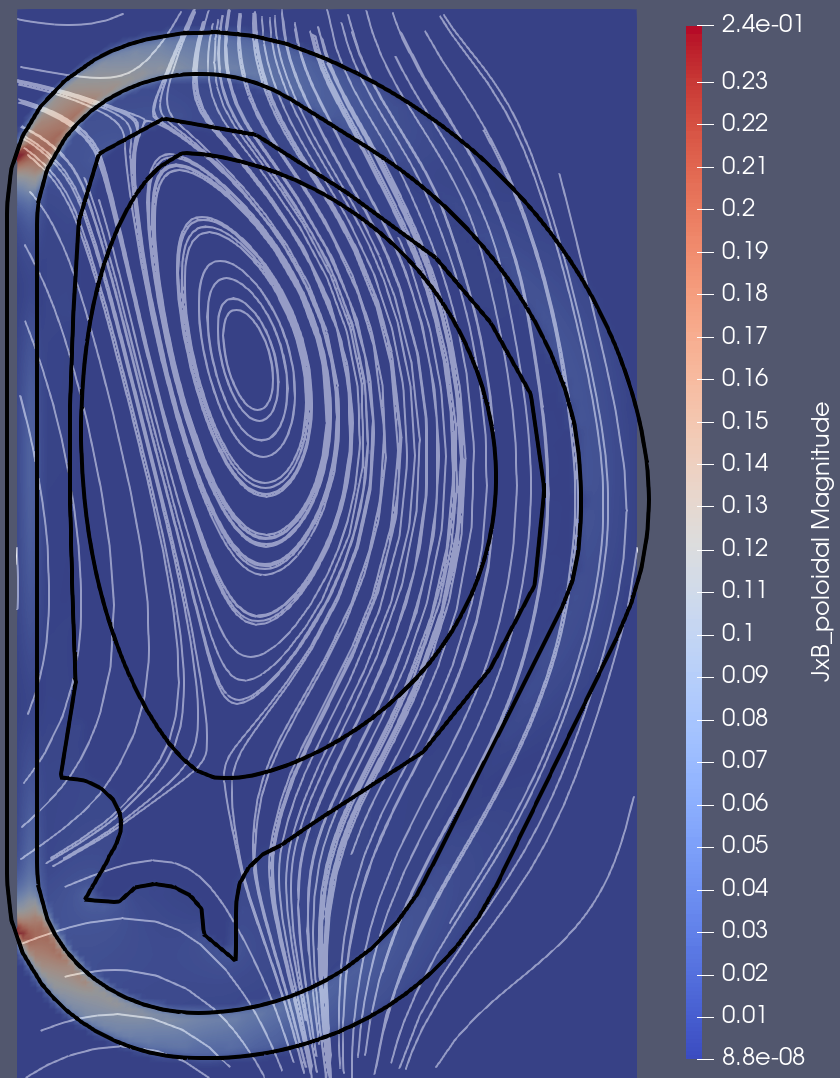}
\caption{$t=\SI{156.14}{\milli\second}$}
\label{fig:jxB_poloidal_halo_53dt}
\end{center}
\end{subfigure}%
~
\begin{subfigure}[b]{0.32\textwidth} 
\begin{center}
\includegraphics[width=\textwidth]{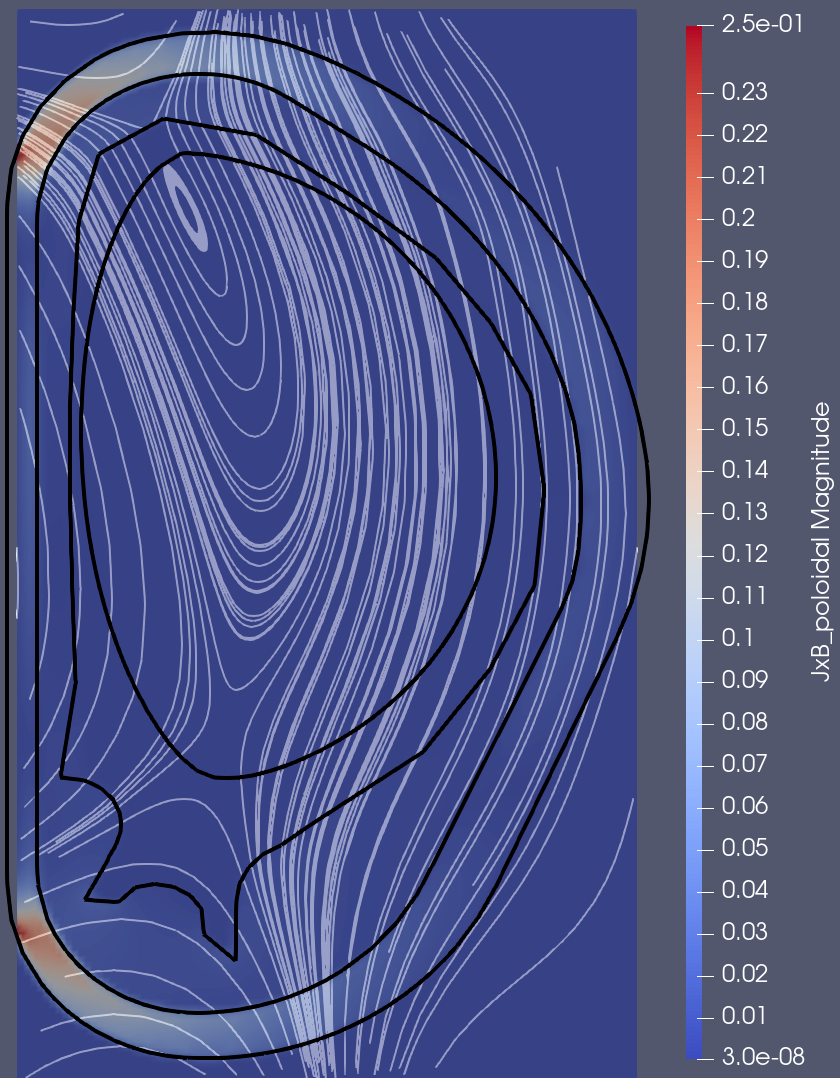}
\caption{$t=\SI{185.60}{\milli\second}$}
\label{fig:jxB_poloidal_halo_63dt}
\end{center}
\end{subfigure}%
~
\begin{subfigure}[b]{0.32\textwidth} 
\begin{center}
\includegraphics[width=\textwidth]{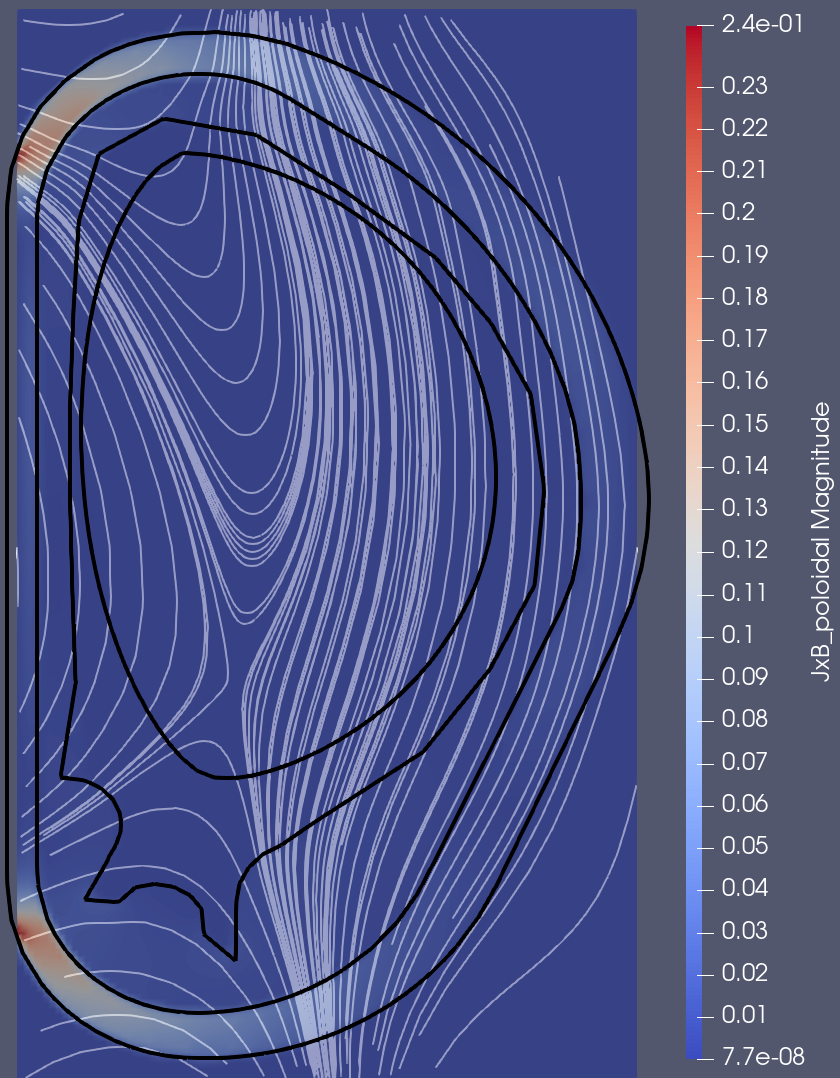}
\caption{$t=\SI{220.96}{\milli\second}$}
\label{fig:jxB_poloidal_halo_75dt}
\end{center}
\end{subfigure}
\caption{The magnitude of the poloidal component of the Lorentz Force $||((\nabla \times \B) \times \B)_{\rm pol}||$ and streamlines of the poloidal magnetic field over time.}
\label{fig:jxB_poloidal_halocurrent_time}
\end{figure}
}

The electromagnetic force loading ($\mathbf{j}\times\mathbf{B}$) due
to the halo and eddy current in the blanket and vacuum vessel during a
VDE can also be quantified from our simulation. The toroidal and
poloidal components of the Lorentz force
($\mathbf{j}\times\mathbf{B}$) are given in
Figures~\ref{fig:jxB_phi_halocurrent_time}
and~\ref{fig:jxB_poloidal_halocurrent_time} respectively, with the
magnetic field lines at six different times of the simulation.  We
notice that the Lorentz force is initially concentrated along the
plasma/wall interface whereas later in time, this force becomes
stronger at the vacuum vessel and almost nil elsewhere.

\subsection{Solver performance}  
\label{sec:performance}

In this section, we vary the resistivity value inside the plasma chamber and check the performance of the solver. More specifically, for each resistivity value, we examine the average numbers of nonlinear and linear iterations. The time step is fixed in each simulation, primarily depending on the plasma resistivity, such that it is always equal to 1\% of the resistive diffusion time, i.e.,
\begin{equation*}
\Delta t = 0.01 \, \tau_\eta = 0.01 \, \frac{a^2 \mu_0}{\eta},
\end{equation*}
where $\mu_0$ denotes the permeability of free space and $a$ is the minor radius which is considered to be equal to 2 for the ITER.
We are interested in verifying that the performance of the solver is independent of $\eta$.

\begin{table}[h!]
\caption{Average number of iterations per time solve for full ITER VDE simulations with different plasma resistivities\label{tab:iterations_varyplasmares}}
\centering
\begin{tabular}{| c | c | c |}
\hline 
$\eta$ (\SI{}{\ohm\meter}) & \# of Newton its./solve  & \# of Krylov its./solve \\
\hline 
\num{9.66e-6} & 3.98 & 10.94 \\
\hline 
\num{4.83e-6} & 3.98 & 12.14 \\
\hline 
\num{2.415e-6} & 4.03 & 12.98 \\
\hline
\end{tabular}
\end{table}

We perform the simulations for 70 time steps and report the average numbers of linear and nonlinear iterations over the entire simulations in Table~\ref{tab:iterations_varyplasmares}.
It is found that the number of nonlinear iterations seems to be stable (around 4 iterations per solve) for the different plasma resistivity values tested, whereas the simulations with lower resistivities seem to be more computationally demanding on the algebraic level as the linear solver seems to require slightly more iterations. The increase remains at a reasonable level for the resistivity range tested though. The default restart value of GMRES in PETSc is 30, which is much larger than this iteration number. 
The algorithm performance confirms our preconditioning strategy is efficient. In the future work, we will further improve the inner sub-block solver by replacing the direct solver for the $\{\V, \B\}$ block with a more scalable option.

\section{Conclusions}
\label{sec:conc}
In this article, we presented a mimetic finite difference discretization of a regularized quasi-static perpendicular plasma dynamics model that is designed to preserve the divergence of the magnetic field. We also proposed a four-level block preconditioner where the factorization and solver strategy is exact for 2 blocks and the overall solver shows good performance. In addition, the properties of the mimetic operators help eliminate sub-blocks in the Schur complement in the third level of this preconditioner. 
The numerical results show that the fluctuations in the divergence remain within the machine precision range and that the performance of the nonlinear solver and the preconditioner allows  large time steps for a range of Lundquist numbers.
Lastly, the full VDE simulation over the actual plasma current diffusion time was shown for ITER configuration. 

\section*{Acknowledgement}
We would like to thank the PETSc team for a number of helpful
discussions.  In particular, we thank Patrick Sanan, the primary
developer of DMStag, for many useful discussions.  This research used
resources provided by the Los Alamos National Laboratory Institutional
Computing Program, which is supported by the U.S. Department of Energy
National Nuclear Security Administration under Contract
No.~89233218CNA000001.  This research also used resources of the
National Energy Research Scientific Computing Center, which is
supported by the Office of Science of the U.S. Department of Energy
under Contract No. DE-AC02-05CH11231.

\appendix
\section{Nonuniqueness of solutions to the model without the viscosity term}
\label{sec:nonuniqueness}
This section addresses the nonuniqueness of the solution to the model without the viscosity regularization term.
Consider the following model in a single domain
\begin{align}
  -\left(  \dfrac{1}{\mu_0} (\nabla\times{\B})\times{\B} \right) \cdot \mathbf{e}_R & = 0 & \qquad \text{in} \quad \Omega^{W}, \label{eqn:ap_v1}  \\
  {\V}_{i\perp} \cdot {\B} & = 0 & \qquad \text{in} \quad \Omega^{W}, \label{eqn:ap_v2}\\
  -\left(   \dfrac{1}{\mu_0} (\nabla\times{\B})\times{\B} \right) \cdot \mathbf{e}_Z & = 0 & \qquad \text{in} \quad \Omega^{W},  \label{eqn:ap_v3}\\
  \nabla^2 \Phi & = - \nabla\cdot\left[ - {\V}_{i\perp}\times {\B} + \frac{\eta}{\mu_0}\left(\nabla\times{\B}\right) \right] &\qquad \text{in} \quad \Omega^{W},    \\
  \boldsymbol{\tau} & = \nabla\Phi - {\V}_{i\perp}\times {\B} + \frac{\eta}{\mu_0}\left(\nabla\times{\B}\right) &\qquad \text{in} \quad \Omega^{W},   \\
  \frac{\partial{\B}}{\partial t} & = - \nabla\times\boldsymbol{\tau} &\qquad \text{in} \quad \Omega^{W},
\end{align}
under some proper boundary condition
\begin{align}
  \mathbf{V}_{i\perp} & = \mathbf{0} &\qquad \text{on} \quad \partial \Omega^{W}, \label{eqn:ap_bc_V}\\
  \Phi & = \Phi_{\rm bc} &\qquad \text{on} \quad \partial\Omega^{W}, \label{eqn:ap_bc_Phi}\\
  \boldsymbol{\tau} & = \boldsymbol{\tau}_{\rm bc} &\qquad \text{on} \quad \partial\Omega^{W}. \label{eqn:ap_bc_tau}
\end{align}
The equation for $n_i$ is dropped as it is not coupled to the four fields considered here. At first glance, it is
tempting to assume the above system is well-posed  since it is a closed system.  
We will however show the form of a possible null space that only needs to satisfy a mild constraint.

Let $(\boldsymbol{\tau} , \B, \V_{i\perp}, \Phi )$ and $(\boldsymbol{\tau} + \mathbf{Z}, \B, \V_{i\perp}  + \mathbf{W}, \Phi + \varphi)$ be two solutions of the above system. We then have:
\begin{align}
\label{eq:WdotB}
\mathbf{W} \cdot {\B} = & ~ 0 & \qquad \text{in} \quad \Omega^{W} ,  \\
\nabla^2 {\varphi} = & ~ \nabla \cdot \left( \mathbf{W}\times {\B} \right) & \qquad \text{in} \quad \Omega^{W},
\label{eq:LaplacianW} \\
\mathbf{Z} = & ~ \nabla \varphi - \mathbf{W}\times {\B} & \qquad \text{in} \quad \Omega^{W}, \\
\nabla\times  \mathbf{Z} = & - \nabla\times (\mathbf{W}\times {\B}) = ~ \mathbf{0} & \qquad \text{in} \quad \Omega^{W}, \label{eq:nil_curl}
\\
\mathbf{W}  = & ~ \mathbf{0} &\qquad \text{on} \quad \partial \Omega^{W}, \label{eqn:ap_bc_W} \\
\varphi = & ~ 0 &\qquad \text{on} \quad \partial\Omega^{W}, \label{eqn:ap_bc_varphi}\\
\mathbf{Z} = & ~ \mathbf{0} &\qquad \text{on} \quad \partial\Omega^{W} \label{eqn:ap_bc_Z}
.
\end{align}
\eqref{eq:nil_curl} implies that there exists a field $g \in \mathrm{H}^{1}(\Omega)$
 such that:
\begin{equation}
\label{eq:conservative_field_2}
\mathbf{W} \times \B = \nabla g ;
\end{equation}
That further implies that
\begin{equation}
\nabla \cdot (\mathbf{W} \times \B) = \nabla^2 g = \nabla^2 \varphi.
\end{equation}
The Dirichlet boundary condition for $\Phi$ suggests that $g$ is equal to $\varphi$, 
and thus $\nabla \varphi = \nabla g = \mathbf{W} \times \B$ and $\mathbf{Z} = 0$.
Moveover, Equations~\eqref{eq:WdotB} 
together with $\nabla \varphi = \mathbf{W} \times \B$ are equivalent to:
\begin{align}
\mathbf{W} & = ||{\B}||^{-2} ({\B} \times \nabla \varphi) \\
\nabla \varphi \cdot {\B} & = \mathbf{0}. \label{eqn:psiconstraint}
\end{align}
To prove the equivalence $\left\{ \begin{aligned} 
  \mathbf{W} \cdot {\B} & = ~ 0 \\
  \nabla \varphi &= \mathbf{W} \times \B
\end{aligned} \right. \Leftrightarrow \left\{ \begin{aligned} 
  \mathbf{W} & = ||{\B}||^{-2} ({\B} \times \nabla \varphi) \\
  \nabla \varphi \cdot {\B} & = \mathbf{0} 
\end{aligned} \right. $, it suffices to compute the dot and cross product with $\B$ for the bottom left side equation and the top right side equation.
This identifies at least one null space for the system, which only needs the scalar function $\varphi$ to satisfy the constraint~\eqref{eqn:psiconstraint}.
As pointed out in Section~\ref{sec:regularization}, such a constraint can be easily satisfied in the axisymmetric tokamak, as the magnetic field is given by~\eqref{eqn:Bfrompsi} 
and any function of $\varphi(\psi)$ satisfies~\eqref{eqn:psiconstraint}.

\section{Regularization through the viscosity term}
\label{sec:uniqueness}

This section show the aforementioned null space is removed through the regularization term.
 For ease of presentation, we consider the full $\V$ formulation. More specifically, 
 we consider to replace the equations~\eqref{eqn:ap_v1}--\eqref{eqn:ap_v3} with the following equation
 \begin{equation}
  \frac{1}{\mu_0} (\nabla\times{\B})\times{\B} = - \nu n_0 m_i \nabla^2 {\V}.
 \end{equation}
Let $(\boldsymbol{\tau} , \B, {\V}, \Phi )$ and $(\boldsymbol{\tau} + \mathbf{Z}, \B, {\V} + \mathbf{W}, \Phi + \varphi)$ again be two solutions of the new model. We then have:
\begin{align}
\nabla^2 \mathbf{W} = & ~ \mathbf{0} , \label{eqn:bndryV} \\
\nabla^2 {\varphi} = & ~ \nabla \cdot \left( \mathbf{W}\times {\B} \right) , \label{eqn:bndryPsi} \\
\mathbf{Z} = & ~ \nabla \varphi - \mathbf{W}\times {\B} , \label{eqn:Z}\\
\nabla\times  \mathbf{Z} = & - \nabla\times (\mathbf{W}\times {\B}) = ~ \mathbf{0} .
\end{align}
The Dirichlet boundary condition implies that $\mathbf{W} = 0$, $\varphi=0$, and  $\mathbf{Z}=0$. Hence, we show the regularization term removes the null space.
This analysis can be easily extended to the case with the $\V_\perp$ case when we consider the axisymmetric case (i.e., $\partial/\partial \phi = 0$). 

\section{Matrix definition of the derived mimetic operators}
\label{sec:der_mim_def}
To obtain a matrix form of the dual operators, we note that 
inner products are represented by symmetric positive definite mass matrices: $\mathbb{M}_n$ for $\mathcal{N}_h$, $\mathbb{M}_e$ for $\mathcal{E}_h$, $\mathbb{M}_f$ for $\mathcal{F}_h$ and $\mathbb{M}_c$ for $\mathcal{C}_h$.
Hence,
\begin{align*}
  (\mathbb{M}_e\,{\mathrm{Grad}_h}\, p_h)^T\, \mathbf{v}_h 
  & = -(\mathbb{M}_n\, p_h)^T\, \widetilde{\mathrm{Div}_h}\,\mathbf{v}_h, \\
  (\mathbb{M}_c\,{\mathrm{Div}_h}\, \mathbf{u}_h)^T\, p_h 
  & = -(\mathbb{M}_f\, \mathbf{u}_h)^T\, \widetilde{\mathrm{Grad}_h}\,p_h, \\
  (\mathbb{M}_f\,{\mathrm{Curl}_h}\, \mathbf{u}_h)^T\, \mathbf{v}_h 
  & = (\mathbb{M}_e\, \mathbf{u}_h)^T\, \widetilde{\mathrm{Curl}_h}\,\mathbf{v}_h.
\end{align*}
These should hold true for any vectors $p_h$, $u_h$ and $\mathbf{v}_h$. 
As a consequence, the matrix definitions are given by
\begin{align}
  \widetilde{\mathrm{Div}_h} & = -\mathbb{M}_n^{-1}\,({\mathrm{Grad}_h})^T\, \mathbb{M}_e, \label{eqn:DDiv-def} \\
  \widetilde{\mathrm{Grad}_h} & = -\mathbb{M}_f^{-1}\,({\mathrm{Div}_h})^T\, \mathbb{M}_c, \label{eqn:DGrad-def} \\
  \widetilde{\mathrm{Curl}_h} & = \mathbb{M}_e^{-1}\,({\mathrm{Curl}_h})^T\, \mathbb{M}_f. \label{eqn:DCurl-def}
\end{align}
The mass matrices are diagonal on structured meshes that are considered in the current work. 

To discretize PDEs, the MFD framework builds upon three primary and three dual operators.
The four discrete spaces along with their primary or derived operators form a discrete de Rham complex. 
The coefficients of the PDEs are usually embedded in the definition of the derived operator. Consider, for example, the following formulas:
\begin{eqnarray}
&&
  \int_\Omega p\,\nabla \cdot {\mathbf u} \,{\rm d}V 
   = -\int_\Omega K^{-1} (K\, \nabla p) \cdot \mathbf{u} \,{\rm d}V 
  + \oint_{\partial \Omega} p\,(\mathbf{u}\cdot \mathbf{n}) \,{\rm d}S, \\
&&
\int_\Omega \mathbf{u} \cdot (\nabla \times \mathbf{v}) \,{\rm d}V 
  = \int_\Omega K^{-1} (K \nabla \times  \mathbf{u}) \cdot \mathbf{v} \,{\rm d}V 
  + \oint_{\partial \Omega} (\mathbf{u}\times \mathbf{v}) \cdot \mathbf{n} \,{\rm d}S, \label{eqn:Kcurl}
\end{eqnarray}
where $K$ is a positive definite tensor. 
These formulas represents the duality between the two first operators, ${\rm div}$ and 
$(K\,{\rm grad})$ for the first equation, ${\rm curl}$ and 
$(K\,{\rm curl})$ for the second one, using a weighted inner product for the vector fields (with $K^{-1}$ 
as the weight). 
The corresponding discrete operators will be in a duality relation with respect to such an inner product.

\section{Magnetic energy dissipation}
\label{sec:magNRJ}
In this section, we show that with a viscosity term and under a proper boundary condition, an energy dissipation law can be achieved if we consider the total magnetic energy.
 Let $(n_i, \boldsymbol{\tau} , \B, {\V}, \Phi )$ be the solution for the five-field model (with full $\V$) with the viscosity term. The total magnetic energy is given by
\begin{equation}
E_{\mathbf{B}} := \frac{1}{2 \mu_0} |\mathbf{B}|^2 .
\end{equation}
We then have:
\begin{align}
  \dfrac{d E_{\mathbf{B}}}{dt} = & \int_{\Omega} \frac{1}{\mu_0} \left( \mathbf{B} \cdot \dfrac{\partial {\B}}{\partial t} \right) dV , \\
   = & - \frac{1}{\mu_0} \int_{\Omega}  \left( \mathbf{B} \cdot (\nabla\times \boldsymbol{\tau}) \right) dV , \\
   = & - \frac{1}{\mu_0} \int_{\Omega}  \left( \boldsymbol{\tau} \cdot (\nabla\times {\B}) \right) dV - \oint_{\partial \Omega} (\mathbf{B}\times \boldsymbol{\tau}) \cdot \mathbf{n} \,{\rm d}S , \\
   = & - \frac{1}{\mu_0} \int_{\Omega}  \left(  ( - {\V} \times {\B} + \frac{\eta}{\mu_0} \nabla \times {\B}) \cdot (\nabla \times {\B} ) \right) dV , \\
   = & ~ \frac{1}{\mu_0} \int_{\Omega}  \left(  (  {\V} \times {\B}) \cdot (\nabla \times {\B} ) \right) dV - \frac{1}{\mu_0^2} \int_{\Omega} \eta |\nabla \times {\B}|^2 dV , \label{eqn:C6} \\ 
   = & - \frac{1}{\mu_0} \int_{\Omega}  \left(  (\nabla \times {\B} ) \times {\B}) \cdot {\V} \right) dV - \frac{1}{\mu_0^2} \int_{\Omega} \eta |\nabla \times {\B}|^2 dV, \label{eqn:C7} \\ 
   = & ~ \nu n_0 m_i \int_{\Omega}  \left(  {\V} \cdot \nabla^2 {\V} \right) dV - \frac{1}{\mu_0^2} \int_{\Omega} \eta |\nabla \times {\B}|^2 dV , \\ 
   = & - \nu n_0 m_i \int_{\Omega} |\nabla {\V}|^2 dV + \nu n_0 m_i \oint_{\partial \Omega}  \left( {\V} \cdot (\nabla {\V} \cdot \mathbf{n}) \right) dS - \frac{1}{\mu_0^2} \int_{\Omega} \eta |\nabla \times {\B}|^2 dV , \\ 
   = & - \nu n_0 m_i \int_{\Omega} |\nabla {\V}|^2 dV - \frac{1}{\mu_0^2} \int_{\Omega} \eta |\nabla \times {\B}|^2 dV , \\
   \leq & ~ 0 .
\end{align}
Here we drop the boundary integral terms assuming the proper boundary condition is imposed. 

\bibliographystyle{elsarticle-num} 
\bibliography{cas-refs}

\end{document}